\documentclass[12pt,epsf,epsfig,psfig]{article}
\usepackage{graphicx}
\usepackage{epsfig}
\def\J{$J/\psi$}
\def\j{J/\psi}
\def\X{$\chi_c$}
\def\x{\chi}
\def\P{$\psi'$}
\def\p{\psi'}
\def\U{$\Upsilon$}

\def\C{c{\bar c}}

\def\q{q{\bar q}}
\def\Q{Q{\bar Q}}
\def\e{\epsilon}
\def\t{\tau}
\def\l{\Lambda_{\rm QCD}}

\def\S{S_{\C}}

\def\NP{{ Nucl.\ Phys.\ }}
\def\PL{{ Phys.\ Lett.\ }}
\def\PR{{ Phys.\ Rev.\ }}

\def\PRL{{ Phys.\ Rev.\ Lett.\ }}

\def\ZP{{ Z.\ Phys.\ }}
\def\EPJ{{Eur.\ Phys.\ J.\ }}

\def\be{\begin{equation}}
\def\ee{\end{equation}}
\def\lsim{\raise0.3ex\hbox{$<$\kern-0.75em\raise-1.1ex\hbox{$\sim$}}}
\def\gsim{\raise0.3ex\hbox{$>$\kern-0.75em\raise-1.1ex\hbox{$\sim$}}}

\begin{document}

BI-TP 2005/51r~\hfill 1.\ 3.\ 06.

\vskip 2cm

\centerline{\Large \bf Colour Deconfinement and Quarkonium Binding}

\vskip1cm

\centerline{\large \bf Helmut Satz} 

\vskip 0.7cm

\medskip

\centerline{Fakult\"at f\"ur Physik, Universit\"at Bielefeld}

\centerline{Postfach 100 131, D-33501 Bielefeld, Germany}
 
\medskip

\centerline{and}

\medskip

\centerline{Centro de F{\'i}sica Te\'orica de Part{\'i}culas (CFTP)}

\centerline{Instituto Superior T\'ecnico, Av. Rovisco Pais, P-1049-001 Lisboa, 
Portugal}

\vskip1cm

\centerline{\bf Abstract:}

\vskip0.5cm

At high temperatures, strongly interacting matter becomes a plasma of
deconfined quarks and gluons. In statistical QCD, deconfinement and the 
properties of the resulting quark-gluon plasma can be investigated by 
studying the in-medium behaviour of heavy quark bound states. In high 
energy nuclear interactions, quarkonia probe different aspects of the 
medium formed in the collision. We survey the results of recent
charmonium production studies in SPS and RHIC experiments.

\vskip1.5cm

\centerline{\bf Contents:}

\leftskip4cm

\vskip1cm

1.\ Introduction

\medskip

2.\ Colour Deconfinement

\medskip

3.\ Quarkonium Binding and Dissociation

\medskip

4.\ Thermal Quarkonium Dissociation  

\medskip

5.\ Quarkonium Production in Nuclear Collisions

\medskip

6.\ Experimental Results from Nucleus-Nucleus Collisions

\medskip

7.\ Outlook

\leftskip0cm

\newpage

\noindent{\large \bf 1.\ Introduction}

\bigskip

The states of matter, their defining features and the transitions between
them have always been among the most challenging problems of physics.
Strongly interacting matter opens a new chapter for such studies. The
elementary particles of strong interaction physics, the hadrons, make up 
most of the observable matter in the universe. They have intrinsic sizes and 
masses, and the ultimate aim of the theory of strong interactions is to 
derive these. Quantum chromodynamics (QCD) does that, in terms of
massless gluons and almost massless quarks -- it is their {\it dynamics} 
which generates the scales. Thus the nucleon is a bound state of three 
very light quarks; the kinetic and potential energy of the system provide 
the nucleon mass, and the binding potential determines the nucleon radius.

\medskip

The interaction between quarks is based on their
intrinsic colour charge, just as that between electrons and protons or
nuclei is determined by their electric charge. The form of the interaction
is quite different, however. The Coulomb potential vanishes for large
separation distance, so that electric charges can be separated and have
an independent existence. In contrast, the potential between quarks 
increases with separation, so that an infinite energy would be needed
to isolate a quark. In other words, the quark constituents of a hadron
are {\sl confined} and not just bound. 

\medskip 

Strong interaction {\sl thermodynamics} shows that quark confinement has its
inherent limits. As bound states of quarks, hadrons have an intrinsic size,
the binding radius. At extreme density, when several hadrons are compressed
into a spatial volume normally occupied by a single hadron, it becomes
impossible to identify among all the overlapping states a specific 
quark-antiquark pair or quark triplet as belonging to a certain hadron. 
The medium becomes instead a dense multi-quark environment in
which any quark can move as far as it wants, since it is always close 
enough to other quarks to satisfy confinement conditions. Confinement as
a large distance feature thus loses its meaning in the short distance
world of matter at extreme density. 

\medskip

It is therefore expected that with 
increasing temperature, strongly interacting matter will undergo a transition 
from a hadronic phase, in which the constituents are colour-neutral bound 
states, to a plasma of deconfined colour-charged quarks and gluons. This
deconfinement transition in QCD is  quite similar to the
insulator-conductor transition in normal condensed matter physics:
it corresponds to the onset of colour conductivity.

\medskip

The thermodynamics of strongly interacting matter can be derived through
first principle calculations in {\sl finite temperature lattice QCD}, and 
we shall return to these studies in the next section. They provide
quite detailed information on the temperature dependence of thermodynamic 
observables, such as pressure or energy density, and they show how these
quantities change as the system passes through the deconfinement transition.
This raises a crucial question: if we were given a box containing strongly
interacting matter, how could we determine the temperature of the medium
and how could we specify the state in which this medium is in? This question
is the central topic of our survey, and we will show that the in-medium
behaviour of {\it heavy quark} bound states provides an excellent tool to 
answer it.

\medskip
  
Since some thirty years it is known that besides the almost massless 
{\it up} and $down$ quarks of the everyday world, and the still relatively 
light {\it strange} quarks required to account for the strange mesons and 
hyperons observed in hadron collisions, there are heavy quarks at the other 
end of the scale, whose bare masses alone are larger than those of most of 
the normal hadrons. These heavy quarks first showed up in the discovery of 
the \J ~meson \cite{Ting}, of mass of 3.1 GeV; it is a bound state of a 
{\it charm} quark ($c$) and its antiquark ($\bar c$), each having a mass 
of some 1.2--1.5 GeV. On the next level there is the \U~meson \cite{Leder}, 
with a mass of about 
9.5 GeV, made up of a {\it bottom} or {\it beauty} 
quark-antiquark pair ($b \bar b$), with each quark here having a mass 
around 4.5 - 4.8 GeV. Both charm and bottom quarks can of course also 
bind with normal light quarks, giving rise to open charm ($D$) and open 
beauty ($B$) mesons. The lightest of these `light-heavy' mesons have 
masses of about 1.9 GeV and 5.3 GeV, respectively. 

\medskip

The bound states of a heavy quark $Q$ and its antiquark $\bar Q$ are 
generally referred to as quarkonia. Besides the initially discovered 
vector ground states \J~and \U, both the $c \bar c$ and the $b \bar b$ 
systems give rise to a number of other {\it stable} bound states of 
different quantum numbers. They are stable in the sense that their mass 
is less than that of two light-heavy mesons, so that strong decays are 
forbidden. The measured stable charmonium spectrum contains the $1S$ 
scalar $\eta_c$ and vector \J, three $1P$ states \X~(scalar, vector 
and tensor), and the $2S$ vector state \P, whose mass is just below 
the open charm threshold. There are further charmonium states above the 
\P; these can decay into $D \bar D$ pairs, and we shall here restrict 
our considerations only to quarkonia stable under strong interactions.
The observed stable charmonium states are summarized in table 1
and the corresponding bottomonium ($b~\bar b$) states in table 2. The binding 
energies $\Delta E$ listed there are the differences between the quarkonium 
masses and the open charm or beauty threshold, respectively.

\vskip0.5cm

\centerline{
\renewcommand{\arraystretch}{1.8}
\begin{tabular}{|c|c|c|c|c|c|c|}
\hline
{\rm state}& $\eta_c$ & $J/\psi$ & $\chi_{c0}$ & 
 $\chi_{c1}$ &  $\chi_{c2}$ & $\psi'$ \\
\hline
{\rm mass~[GeV]}&
2.98&
3.10&
3.42&
3.51&
3.56&
3.69 \cr
\hline
$\Delta E$ {\rm[GeV]}&
0.75&
0.64&
0.32&
0.22&
0.18&
0.05\cr
\hline
\end{tabular}}

\vskip0.5cm

\centerline{~~Table 1: Charmonium states and binding energies}

\vskip0.8cm

\centerline{
\renewcommand{\arraystretch}{1.8}
\begin{tabular}{|c|c|c|c|c|c|c|c|c|c|}
\hline
{\rm state}
& $\Upsilon$ 
& $\chi_{b0}$
& $\chi_{b1}$ 
& $\chi_{b2}$ 
& $\Upsilon'$
& $\chi'_{b0}$ 
& $\chi'_{b1}$ 
& $\chi'_{b2}$ 
& $\Upsilon''$ \\
\hline
{\rm mass~[GeV]}&
9.46&
9.86&
9.89&
9.91&
10.02&
10.23&
10.26&
10.27&
10.36\cr
\hline
$\Delta E$ {\rm[GeV]}&
1.10&
0.70&
0.67&
0.64&
0.53&
0.34&
0.30&
0.29&
0.20\cr
\hline
\end{tabular}}

\bigskip

\centerline{~~Table 2: Bottomonium states and binding energies}

\vskip0.5cm

Quarkonia are rather unusual hadrons. The masses of the light hadrons,
in particular those of the non-strange mesons and baryons, arise almost
entirely from the interaction energy of their nearly massless quark 
constituents. In contrast, the quarkonium masses are largely determined
by the bare charm and bottom quark masses. These large quark masses allow 
a very straightforward calculation of many basic quarkonium properties, 
using non-relativistic potential theory. It is found that, in particular,
the ground states and the lower excitation levels of quarkonia are very
much smaller than the normal hadrons, and that they are very tightly bound.
Now deconfinement is a matter of scales: when the separation between normal 
hadrons becomes much less than the size of these hadrons, they melt to form 
the quark-gluon plasma. What happens at this point to the much smaller 
quarkonia?  When do they become dissociated?
That is the central theme of this report seen from another angle.    
We shall show that the disappearance of specific quarkonia signals the 
presence of a deconfined medium of a specific temperature \cite{M-S}.
Thus the study of the quarkonium spectrum in a given medium 
is akin to the spectral analysis of stellar media, where the
presence or absence of specific excitation lines allows a determination
of the temperature of the stellar interior.

\medskip

The prediction of a new state of strongly interacting matter led quite
naturally to the question of where and how to observe it. The
very early universe is an obvious case: according to the usual evolution
equations, giving its energy density as function of age, 
it must have been a quark-gluon plasma in the first 10 $\mu s$ after
the big bang. The core of neutron stars is another possibility, and
they may in fact consist of deconfined quarks. But the essential
boost for the field came from the idea to use high energy nuclear
collisions to form small and short-lived bubbles of the quark-gluon plasma
in the laboratory, making them accessible to terrestial experimental
study. This idea clearly contains assumptions to be verified, and
we shall return to its basis later on. Here we only note that again
an absolutely crucial question is how to probe the thermal conditions
of the medium formed in such collisions. In view of the evolution
of such a medium, this is a much more complex question than that 
addressed in the case of a box containing equilibrated strongly interacting
matter. Nevertheless, as we shall see, quarkonia may well provide a
tool for this more general task as well.

\medskip

After this introduction, we shall first review the essential aspects 
of deconfinement (section 2.) and then summarize the basic properties of
quarkonia and the dynamics of their dissociation (section 3.). Following 
this, we survey different approaches to quarkonium binding in QCD 
thermodynamics (section 4.). We then review the main theoretical
features of quarkonium production in elementary as well as $p-\!A$ and
nucleus-nucleus collisions (section 5.), and finally we try to see what 
one can learn from the present experimental results (section 6.).

\vskip0.5cm

\noindent{\large \bf 2.\ Colour Deconfinement}

\bigskip

\begin{figure}[h]
\vskip0.5cm
\begin{minipage}[t]{8cm}
\hskip0.1cm 
\epsfig{file=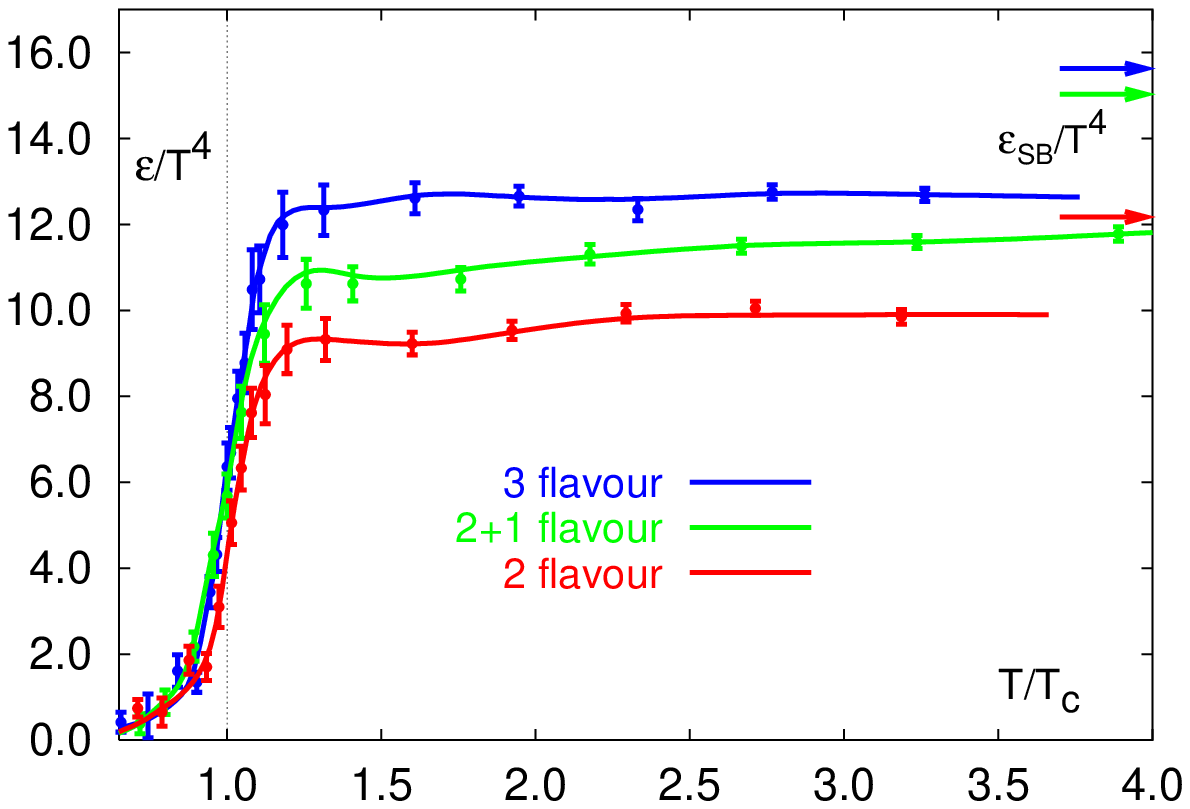,width=7.5cm}
\caption{Energy density vs.\ $T$ \cite{K-L-P,schlad}}
\label{edens}
\end{minipage}
\begin{minipage}[t]{8cm}
\vspace*{-5.3cm}
\hskip0.3cm
\epsfig{file=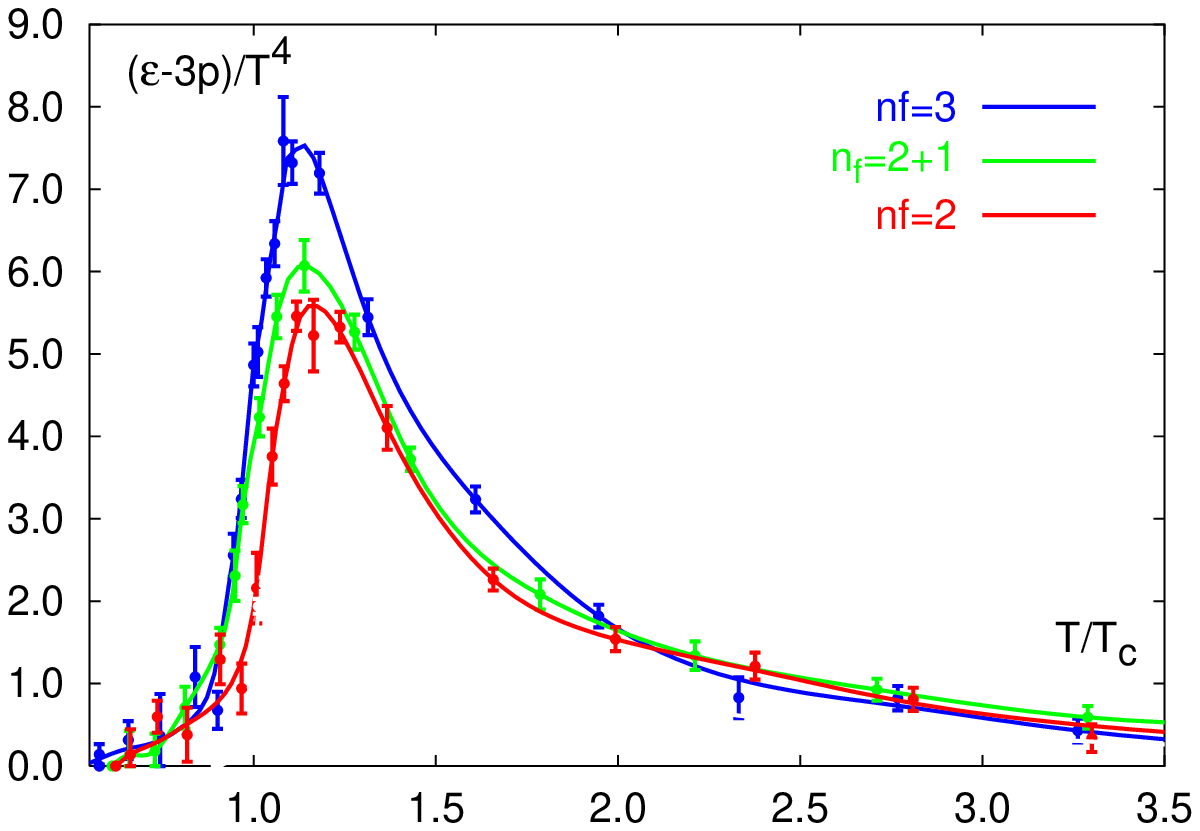,width=7.5cm}
\caption{$\Delta=(\e - 3P)/T^4$ vs.\ $T$ \protect\cite{delta}} 
\label{inter}
\end{minipage}\end{figure}

The transition from hadronic matter to a plasma of deconfined quarks and 
gluons has been studied extensively in finite temperature lattice QCD.
We shall consider here the case of vanishing baryon-number density (baryons
and antibaryons in equal numbers) and comment only briefly on the situation
at finite baryon density at the end of this section.
At the transition point, the energy density of the system increases by 
the latent heat of deconfinement, i.e., it grows from a value determined 
by the degrees of freedom of a hadron gas to a much higher one governed 
by the degrees of freedom of a quark-gluon plasma. To illustrate this, 
we recall that the energy density of an ideal gas of massless pions is 
\be
\e_{\pi} = 3~{\pi^2 \over 30}~ T^4 \simeq ~ T^4,
\label{piongas}
\ee
corresponding to the three possible pion charges, while an ideal 
quark-gluon plasma, with two massless quark flavours, gives
\be
\e_{QGP} = \{~\!2\times 8 + {7 \over 8}~ [ 2 \times 2 \times 2 \times 3 ]~\!\}
~ {\pi^2 \over 30}~T^4 = 37~{\pi^2 \over 30}~ T^4 \simeq ~12~T^4, 
\label{qgp}
\ee
as determined by the 16 gluonic and 24 quark-antiquark degrees of freedom. 
Hence in the vicinity of $T=T_c$, the energy density increases by more than 
a factor ten. The actual behaviour, as obtained in finite temperature 
lattice QCD for different flavour compositions \cite{K-L-P,schlad}, 
is shown in Fig.\ \ref{edens}.

\vskip-0.5cm

Above $T_c$, the medium consists of deconfined quarks and gluons.
We emphasize that deconfinement does not imply the absence of 
interaction -- it is only the requirement to form colour neutral bound 
states that has been removed. In Fig.\ \ref{inter} it is seen that 
up to $T \sim 3 ~T_c$ the `interaction measure' $\Delta=(\e - 3P)/T^4$ 
remains sizeable and does not vanish \cite{delta}, as it would for an 
ideal gas of 
massless constituents. This strong interaction in the quark-gluon plasma
has been interpreted in various ways, as gradual onset of deconfinement 
starting from high momenta \cite{interaction}, as difference
between physical and perturbative vacua \cite{A-H}, and even
in terms of a large number of coloured ``resonance'' states 
\cite{Shuryak}. Recent studies of the relevant degrees of freedom in the 
quark-gluon plasma \cite{degree-f} seem to indicate, however, that
a description based on quarks having some kind of thermal mass is
more likely to be correct. 

\medskip

To specify the nature of the critical behaviour observed at the transition, 
recall that in the limit of large quark mass ($m_q\to \infty$) we
recover pure gauge theory and the average value of the Polyakov loop, 
\be
L(T) \sim \exp\{-F_{Q\bar Q}/T\},
\label{polya}
\ee
serves as order parameter \cite{L}. Here 
\be
F_{Q\bar Q} = \lim_{r\to \infty} F_{Q\bar Q}(r,T)
\label{F}
\ee
denotes the large distance limit of the free energy of a heavy quark-antiquark 
pair, which diverges in the confined regime and becomes finite through
colour screening in the quark-gluon plasma, 
\be
L(T) 
~\left\{\matrix{=0 & T \leq T_L~~{\rm confinement}\cr
~&~\cr
\not= 0 &  ~~T > T_L~~{\rm deconfinement}\cr}
\right.
\label{confinement}
\ee

\medskip

This defines a critical deconfinement temperature $T_L$, separating the
confined and deconfined phases. The underlying mechanism for such behaviour 
is a global $Z_3$ invariance of the Lagrangian of pure $SU(3)$ gauge theory, 
which becomes spontaneously broken in the deconfined phase, for $T \geq T_L$.  

\medskip

In the other extreme, the chiral limit of massless quarks ($m_q \to 0$),
the QCD Lagrangian is chirally symmetric, and the chiral condensate
\be
\chi(T) \equiv \langle \bar{\psi} \psi \rangle 
\label{chiralcond}
\ee
serves as order parameter. At low temperatures, the massless quarks
acquire an effective mass by dressing with gluons, so that the chiral 
symmetry is spontaneously broken. At high temperatures, thermal motion
removes the dressing, so that the behaviour of $\chi(T)$,
\be
\chi(T) 
~\left\{\matrix{\not=0 & T<T_{\chi}~~ {\rm  chiral~symmetry~broken}\cr
~&~\cr
=0 & ~T \geq T_{\chi} ~~{\rm chiral~symmetry~restored,}\cr}
\right.
\label{chiral}
\ee
defines a chiral restoration temperature $ T_{\chi}$ separating the
phases of broken and restored chiral symmetry.

\medskip

\begin{figure}[htb]
\vspace*{-1.8cm}
\mbox{
\epsfig{file=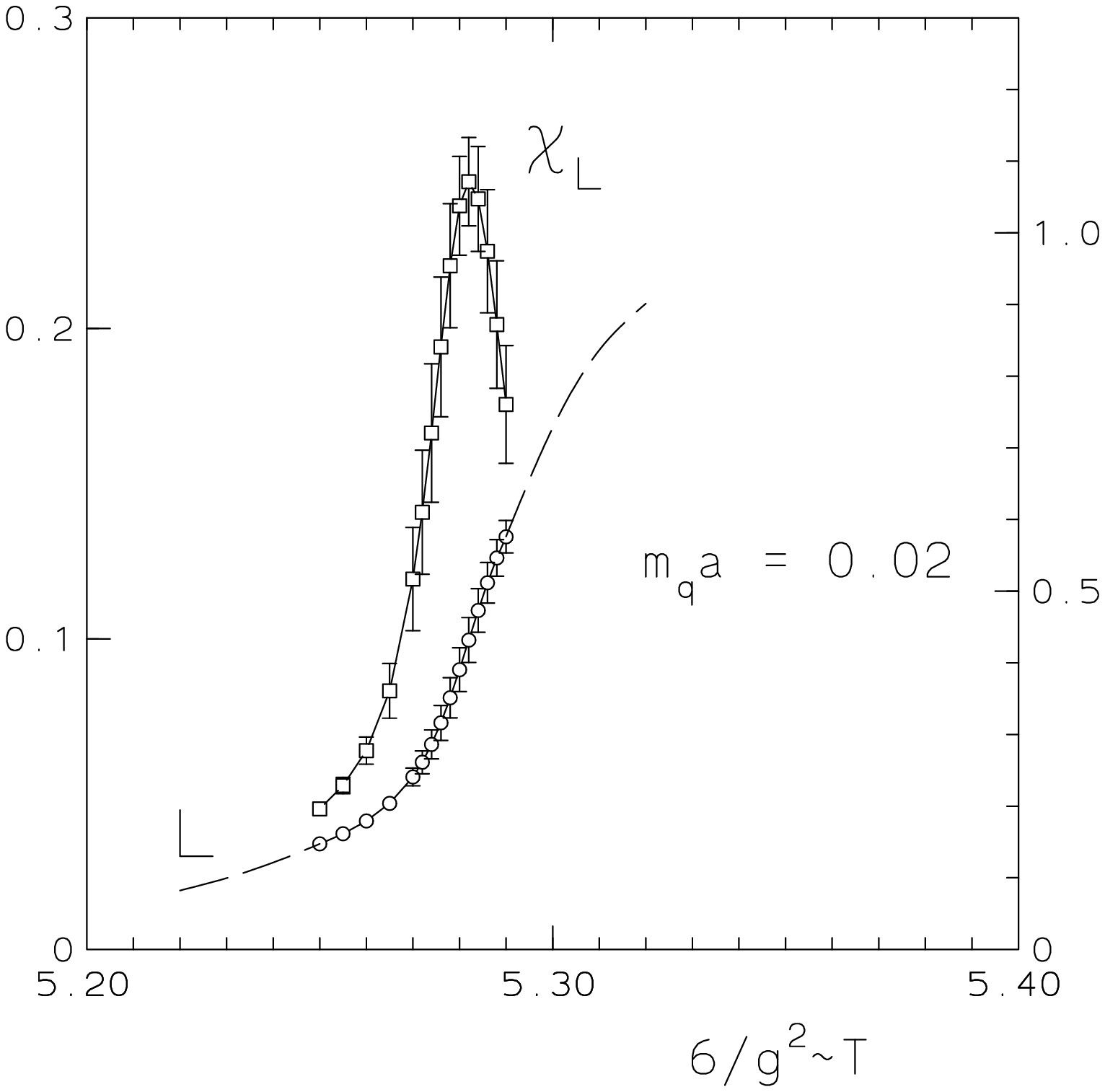,width=7.5cm,height=9.5cm}
\hskip0.5cm
\epsfig{file=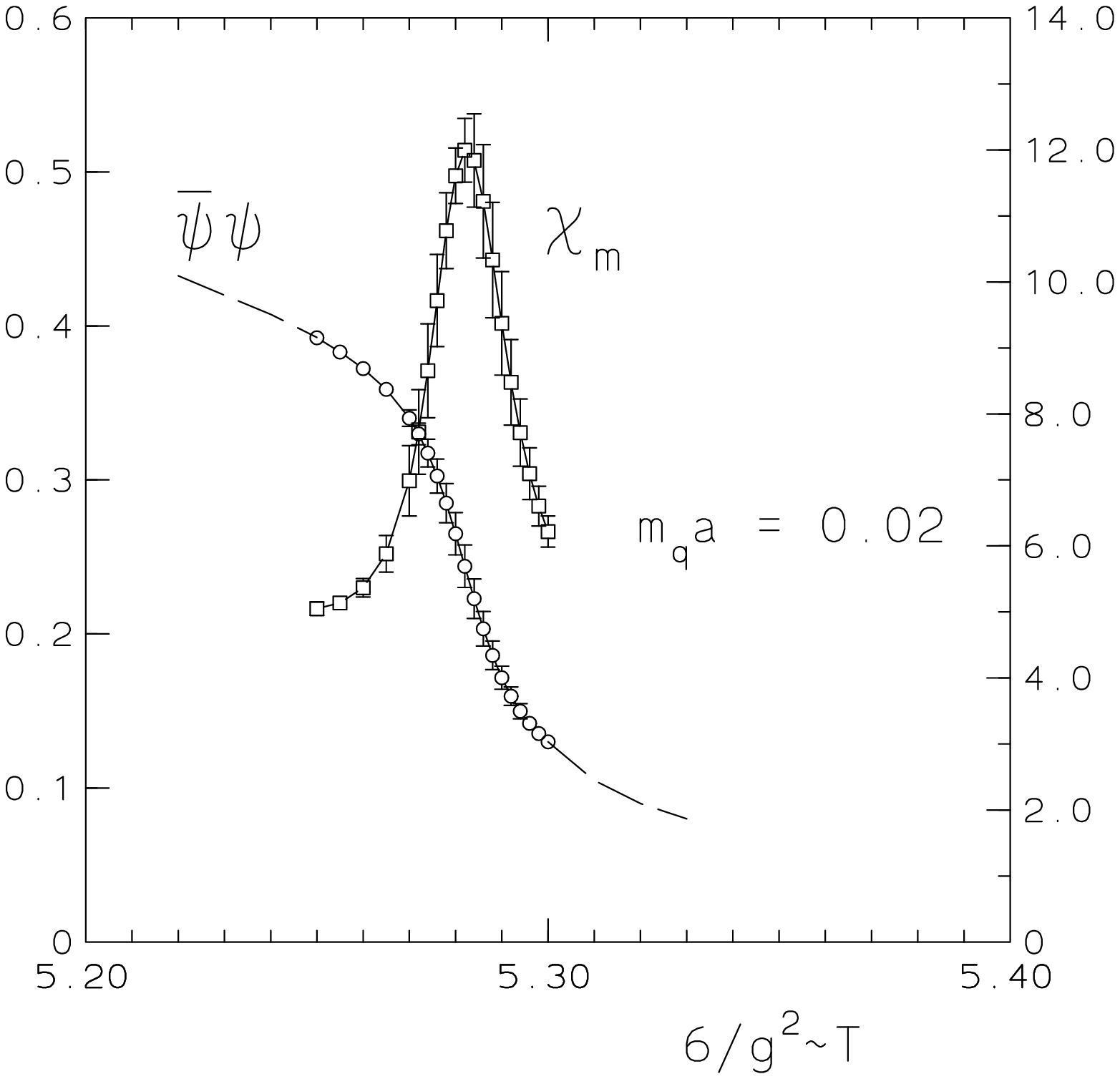,width=7.5cm, height=9.5cm}}
\vspace*{-1.9cm}
\caption{Polyakov loop and chiral condensate vs.\ temperature 
\protect\cite{K-L}}
\label{rapid}
\end{figure}

While there thus exists well-defined thermal critical behaviour in the
two limits of large and small quark mass, the QCD Lagrangian has for
general finite quark masses neither a global $Z_3$ (deconfinement) nor
a chiral symmetry. Nevertheless it is found that both the Polyakov
loop and the chiral condensate continue to vary rapidly in the same narrow
temperature interval, as shown in Fig.\ \ref{rapid} for the case of two
relatively light quark flavours \cite{K-L}. We can thus 
still identify an ``almost critical'' behaviour, the so-called ``rapid
cross-over''. Since this occurs for $L(T)$ and $\chi(T)$ at the
same temperature, deconfinement and chiral symmetry restoration are said
to coincide.

\begin{figure}[htb]
\centerline{\epsfig{file=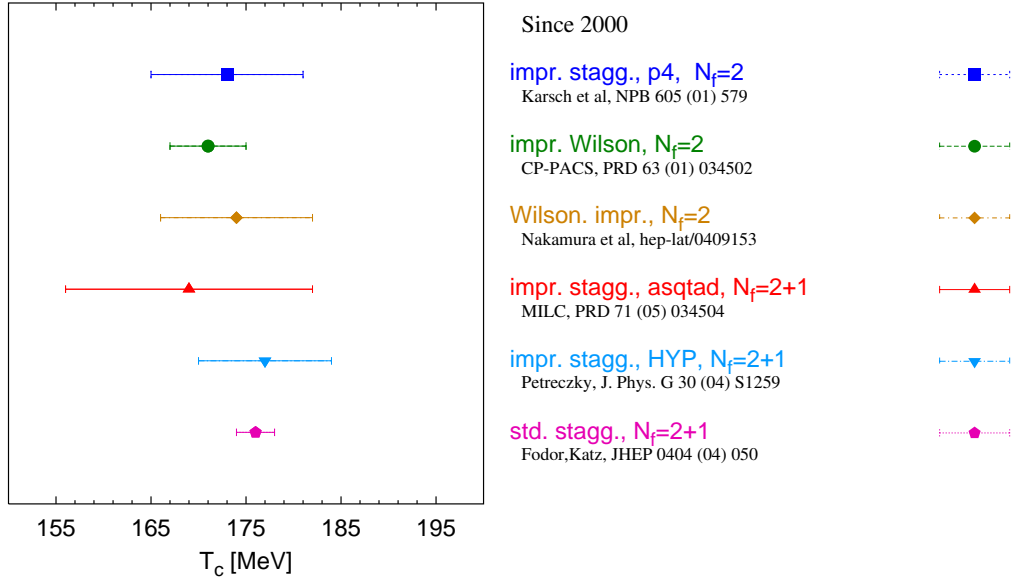,width=8cm,angle=-90}}
\caption{Critical temperature results from different lattice 
studies \protect\cite{Peter-Moriond}}
\label{tc}
\end{figure}

The ``critical temperature'' determined in this way depends on 
the masses and the number of quark flavours. For two light as well as 
for two light plus one heavier quark flavour, most studies 
\cite{Peter-Moriond} indicate $T_c = 175 \pm 10$ MeV, as shown in 
Fig.\ \ref{tc}, although some indications for a higher value were 
recently reported \cite{katz}.

\medskip

\begin{figure}[htb]
\centerline{\epsfig{file=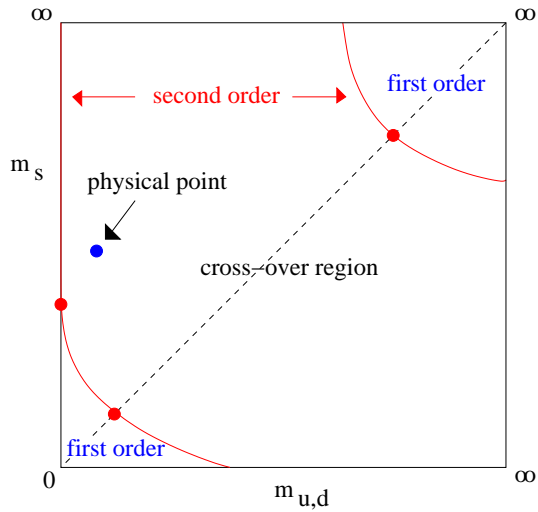,width=7cm}}
\caption{Critical behaviour as function of quark masses} 
\label{nature}
\end{figure}

The precise form of critical behaviour in the limiting cases also
depends on the number of colours and flavours. The situation isillustrated 
in Fig.\ \ref{nature} for colour $SU(3)$ and three
quark flavours ($u,d,s$). In the gauge theory limit $m_q \to \infty$
for all flavours, the transition is discontinuous (first order). 
Decreasing the quark mass decreases the discontinuity, until it
vanishes on a line of second order transitions, beyond which there
is the rapid cross-over region. For three massless flavours, the
chiral transition is also first order, and increasing the quark mass
leads again to a decreasing discontinuity and a line of second order
transitions. For two massless quark flavours and a sufficiently massive
third flavour, the transition is of second order up to the two-flavour
limit. The ``physical point'' corresponding to two light $u$ and $d$ 
quarks and one heavier $s$ quark has now been established with some
certainty and appears to fall into the cross-over region. 

\begin{figure}[htb]
\centerline{\epsfig{file=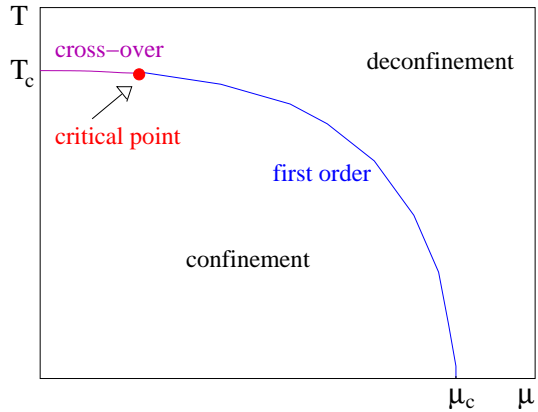,width=7cm}}
\caption{Critical behaviour for two light quark flavours,
as function of $T$ and $\mu$}
\label{baryo}
\end{figure}

The phase structure at non-vanishing baryon density has stimulated
much interest \cite{stephanov}, but it encounters 
calculational problems in the computer simulation. Quite recently,
a number of methods to overcome these have been put forward 
\cite{mub}, and today we believe that the phase diagram in $T, \mu$ has 
the form shown in Fig.\ \ref{baryo},
with $\mu$ denoting the baryochemical potential. 

\medskip

We can thus conclude this section by noting that the thermodynamics of
strongly interacting matter is today known in considerable quantitative
detail, with deconfinement for $\mu=0$
setting in at $T_c=175 \pm 10$ MeV. The
actual nature of the ``transition'' remains, however, quite enigmatic,
since it occurs as ``rapid cross-over'' and not as a genuine thermal
phase transition. Further clarification could come from studies in
the context of geometric critical behaviour, relating the onset of
deconfinement to cluster formation and percolation. 

\medskip

Hence we know that confined hadronic and deconfined quark-gluon states of
matter exist. How can we probe strongly interacting matter through 
specific observables, in order to determine its thermal parameters   
(temperature, energy density) as well as to identify its confinement
status? Since we want to show that this can be done with the help of
quarkonium states, we first summarize in the next section the 
essentials of quarkonium binding and break-up. 

\vskip0.5cm

\noindent{\large \bf 3.\ Quarkonium Binding and Dissociation}

\bigskip

We had defined quarkonia as bound states of heavy quarks which are 
stable under strong decay, i.e., $M_{c\bar c} \leq 2 M_D$ for charmonia 
and $M_{b\bar b} \leq 2 M_B$ for bottomonia. Since the quarks are heavy,
with $m_c \simeq 1.2 - 1.5 $ GeV for the charm and $m_b \simeq 4.5 - 4.8$ 
GeV for the bottom quark, quarkonium spectroscopy can be studied quite well 
in non-relativistic potential theory. The simplest (``Cornell'') confining
potential \cite{Cornell} for a $\Q$ at separation distance $r$ has the form
\be
V(r) = \sigma ~r - {\alpha \over r}
\label{cornell}
\ee
with a string tension $\sigma \simeq 0.2$ GeV$^2$ and a Coulomb-like term
with a gauge coupling $\alpha \simeq \pi/12$. The corresponding 
Schr\"odinger equation
\be
\left\{2m_c -{1\over m_c}\nabla^2 + V(r)\right\} \Phi_i(r) = M_i \Phi_i(r)
\label{schroedinger}
\ee
then determines the bound state masses $M_i$ and the wave functions 
$\Phi_i(r)$, and with
\be
\langle r_i^2 \rangle = \int d^3r~ r^2 |\Phi_i(r)|^2 / 
\int d^3r~|\Phi_i(r)|^2 .
\label{radii}
\ee
the latter in turn provide the (squared) average bound state radii.

\medskip

To obtain a first idea of the results, we consider the semi-classical 
solution obtained by making use of the uncertainty relation 
$ \langle p^2 \rangle \langle r^2 \rangle  
\simeq 1$ in the expression for the $\Q$ energy 
\be
E(r) = 2m + {p^2 \over m} + V(r) \simeq 2m + {1 \over m r^2} + V(r).
\label{semi1}
 \ee
Minimizing $E(r)$ with respect to $r$, $dE/dr = 0 $, gives 
\be
r_0 \simeq 0.44~{\rm fm}
~~{\Rightarrow}~~ M_0 = E(r_0) \simeq 3.1~ {\rm GeV}
\label{semi-c}
\ee
for the generic charmonium ground state, and
\be
r_0 \simeq 0.33~{\rm fm}
~~{\Rightarrow}~~ M_0 = E(r_0) \simeq 9.6~ {\rm GeV}
\label{semi-b}
\ee
and for the corresponding bottomonium state. Here $r_0$ specifies the
$\Q$ separation distances, i.e., it gives twice the radius. From Tables 1 
and 2 we see that these first estimates already reproduce quite well the 
spin-averaged ground states.

\medskip

The exact solution of eq.\ (\ref{schroedinger}) gives in fact a very good 
account of the full (spin-averaged) quarkonium spectroscopy, as seen in
Table 3 \cite{D-K-S}. The line labelled $\Delta M$ shows the differences 
between the experimental and the calculated values; they are in all cases 
less than 1 \%. Again $r_0$ gives the $\Q$ separation for the state in 
question. The input parameters for these results are $m_c=1.25$ GeV,
$m_b=4.65$ GeV, $\sqrt \sigma = 0.445$ GeV, $\alpha=\pi/12$.

\vskip0.5cm

\hskip1.5cm
\renewcommand{\arraystretch}{1.8}
\begin{tabular}{|c|c|c|c|c|c|c|c|c|}
\hline
{\rm state}& $J/\psi$ & $\chi_c$ & $\psi'$  & $\Upsilon$
 & $\chi_b$ & 
$\Upsilon'$ & $\chi_b'$ & $\Upsilon''$ \\
\hline
{\rm mass~[GeV]}&
3.10&
3.53&
3.68&
9.46&
9.99&
10.02&
10.26&
10.36 \\
\hline
$\Delta E$ {\rm[GeV]}&0.64&0.20&0.05 &1.10&
0.67&0.54&0.31&0.20 \cr
\hline
$\Delta M$ {\rm[GeV]}&0.02&-0.03&0.03 & 0.06&
-0.06&-0.06&-0.08&-0.07 \cr
\hline
{$r_0$ \rm [fm]}&0.50&0.72&0.90&
0.28&0.44& 0.56& 0.68 &0.78 \cr
\hline
\end{tabular}

\bigskip

\centerline{Table 3:
Quarkonium Spectroscopy from Non-Relativistic Potential Theory \cite{D-K-S}}

\bigskip

We thus see that in particular the \J~ and the lower-lying bottomonium
states are very tightly bound ($ 2M_{D,B} - M_0 \gg \l)$ and of very
small spatial size ($r_0 \ll 2 r_h \simeq 2$ fm). Through what kind of
interaction dynamics can they then be dissociated? 

\medskip

\begin{figure}[htb]
\centerline{\epsfig{file=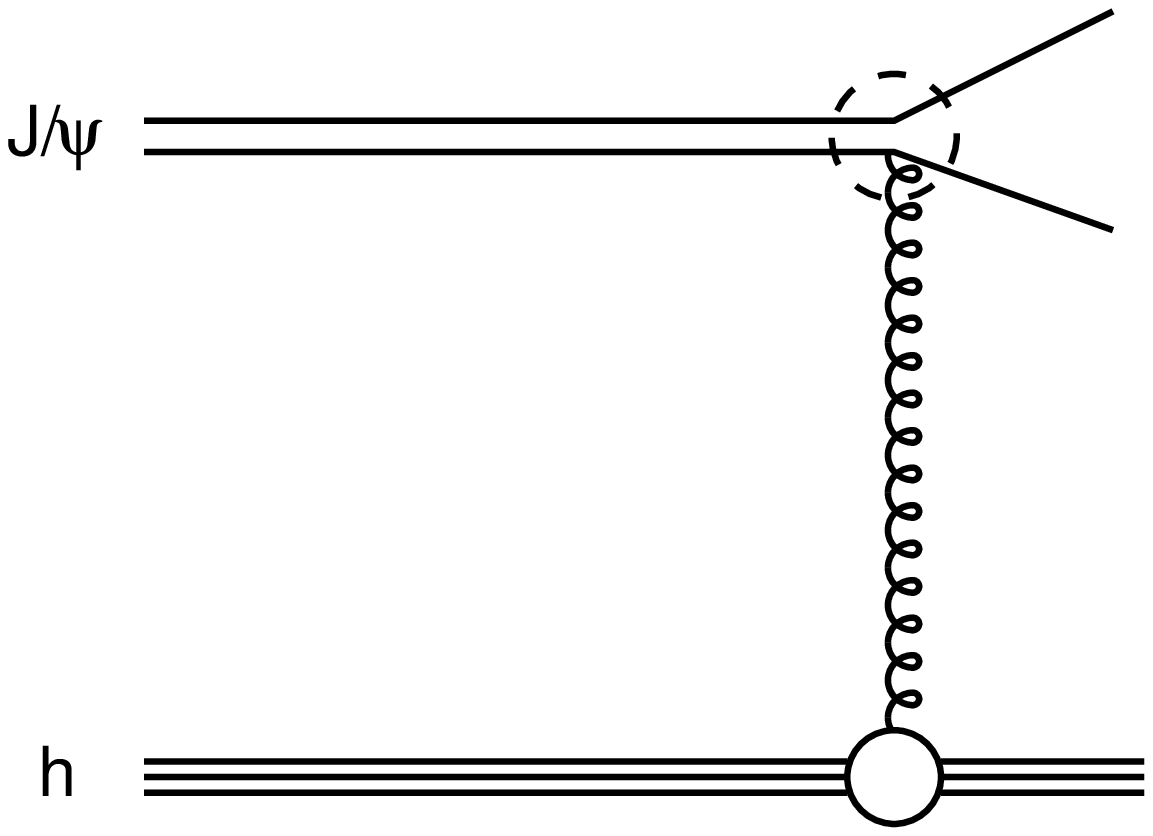,width=5cm}}
\caption{\J-hadron interaction}
\label{hadrodis}
\end{figure}

As illustration, we consider the collision dissociation of a \J. 
It is very small ($r_{\j} \sim 0.25$ fm) and hence can only be resolved
by a sufficiently hard probe. It is moreover tightly bound
($2M_D-M_{\j} \sim 0.6$ GeV), so that only a sufficiently energetic
projectile can break the binding. Hence in a collision with a normal hadron, 
the \J~can only be dissociated by an interaction with a hard gluon 
contituent of the hadron, not with the hadron as a whole (see Fig.\ 
\ref{hadrodis}).  


\medskip

The gluon momentum distribution $g(x)$ in a hadron is determined in deep
inelastic lepton-hadron scattering; with $k_h$ denoting the gluon momentum,
$x = k_h/p_h$ specifies the fraction of the incident hadron momentum $p_h$
carried by the gluon. For mesons, one finds for large momenta
\be
g(x)  \sim (1-x)^3,
\label{gluon-pdf} 
\ee
so that the average momentum of a hadronic gluon is
\be
\langle k \rangle_h = {1\over 5} \langle p_h \rangle.
\label{gluon-mom}
\ee
For thermal hadrons in confined matter, 
$\langle p_h \rangle \sim 3T$, with $T < 175$ MeV,
so that with
\be
\langle k \rangle_h = {3\over 5} T \leq 0.1~{\rm GeV} \ll 0.6~{\rm GeV}
\label{gluonT}
\ee
the gluon momentum is far too low to allow a dissociation of the 
\J. 

\medskip

On the other hand, the average momentum of a deconfined thermal
gluon in a quark-gluon plasma will be
\be
\langle k_g \rangle \simeq 3~T 
\label{gluon-dec}
\ee
and for $T > 1.15~ T_c$, this provides enough energy to overcome
the \J~binding.  We thus expect that a hot deconfined medium can
lead to \J~dissociation, while the gluons available in a confined
medium are too soft to allow this. 

\medskip

\begin{figure}[htb]
\centerline{\epsfig{file=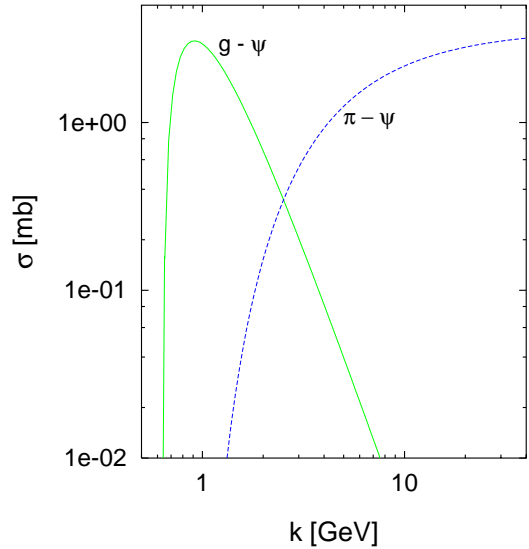,width=7.5cm,angle=-90}}
\caption{Gluon and hadron \J~dissociation cross-sections 
\protect\cite{K-S-OP}}
\label{g-h-dis}
\end{figure}

To make these considerations quantitative, one first has to calculate
the cross-section for the gluon-dissociation of a \J, a QCD analogue
of the photo-effect. This can be carried out using the operator product
expansion \cite{Bhanot,K-S-OP}, and the result is
\be
\sigma_{g-\j} \sim {1\over m_c^2}~{(k/\Delta E_{\psi}-1)^{3/2}\over
 (k/\Delta E_{\psi})^{5}}
\label{gluo-effect}
\ee
with $\Delta E_{\j} = 2M_D - M_{\j}$. The corresponding cross-section
for the hadron dissociation is then obtained by convoluting the
gluon-dissociation cross-section (\ref{gluo-effect}) with the
hadronic gluon distribution function $g(x)$, which for \J-meson
interactions leads to 
\be
\sigma_{h-\j} \simeq \sigma_{\rm geom} (1 - \lambda_0/\lambda)^{5.5}
\label{hadro-J}
\ee
with $\lambda \simeq (s-M_{\psi}^2)/M_{\psi}$ and 
$\!\lambda_0 \simeq (M_h + \Delta E_{\psi}$). Here $\sigma_{\rm geom}
\simeq \pi r_{\j}^2 \simeq 2$ mb is the geometric \J~cross-section and
$M_h$ denotes the mass of the incident meson. In Fig.\ \ref{g-h-dis},
we compare the two dissociation cross-sections (\ref{gluo-effect})
and (\ref{hadro-J}) as function of the incident projectile momentum.


\medskip

This result confirms the qualitative arguments given at the beginning
of this section: typical thermal gluon momenta near 1 GeV produce a large 
dissociation cross-section, whereas hadron momenta in a thermal range (up 
to 2 - 3 GeV) still lead to a vanishingly small cross-section. In other 
words, the \J~should survive in confining media, but become dissociated 
in a hot quark-gluon plasma. 

\medskip

The calculations leading to eq.\ (\ref{gluo-effect}) are exact in the
large quark mass limit $m_Q \to \infty$. It is not clear if the charm 
quark mass really satisfies this condition, and so the dissociation cross 
section in \J-hadron collisions has been discussed in other approaches, 
some of which lead to much larger values \cite{J-dis}. The basic 
question is apparently whether or not the charmonium wave function
has enough overlap with that of the usual hadrons to lead to a sizeable
effect. It may well be that this question is resolvable only
experimentally, by testing if slow charmonia in normal nuclear matter
suffer significant dissociation. Such experiments are definitely
possible \cite{inv-kine}. 
    
\vskip0.8cm

\noindent{\large \bf 4.\ Thermal Quarkonium Dissociation} 

\bigskip

We shall here address the first of our ``basic'' questions: how can we 
identify through measurements the states of matter in a fully equilibrated 
QCD medium? Assume that we are given a box of such strongly interacting
matter, in a stationary state of thermal equilibrium at a given temperature.
Is it possible to specify the state of the given matter and transitions 
between different states in terms of observable quantities calculated in QCD?

\bigskip

\noindent{\bf 4.1 Interaction Range and Colour Screening}

\bigskip

\begin{figure}[htb]
\hspace*{0.1cm}
\centerline{\epsfig{file=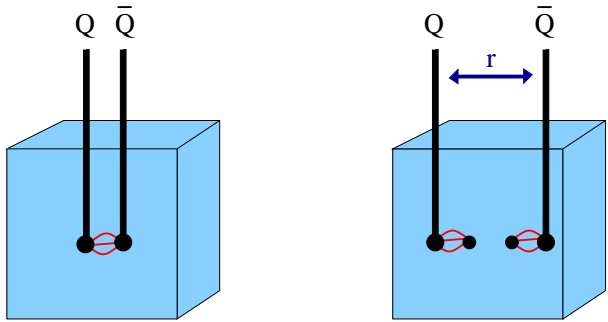,width=8cm}}
\caption{String breaking for a $\Q$ system}
\label{string-b}
\end{figure}

Consider a colour-singlet bound state of a heavy quark $Q$ and its 
antiquark $\bar Q$, put into the medium in such a way that we can measure 
the energy of the system as function of the $\Q$ separation $r$ (see Fig.\ 
\ref{string-b}). The quarks are assumed to be heavy so that they are static 
and any energy changes indicate changes in the binding energy. We consider 
first the case of vanishing baryon density; at $T=0$, the box is therefore 
empty. 

\medskip

\medskip

In vacuum, i.e., at $T=0$, the free energy of the $\Q$ pair is
assumed to have the string form \cite{Cornell} 
\be
F(r) \sim \sigma r
\label{string}
\ee
where $\sigma \simeq 0.16$ GeV$^2$ is the string tension as 
determined in the spectroscopy of heavy quark resonances (charmonium
and bottomonium states). Thus $F(r)$ increases with separation
distance; but when it reaches the value of a pair of dressed
light quarks (about the mass of a $\rho$ meson), it becomes energetically
favorable to produce a $\q$ pair from the vacuum, break the string
and form two light-heavy mesons ($Q\bar q$) and ($\bar Q q$). These
can now be separated arbitrarily far without changing the energy of the
system (Fig.\ \ref{string-b}). 

\medskip

The string breaking energy for charm quarks is found to be
\be
F_0 = 2(M_D - m_c) \simeq 1.2~{\rm GeV};
\label{charm-break}
\ee
for bottom quarks, one obtains the same value,
\be
 F_0 = 2(M_B - m_b) \simeq 1.2~{\rm GeV},
\label{bottom-break}
\ee
using in both cases the quark mass values obtained in the solution
leading to Table 3. Hence the onset of string breaking is evidently 
a property of the vacuum as a medium. It occurs when the two heavy 
quarks are separated by a distance
\be
r_0 \simeq 1.2~{\rm GeV}/\sigma \simeq 1.5~{\rm fm},
\label{stringbreak}
\ee
independent of the mass of the (heavy) quarks connected by the string.

\medskip

\begin{figure}[h]
\hspace*{0.1cm}
\centerline{\epsfig{file=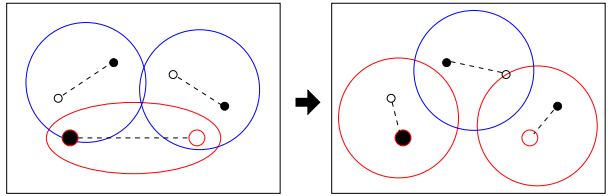,width=8cm}}
\caption{In-medium string breaking through recoupling}
\label{flip}
\end{figure}

If we heat the system to get $T>0$, the medium begins to contain light 
mesons, and the large distance $\Q$ potential $F(\infty,T)$ decreases, since 
we can use these light hadrons to achieve an earlier string breaking through 
a kind of flip-flop recoupling of quark constituents \cite{miya}, resulting 
in an 
effective screening of the interquark force (see Fig.\ \ref{flip}).
Near the deconfinement point, the hadron density increases rapidly, and
hence the recoupling dissociation becomes much more effective, causing
a considerable decrease of $F(\infty,T)$.

A further increase of $T$ will eventually bring the medium to the 
deconfinement point $T_c$, at which chiral symmetry restoration causes 
a rather abrupt drop of the light quark dressing (equivalently, of the 
constituent quark mass), increasing strongly the density of constituents. 
As a consequence, $F(\infty,T)$ now continues to drop sharply.   
Above $T_c$, light quarks and gluons become deconfined colour charges, and 
this quark-gluon plasma leads to a colour screening, which limits the range
of the strong interaction. The colour screening radius $r_D$, which determines 
this range, is inversely proportional to the density of charges, so that
it decreases with increasing temperature. As a result, the $\Q$ interaction
becomes more and more short-ranged. 

\medskip

\begin{figure}[htb]
\hspace*{-0.3cm}
\centerline{\epsfig{file=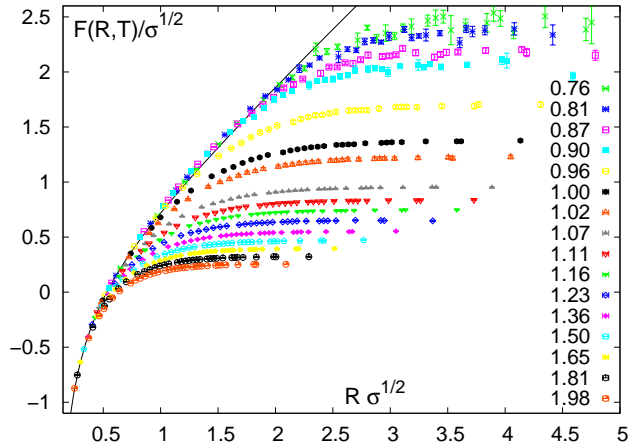,width=9cm}}
\caption{The colour singlet $\Q$ free energy $F(r,T)$ vs.\ $r$ at 
different $T$ \cite{K-Z}}
\label{OKfree}
\end{figure}

In summary, starting from $T=0$, the $\Q$ probe first tests vacuum string 
breaking, then a screening-like
dissociation through recoupling of constituent quarks, and 
finally genuine colour screening. In Fig.\ \ref{OKfree}, we show the 
behaviour obtained in full two-flavour QCD for the colour-singlet
$\Q$ free energy as a function of $r$ for different $T$ \cite{K-Z}. 

\medskip

\begin{figure}[htb]
\mbox{
\hskip1cm
\epsfig{file=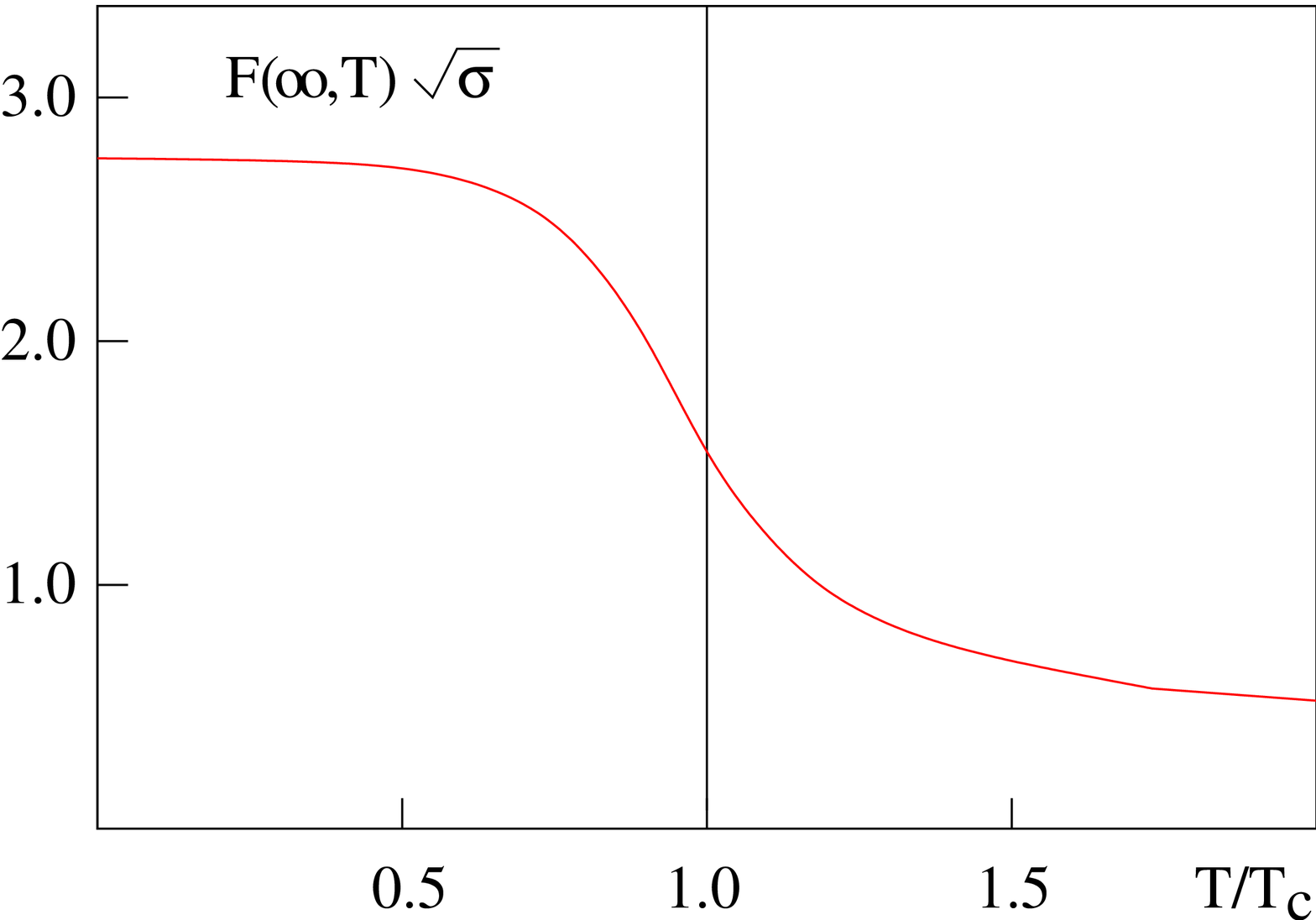,width=5cm}
\hskip3cm
\epsfig{file=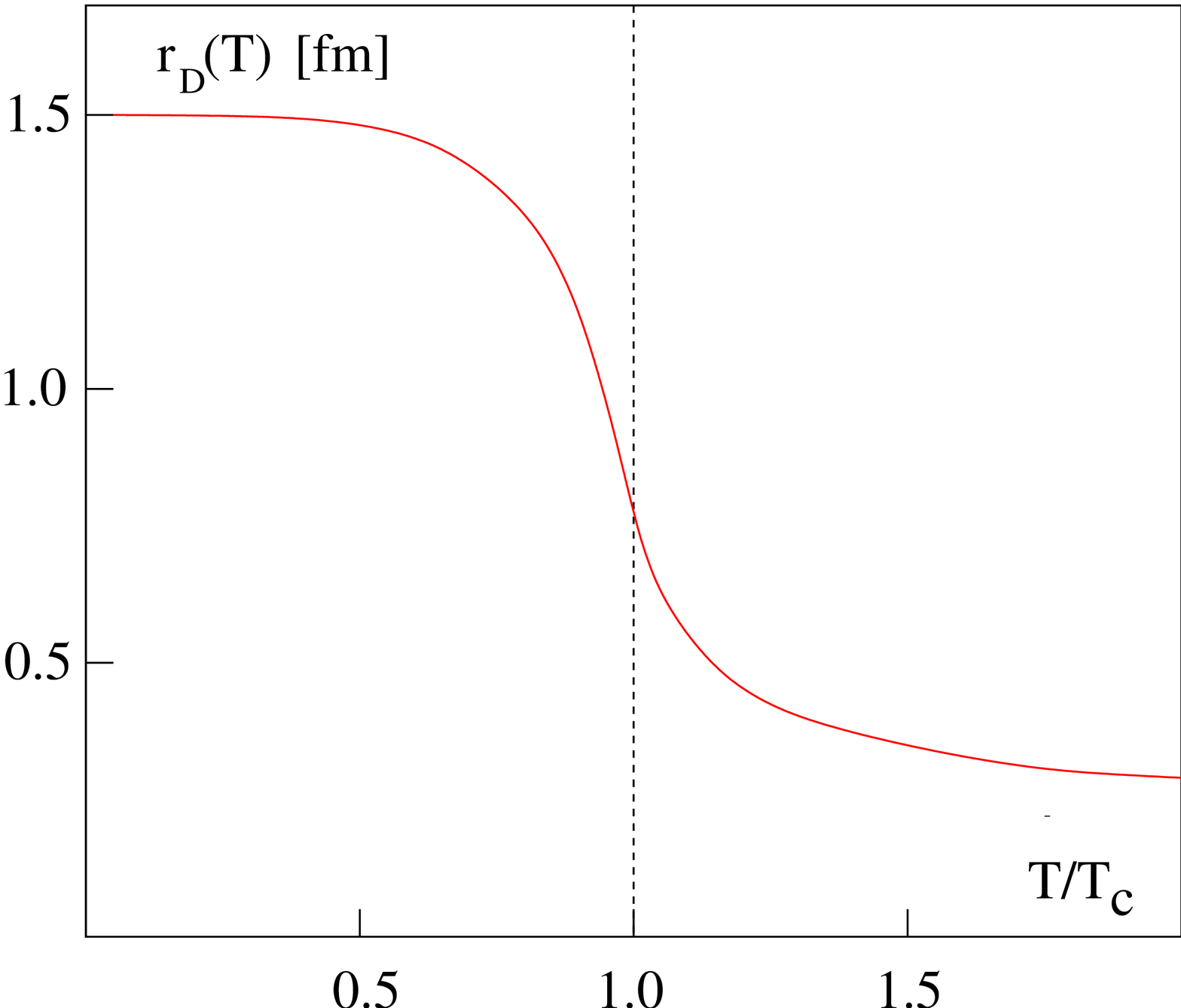,width=5cm,height=3.6cm}}
\caption{String breaking potential and interaction range at different 
temperatures}
\label{free-T}
\end{figure}

It is evident in Fig.\ \ref{OKfree} that the asymptotic value $F(\infty,T)$, 
i.e., the energy needed to separate the $\Q$ pair, decreases with increasing 
temperature, as does the separation distance at which ``the string breaks''.
For the moment we consider the latter to be defined by the point beyond which 
the free energy remains constant within errors, returning in section 4.3 to a 
more precise definition. The behaviour of both quantities is shown in Fig.\ 
\ref{free-T}. Deconfinement is thus reflected very clearly in the temperature
behaviour of the heavy quark potential: both the string breaking energy 
and the interaction range drop sharply around $T_c$. The latter decreases 
from hadronic size in the confinement region to much smaller values in the 
deconfined medium, where colour screening is operative. 

\medskip

\begin{figure}[htb]
\vspace*{-6.3cm}
\hspace*{8.5cm}
{\epsfig{file=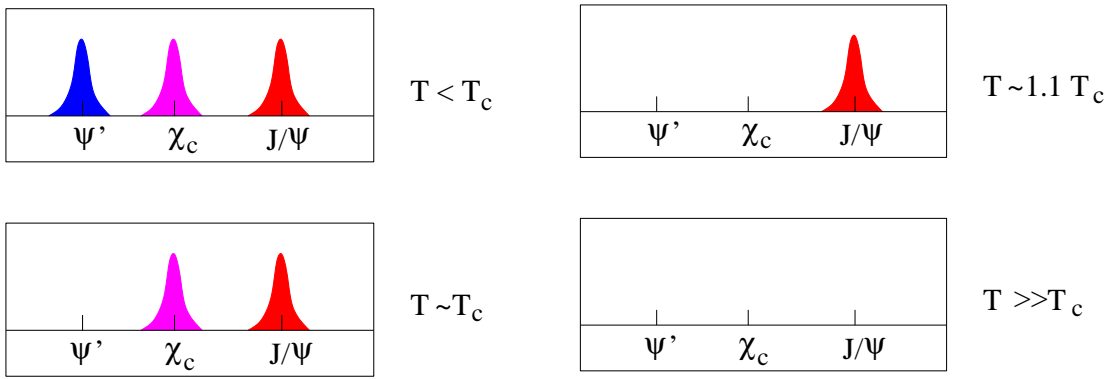,width=7cm}}
\vspace*{0.5cm}
\caption{Charmonium spectra at different temperatures}
\label{melt-t}
\end{figure}

The in-medium behaviour of heavy quark bound states thus does serve 
quite well as probe of the state of matter in QCD thermodynamics. We
had so far just considered $\Q$ bound states in general. Let us now
turn to a specific state such as the \J. What happens when the range 
of the binding force becomes smaller than the radius of the state?
Since the $c$ and the $\bar c$ can now no longer see each other, the
\J~ must dissociate for temperatures above this point. Hence the
dissociation points of the different quarkonium states provide a 
way to measure the temperature of the medium. The effect is illustrated
schematically in Fig. \ref{melt-t}, showing how with increasing 
temperature the different charmonium states ``melt'' sequentially
as function of their binding strength; the most loosely bound state 
disappears first, the ground state last.

\begin{figure}[htb]
\hskip1cm
{\epsfig{file=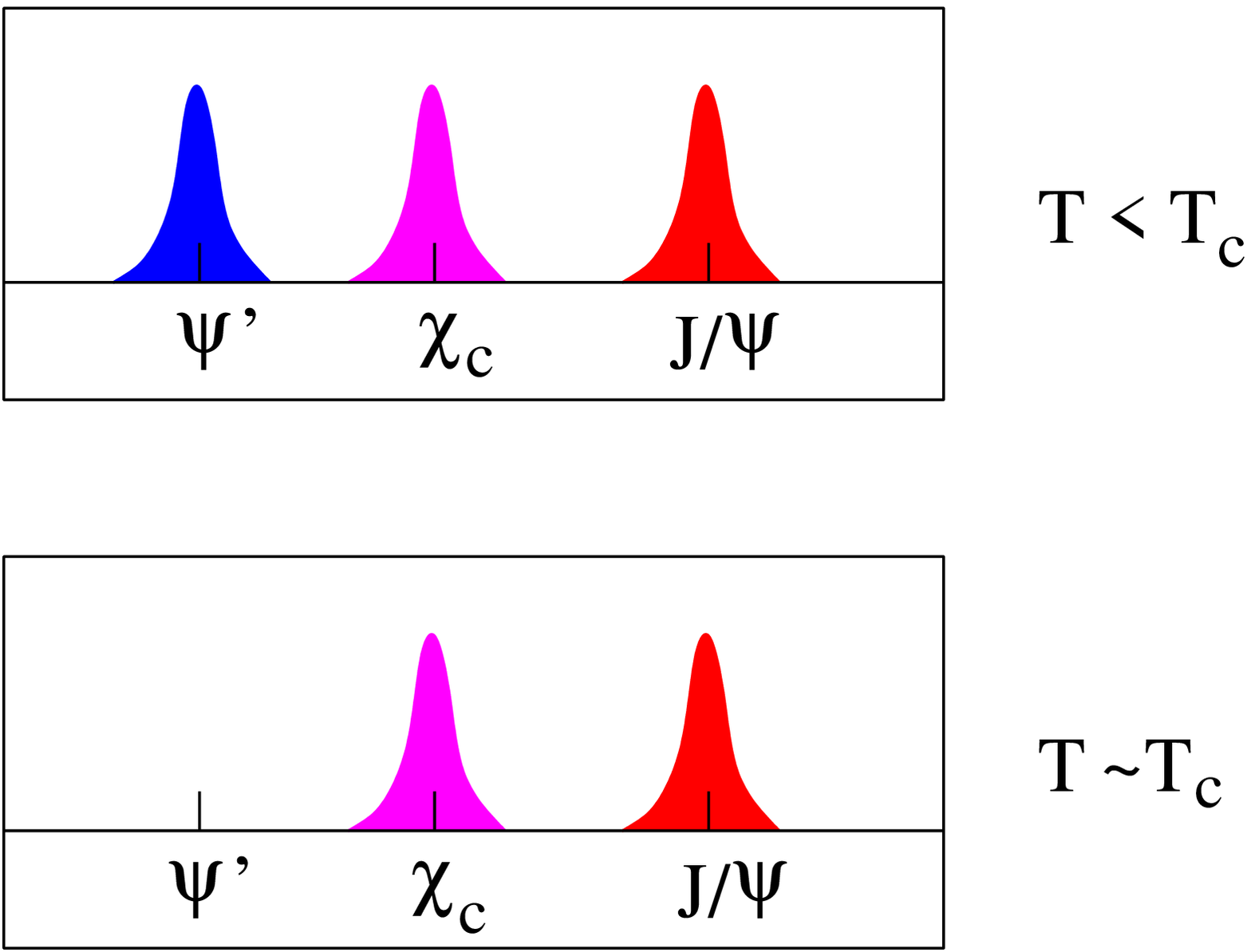,width=6.5cm}}
\end{figure}
\hskip4cm

\begin{figure}[htb]
\hspace*{-0.3cm}
\centerline{\epsfig{file=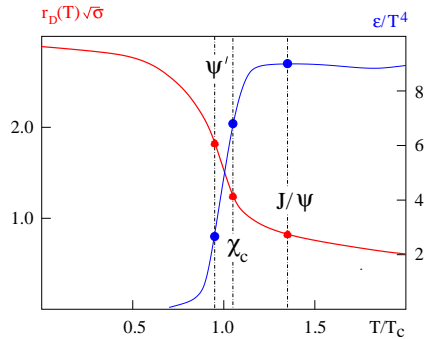,width=5.5cm}}
\caption{Charmonium dissociation vs.\ temperature and energy density}
\label{melt-e}
\end{figure}

Moreover, since finite temperature lattice QCD also provides the
temperature dependence of the energy density, the melting of the
different charmonia or bottomonia can be specified as well in terms
of $\e$. In Fig\ \ref{melt-e}, we illustrate this, combining the 
the energy density from Fig.\ \ref{edens} and the force radii from 
Fig.\ \ref{free-T}. It is evident that although \P~and \X~are expected
to melt around $T_c$, the corresponding dissociation energy densities 
will presumably be quite different.

\medskip

\vspace*{-0.3cm}

\medskip

To make these considerations quantitative, we thus have to find a way
to determine the in-medium melting points of the different quarkonium
states. This problem has been addressed in three different approaches:
\vskip-0.2cm
\begin{itemize}
\vskip-0.2cm
\item{Model the heavy quark potential as function of the temperature,
$V(r,T)$, and solve the resulting Schr\"odinger equation
(\ref{schroedinger}).}
\vspace*{-0.2cm}
\item{Determine the internal energy $U(r,T)$ of a $\Q$ pair at separation 
distance $r$ from lattice results for the corresponding free energy $F(r,T)$, 
using the thermodynamic relation 
\vskip-0.5cm
\be
U(r,T) = -T^2 \left( {\partial [F(r,T)/T] \over \partial T}\right)
= F(r,T) - T \left({\partial F(r,T) \over \partial T}\right),
\label{intern}
\ee
and solve the Schr\"odinger equation with $V(r,T)=U(r,T)$ as the binding 
potential.}
\vskip-0.5cm
\item{Calculate the quarkonium spectrum directly in finite temperature
lattice QCD.}
\end{itemize}
Clearly the last is the only model-independent way, and it will
in the long run provide the decisive determination. However, the direct
lattice study of charmonium spectra has become possible only quite
recently, and so far only in quenched QCD (no dynamical light quarks);
corresponding studies for bottomonia are still more difficult.
Hence much of what is known so far is based on Schr\"odinger equation
studies with different model inputs. To illustrate the model-dependence
of the dissociation parameters, we first cite some early work using different
models for the temperature dependence of $V(r,T)$, then some recent 
studies based on lattice results for $F(r,T)$, and finally summarize
the present state of direct lattice calculations of charmonia in finite
temperature media.

\vskip0.5cm

\noindent{\bf 4.2 Potential Model Studies}

\bigskip

The first quantitative studies of finite temperature charmonium dissociation
\cite{K-M-S} were based on screening in the form obtained in one-dimensional 
QED, the so-called Schwinger model. The confining part of the Cornell
potential (\ref{cornell}), $V(r) \sim \sigma r$, is the solution of the 
Laplace equation in one space dimension. In this case, Debye-screening leads 
to \cite{Dixit}
\be
V(r,T) \sim \sigma r \left\{ {1-e^{-\mu r} \over \mu r} \right\}, 
\label{schwinger-e}
\ee
where $\mu(T)$ denotes the screening mass (inverse Debye radius) for the
medium at temperature $T$. This form reproduces at least qualitatively 
the convergence to a finite large distance value $V(\infty,T) = 
\sigma/\mu(T)$, and since $\mu(T)$ increases with $T$, it also gives
the expected decrease of the potential with increasing temperature.
Combining this with the usual Debye screening for the $1/r$ part of
eq.\ (\ref{cornell}) then leads to  
 \be
V(r,T) \sim \sigma r \left\{ {1-e^{-\mu r} \over \mu r} \right\} 
- {\alpha \over r} e^{-\mu r} = {\sigma \over \mu} 
\left\{ 1-e^{-\mu r} \right\} - {\alpha \over r} e^{-\mu r}
\label{schwinger}
\ee
for the screened Cornell potential. In \cite{K-M-S}, the screening mass 
was assumed to have the form $\mu(T) \simeq 4~T$, as obtained in first
lattice estimates of screening in high temperature $SU(N)$ gauge theory.  
Solving the Schr\"odinger equation with these inputs, one found that
both the \P~and the \X~become dissociated essentially at $T \simeq T_c$,
while the \J~persisted up to about $1.2~T_c$. Note that as function of
the energy density $\e \sim T^4$, this meant that the \J~really survives
up to much higher $\e$.

\medskip

This approach has two basic shortcomings:
\vspace*{-0.2cm}
\begin{itemize}
\item{The Schwinger form (\ref{schwinger-e}) corresponds to the screening
of $\sigma r$ in one space dimension; the correct result in three space
dimensions is different \cite{Dixit}.}
\vspace*{-0.2cm}
\item{The screening mass $\mu(T)$ is assumed in its high energy form;
lattice studies show today that its behaviour near $T_c$ is quite 
different \cite{D-K-K-S}.}
\end{itemize}

While the overall behaviour of this approach provides some first insight
into the problem, quantitative aspects require a more careful treatment.

\medskip

\begin{figure}[htb]
\centerline{\epsfig{file=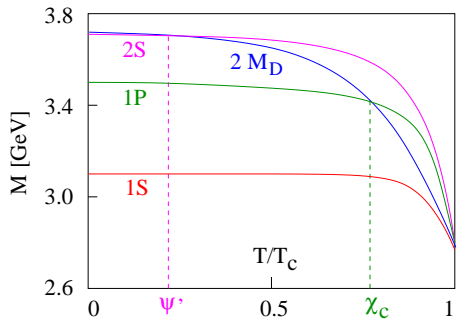,width=6cm}}
\caption{Temperature dependence of open and hidden charm masses \cite{D-P-S1}}
\label{masses}
\end{figure}

When lattice results for the heavy quark free energy as function of the 
temperature first became available, an alternative description appeared
\cite{D-P-S1}. It assumed that in the thermodynamic relation (\ref{intern})
the entropy term $-T(\partial F /\partial T)$ could be neglected, thus
equating binding potential and free energy,
\be
V(r,T) = F(r,T) -T(\partial F /\partial T) \simeq F(r,T).
\label{DPS}
\ee
Using this potential in the Schr\"odinger equation (\ref{schroedinger})
specifies the temperature dependence of the different charmonium 
masses. On the other hand, the large distance limit of $V(r,T)$ 
determines the temperature variation of the open charm meson $D$,
\be
2 M_D(T) \simeq 2 m_c + V(\infty,T)
\label{D-mass}
\ee
In fig.\ \ref{masses}, we compare the resulting open and hidden charm 
masses as function of temperature. It is seen that the \P~mass falls
below $2~M_D$ around 0.2 $T_c$, that of the \X~at about 0.8 $T_c$;
hence these states disappear by strong decay at the quoted temperatures.
Only the ground state \J~survives up to $T_c$ and perhaps
slightly above; the lattice data available at the time did not extend 
above $T_c$, so further predictions were not possible. 

The main shortcoming of this approach is quite evident. 
The neglect of the entropy term in the potential reduces $V(r,T)$ 
and hence the binding. As a result, the $D$ mass drops faster with
temperature than that of the charmonium states, and it is this effect
which leads to the early charmonium dissociation. Moreover, 
in the lattice studies used here, only the colour averaged free energy
was calculated, which leads to a further reduction of the binding
force.

\medskip

We conclude from these attempts that for a quantitative potential theory 
study, the free energy has to be formulated in the correct
three-dimensional screened Cornell form, and it then has to be
checked against the space- and temperature-dependence of the corresponding
colour singlet quantity obtained in lattice QCD. 

\vskip0.5cm

\noindent{\bf 4.3 Screening Theory}

\bigskip

The modification of the interaction between two charges immersed in a 
dilute medium of charged constituents is provided by Debye-H\"uckel theory, 
which for the Coulomb potential in three space dimensions leads to the 
well-known Debye screening, 
\be
{1\over r} ~\to~{1\over r} \e^{-\mu r},
\label{debye}
\ee
where $r_D=1/\mu$ defines the screening radius \cite{Landau}. 
Screening can be evaluated more generally \cite{Dixit}
for a given free energy $F(r) \sim r^q$ in $d$ space dimensions, with 
an arbitrary number $q$. We shall here apply this to the two terms
of the Cornell form, with $q=1$ for the string term, $q=-1$ for the gauge 
term, in $d=3$ space dimensions \cite{D-K-K-S}.

\medskip

We thus assume that the screening effect can be calculated separately for 
each term, so that the screened free energy becomes
\be
F(r,T) =  F_s(r,T) + F_c(r,T) =
\sigma r f_s(r,T) - {\alpha \over r} f_c(r,T).
\label{screen-cornell}
\ee
The screening functions $f_s(r,T)$ and $f_c(r,T)$ must satisfy
\begin{eqnarray}
f_s(r,T)&=&f_c(r,T)= 1 ~~{\rm for}~ T \rightarrow 0,\nonumber\\
f_s(r,T)&=&f_c(r,T)= 1 ~~{\rm for}~ r \!\rightarrow 0,
\label{screen-f}
\end{eqnarray}

since at $T=0$ there is no medium, while in the short-distance limit
$T^{-1} \gg r \to 0$, the medium has no effect. The resulting forms are
\cite{Dixit}
\be
F_c(r,T) = -{\alpha \over r}\left[e^{-\mu r} + \mu r\right]
\label{screen-gauge}
\ee
for the gauge term, and
\begin{eqnarray}
\hspace{-0.4cm}F_s(r,T) = {\sigma \over \mu}
\left[ {\Gamma\left(1/4\right) \over 2^{3/2}
\Gamma\left(3/4\right)}-{\sqrt{\mu r} \over 2^{3/4}\Gamma\left(3/4\right)}
K_{1/4}[{(\mu r)}^2]\right]
\label{screen-string}
\end{eqnarray}

for the string term. The first term in eq.\ (\ref{screen-string}) 
gives the constant large
distance limit due to colour screening; the second provides a Gaussian 
cut-off in $x=\mu r$, since $K_{1/4}(x^2) \sim \exp\{-x^2))$, which is in 
contrast to the exponential cut-off given by the Schwinger form 
(\ref{schwinger}).

\begin{figure}[htb]
\hspace*{-0.3cm}
{\epsfig{file=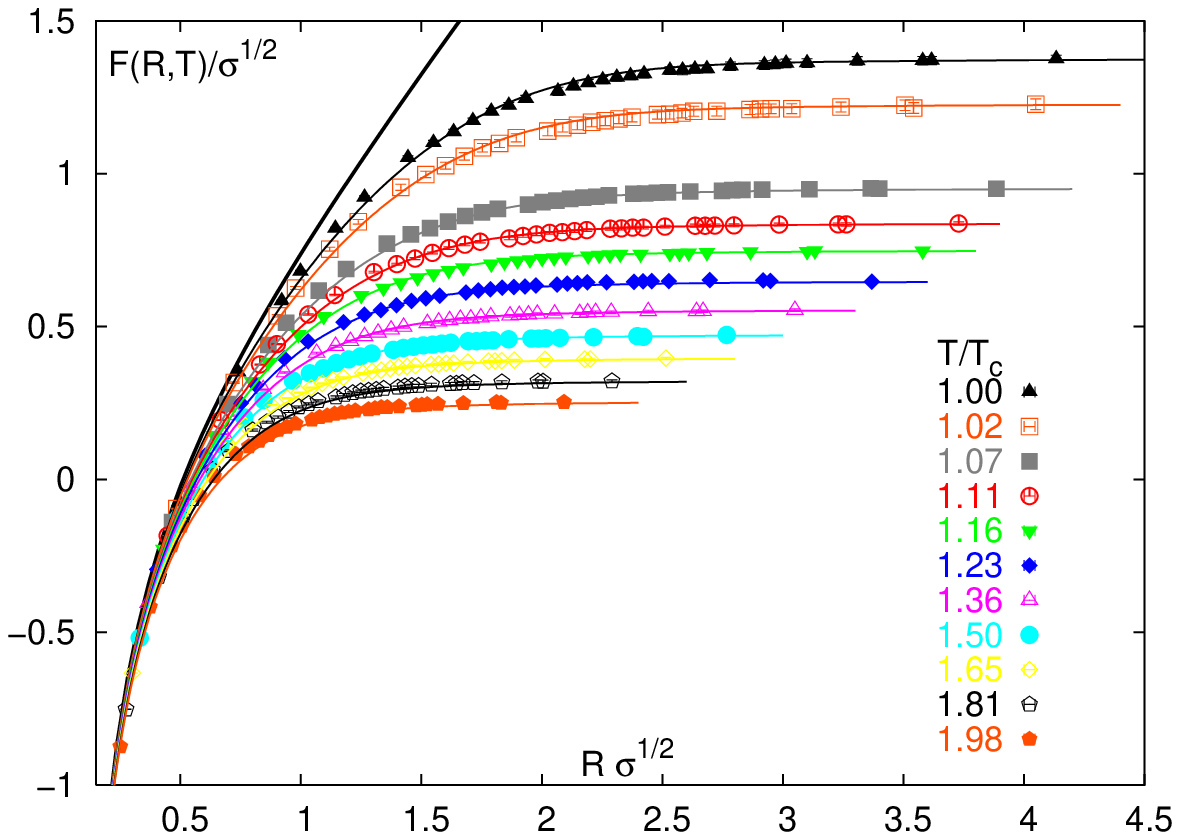,width=8cm}}
{\epsfig{file=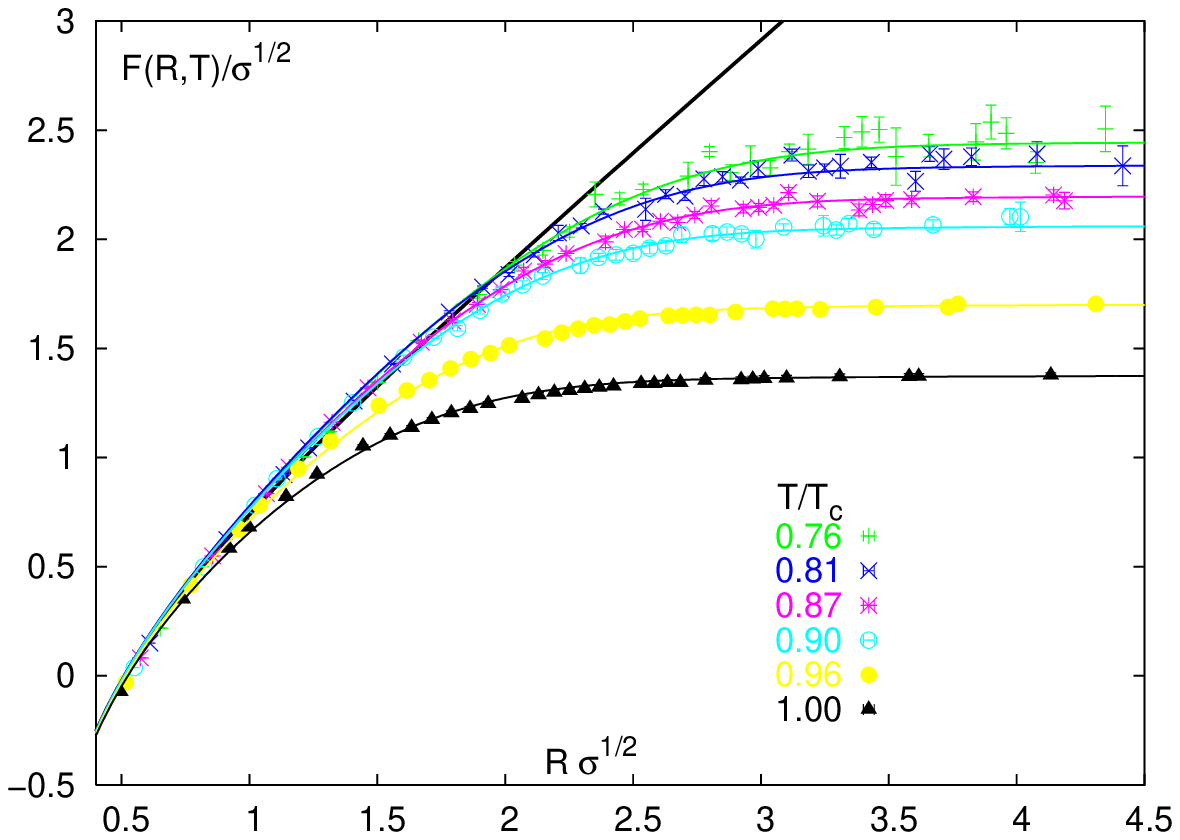,width=8cm}}
\hspace*{4cm}(a)\hspace*{7.5cm}(b)
\caption{Screening fits to the $\Q$ free energy $F(r,T)$ for $T\geq T_c$
(left) and $T\leq T_c$ (right) \cite{D-K-K-S}}
\label{free}
\end{figure}

\begin{figure}[htb]
\hspace*{-0.3cm}
\centerline{\epsfig{file=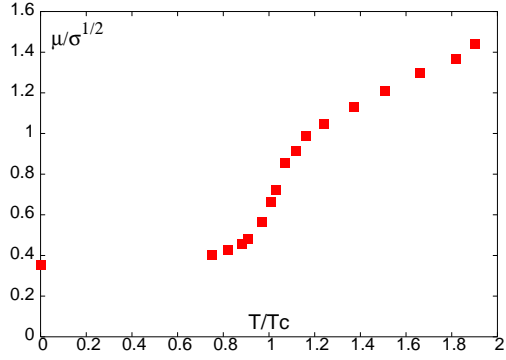,width=7cm}}
\caption{The screening mass $\mu(T)$ vs.\ $T$ \protect\cite{D-K-K-S,D-K-S}}
\label{sc-mass}
\end{figure}

At temperatures $T > T_c$, when the medium really consists of unbound
colour charges, we thus expect the free energy $F(r,T)$ to have the form
(\ref{screen-cornell}), with the two screened terms given by eqs.\ 
(\ref{screen-gauge}) and (\ref{screen-string}). In Fig.\ \ref{free}a,
it is seen that the results for the colour singlet free energy calculated
in two-flavour QCD \cite{OK} are indeed 
described very well by this form. The only parameter, the screening mass
$\mu$, is shown in  Fig.\ \ref{sc-mass}, and as expected, it first increases
rapidly in the transition region and then turns into the perturbative
form $\mu \sim T$.

\medskip

The behaviour of $\Q$ binding in a plasma of unbound quarks and gluons is
thus well described by colour screening. Such a description is in fact 
found to work well also for $T<T_c$, when quark recombination leads to an
effective screening-like reduction of the interaction range, provided one
allows higher order contributions in $x=\mu r$ in the Bessel function
$K_{1/4}(x^2)$ governing string screening \cite{D-K-K-S}. The resulting fit 
to the two-flavour colour singlet free energy below $T_c$ is shown in 
Fig.\ \ref{free}b, using
$K_{1/4}(x^2+ (1/2) x^4)$ in eq.\ (\ref{screen-string}); the corresponding 
values for the screening mass are included in Fig.\ \ref{sc-mass}.   

\medskip

With the free energy $F(r,T)$ of a heavy quark-antiquark pair given in 
terms of the screening form obtained from Debye-H\"uckel theory, the 
internal energy $U(r,T)$ can now be obtained through the thermodynamic 
relation (\ref{intern}), and this then provides the binding potential
$V(r,T)$ for the temperature dependent version of the Schr\"odinger 
equation (\ref{schroedinger}). The resulting solution then specifies
the temperature-dependence of charmonium binding as based on the 
correct heavy quark potential \cite{D-K-S}.
Let us briefly comment on the thermodynamic
basis of this approach. The pressure $P$ of a thermodynamic system is
given by the free energy, $P=-F=-U+TS$; it is determined by the kinetic
energy $TS$ at temperature $T$ and entropy $S(T)$, reduced by the potential
energy $U(T)$ between the constituents. In our case, all quantities give
the difference between a thermodynamic system containing a $\Q$ pair 
and the corresponding system without such a pair. The potential energy
of the $\Q$ pair, due both to the attraction of $Q$ and $\bar Q$ and
to the modification which the pair causes to the internal energy of 
the other constituents of the medium, is therefore given by $U$.
To determine the dissociation points for the different quarkonium
states, we thus have to solve the Schr\"odinger equation (\ref{schroedinger})
with $V(r,T)=U(r,T)$.

\medskip

\begin{figure}[h]
\vskip0.5cm
\begin{minipage}[t]{7cm}
\epsfig{file=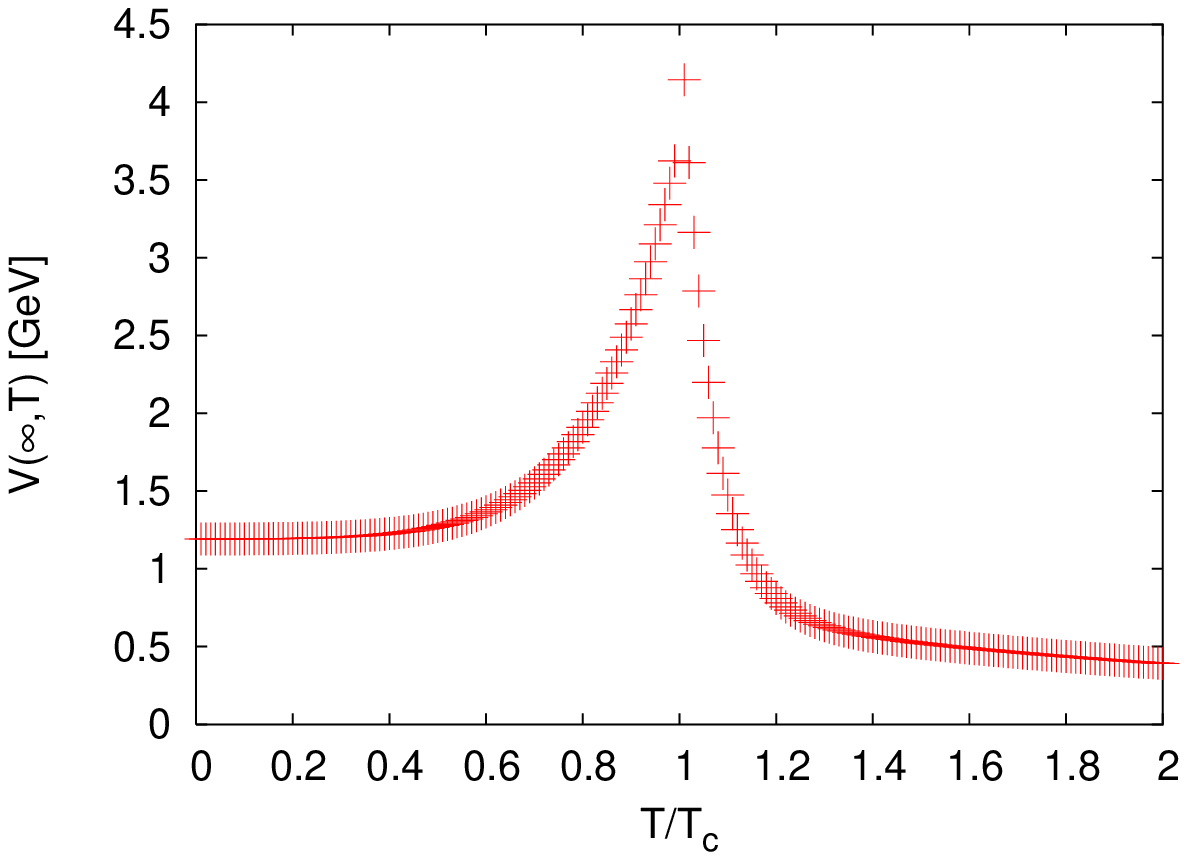,width=7.5cm}
\caption{The large distance limit of the quarkonium potential $V(r,T)$
\protect\cite{D-K-S}}
\label{vinf}
\end{minipage}
\hspace{1.3cm}
\begin{minipage}[t]{7cm}
\vspace*{-5.3cm}
\hskip-0.3cm
\epsfig{file=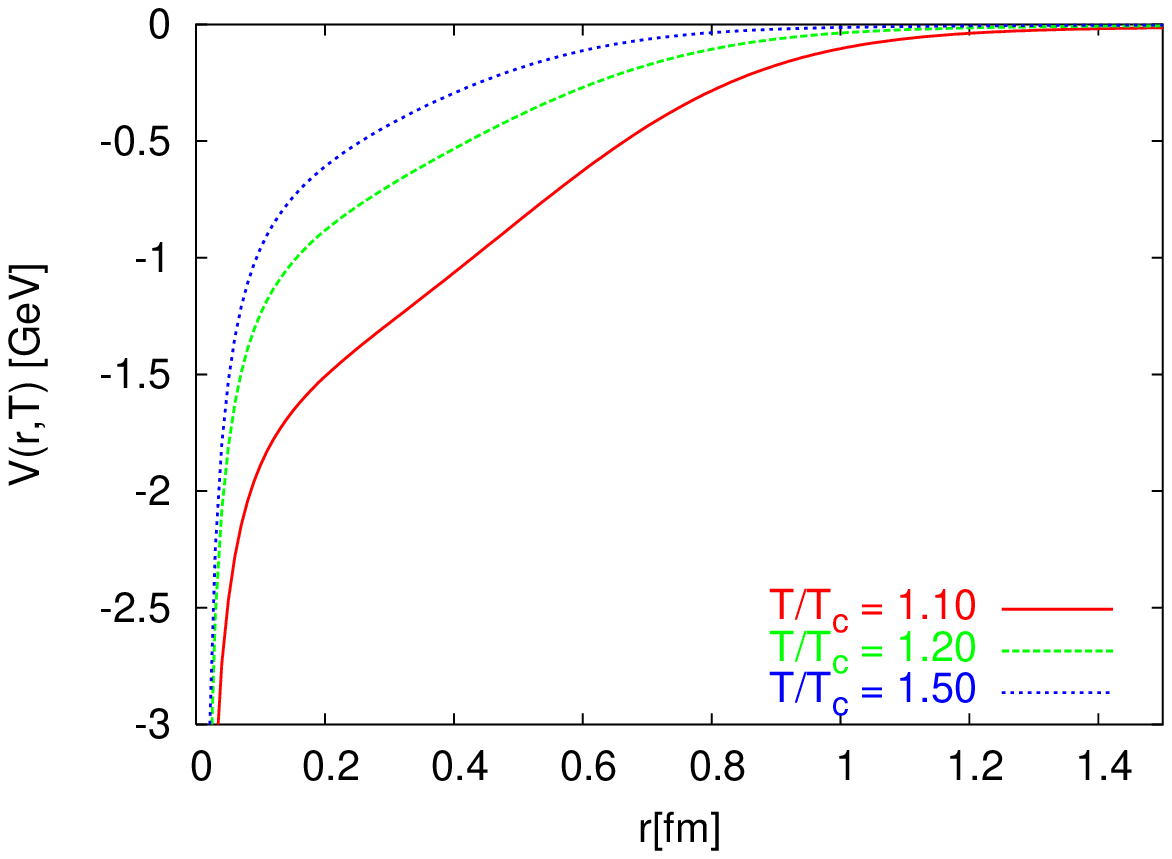,width=7.5cm}
\caption{Variation of $\tilde V(r,T)$ with $r$ for different $T$
\protect\cite{D-K-S}}
\label{vr}
\end{minipage}
\end{figure}

From eqs.\ (\ref{screen-gauge}) and (\ref{screen-string}) we 
get for the $\Q$ potential
\be
V(r,T)= V(\infty,T) + \tilde V(r,T),
\ee
with 
\be
V(\infty,T) = c_1{\sigma \over \mu} - \alpha \mu
+ T{d\mu \over dT}[c_1 {\sigma \over \mu^2} + \alpha],
\ee
and $c_1=\Gamma(1/4)/2^{3/2} \Gamma(3/4)$, and where $\tilde V(r,T)$ 
contains the part of the potential which vanishes for $r\to \infty$.
The behaviour of $V(\infty,T)$ as function of the temperature is
shown in Fig.\ \ref{vinf}. It measures (twice) the energy of the cloud
of light quarks and gluons around an isolated heavy quark, of an extension 
determined by the screening radius, relative to the energy contained in
such a cloud of the same size without a heavy quark. This energy difference 
arises from the interaction of the heavy quark with light quarks and gluons
of the medium, and from the modification of the interaction between the light 
constituents themselves, caused by the presence of the heavy charge. --
The behaviour of $\tilde V(r,T)$ is shown in Fig.\ \ref{vr} for three
different values of the temperature. It is seen that with increasing $T$,
screening reduces the range of the potential. 

\medskip

\begin{figure}[htb]
\hspace*{-1cm}
{\epsfig{file=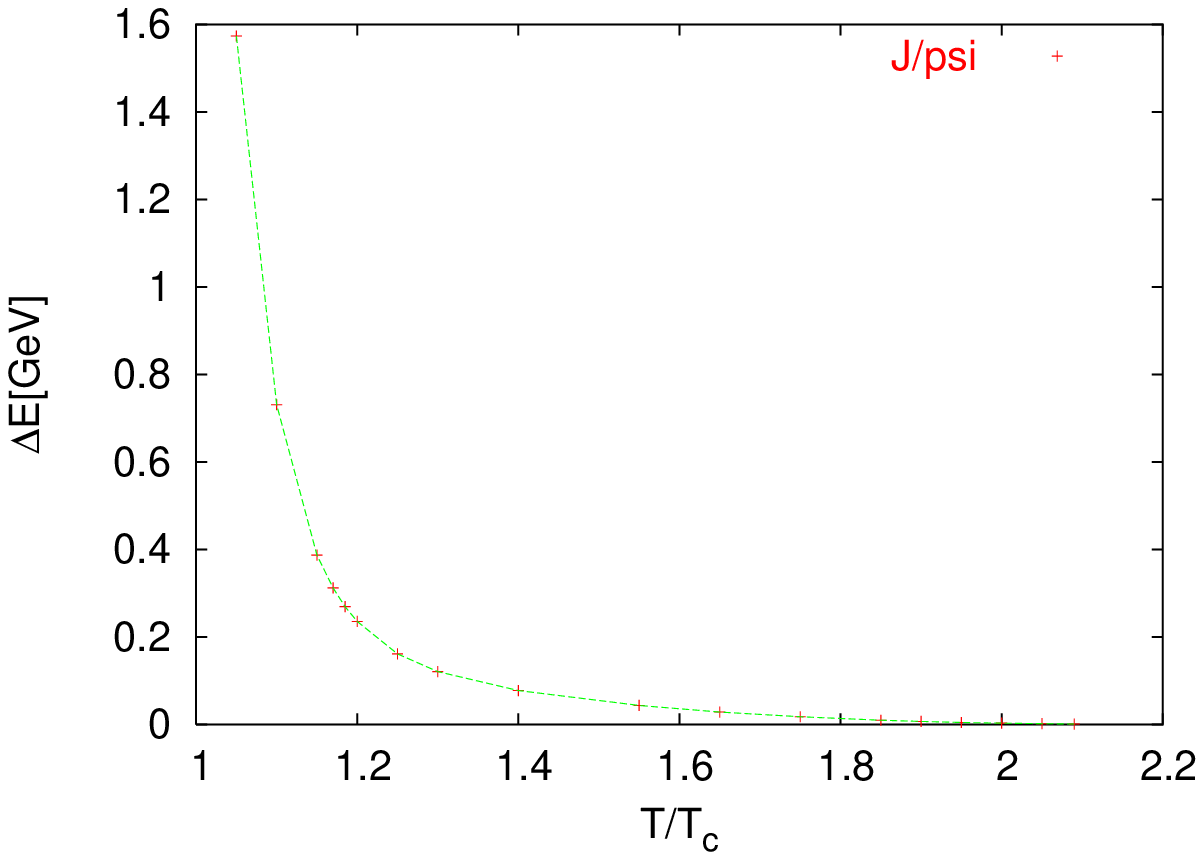,width=7.5cm}}
{\epsfig{file=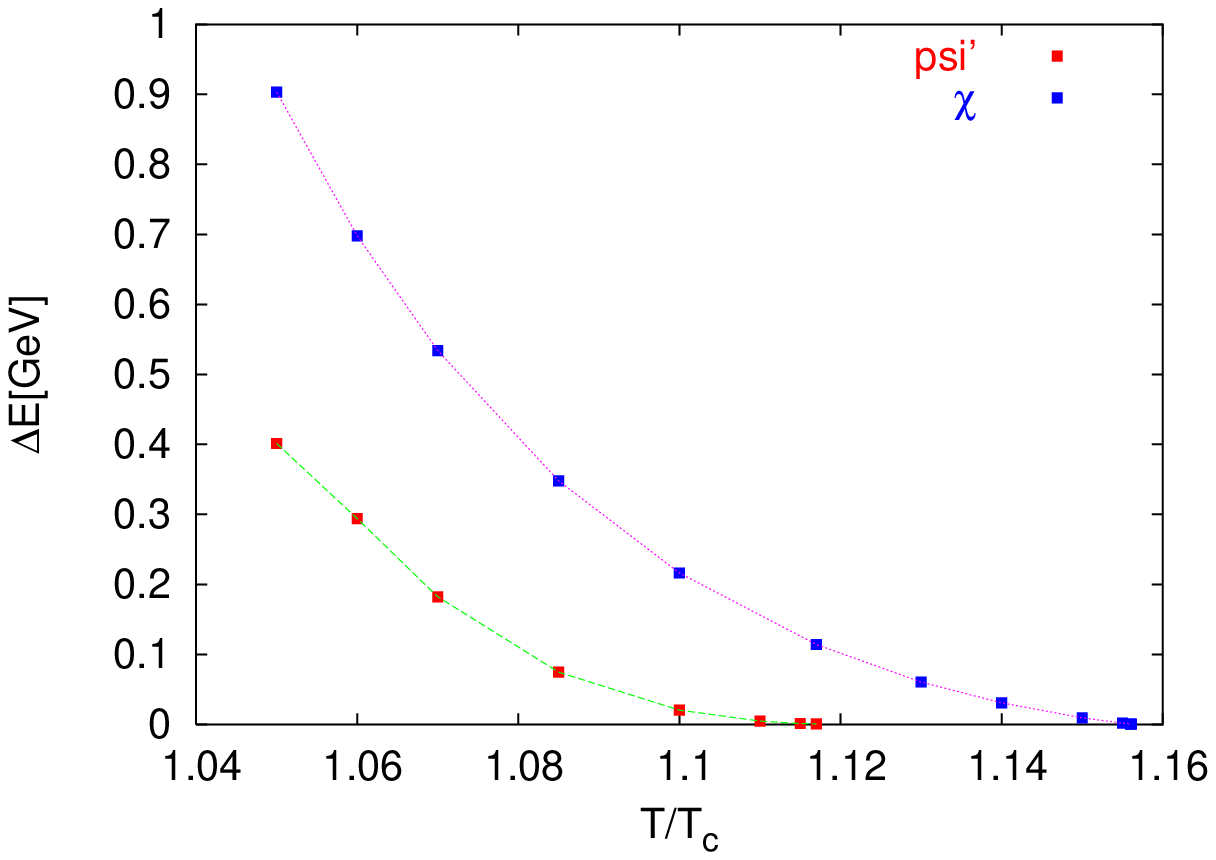,width=7.5cm}}
\hspace*{3.5cm}(a)\hspace*{7.5cm}(b)
\caption{$T$-dependence of binding energy for 
\J~(a) and for \X~/\P~(b) \protect\cite{D-K-S}}
\label{c-binding}
\end{figure}

\begin{figure}[htb]
\hspace*{-1cm}
{\epsfig{file=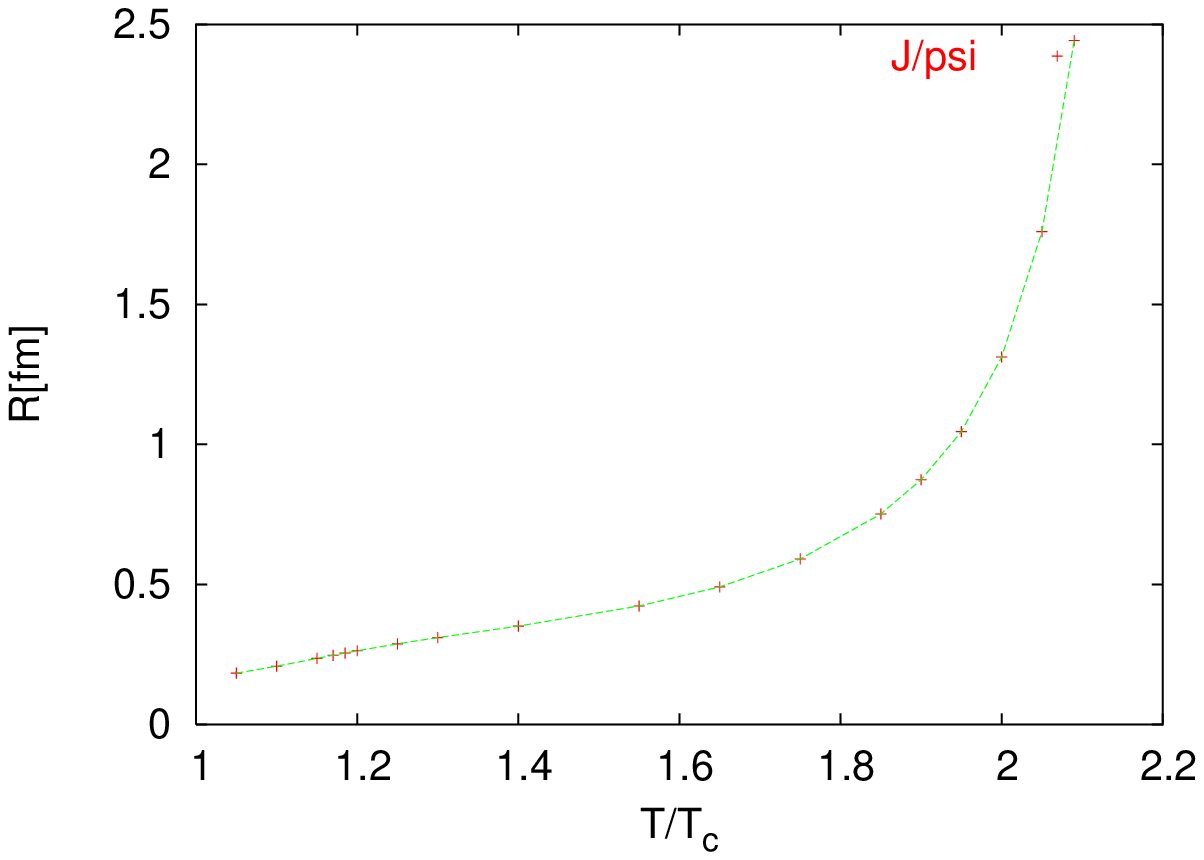,width=7.5cm}}
{\epsfig{file=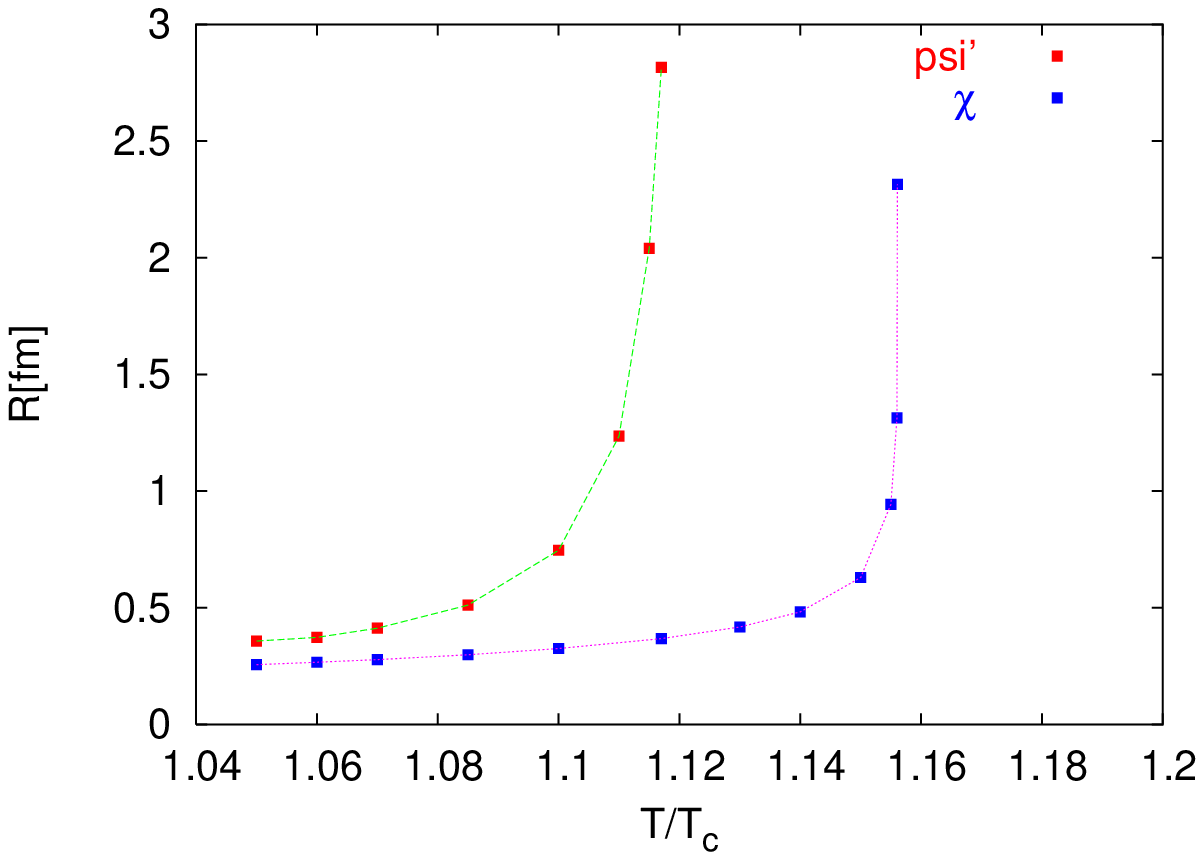,width=7.5cm}}
\hspace*{3.5cm}(a)\hspace*{7.5cm}(b)
\caption{$T$-dependence of bound state radii for 
\J~(a) and for \X~/\P~(b) \protect\cite{D-K-S}}
\label{c-radius}
\end{figure}

The relevant Schr\"odinger equation now becomes
\be
\left\{{1\over m_c}\nabla^2 - \tilde V(r,T)\right\} \Phi_i(r) = 
\Delta E_i(T) \Phi_i(r)
\label{T-schroedinger}
\ee 
where
\be
\Delta E_i(T)=V(\infty,T) - M_i - 2m_c 
\ee
is the binding energy of charmonium state $i$ at temperature $T$.
When it vanishes, the bound state $i$ no longer exists, so that 
$\Delta E_i(T)=0 $ determines the dissociation temperature $T_i$ 
for that state. The temperature enters only through the $T$-dependence
of the screening mass $\mu(T)$, as obtained from the analysis of the 
lattice results for $F(r,T)$. In Fig.\ \ref{c-binding}, we show the 
resulting binding energy behaviour for the different charmonium states,
obtained with $m_c=1.25$ GeV and $\sqrt \sigma=0.445$ GeV;
in Fig.\ \ref{c-radius}, we show the corresponding 
bound state radii \cite{D-K-S}. It is seen that in particular 
the divergence of the 
radii defines quite well the different dissociation points. 

The same formalism, with $m_b=4.65$ GeV replacing $m_c$, leads to the
bottomonium dissociation points. The combined quarkonium results are 
listed in Table 4. They agree quite well with those obtained in a
very similar study based on corresponding free energies obtained
in quenched lattice QCD \cite{Wong}, indicating that above $T_c$ gluonic 
effects dominate. Using a parametrically generalized screened Coulomb
potential obtained from lattice QCD results also leads to very compatible 
results for the $N_f=2$ and quenched cases\cite{Alberico}. We recall
here that the main underlying change, which is responsible for the much
higher dissociation temperatures for the quarkonium ground states, is the 
use of the full internal energy (\ref{intern}), including the entropy term,
as potential in the Schr\"odinger equation: this makes the binding much
stronger. 

\medskip

We should note, however, that in all such potential studies
it is not so clear what binding energies of less than a few MeV 
or bound state radii of several fermi can mean in a medium whose
temperature is above 200 MeV and which leads to screening radii 
of less than 0.5 fm. In such a situation, thermal activation \cite{K-ML-S}
can easily dissciate the bound state.

\medskip

\begin{center}
\renewcommand{\arraystretch}{2.0}
\begin{tabular}{|c||c|c|c||c|c|c|c|c|}
\hline
 state & J/$\psi(1S)$ & $\chi_c$(1P) & $\psi^\prime(2S)$&$\Upsilon(1S)$&
$\chi_b(1P)$&$\Upsilon(2S)$&$\chi_b(2P)$&$\Upsilon(3S)$\\
\hline
\hline
$T_d/T_c$ & 2.10  & 1.16 & 1.12 & $>4.0$ & 1.76 & 1.60& 1.19 & 1.17 \\
\hline
\end{tabular}\end{center}

\medskip

\centerline{Table 4: Quarkonium Dissociation Temperatures \cite{D-K-S}}

\vskip0.5cm

\noindent{\bf 4.4 Charmonium Correlators}

\bigskip

\begin{figure}[htb]
\centerline{\epsfig{file=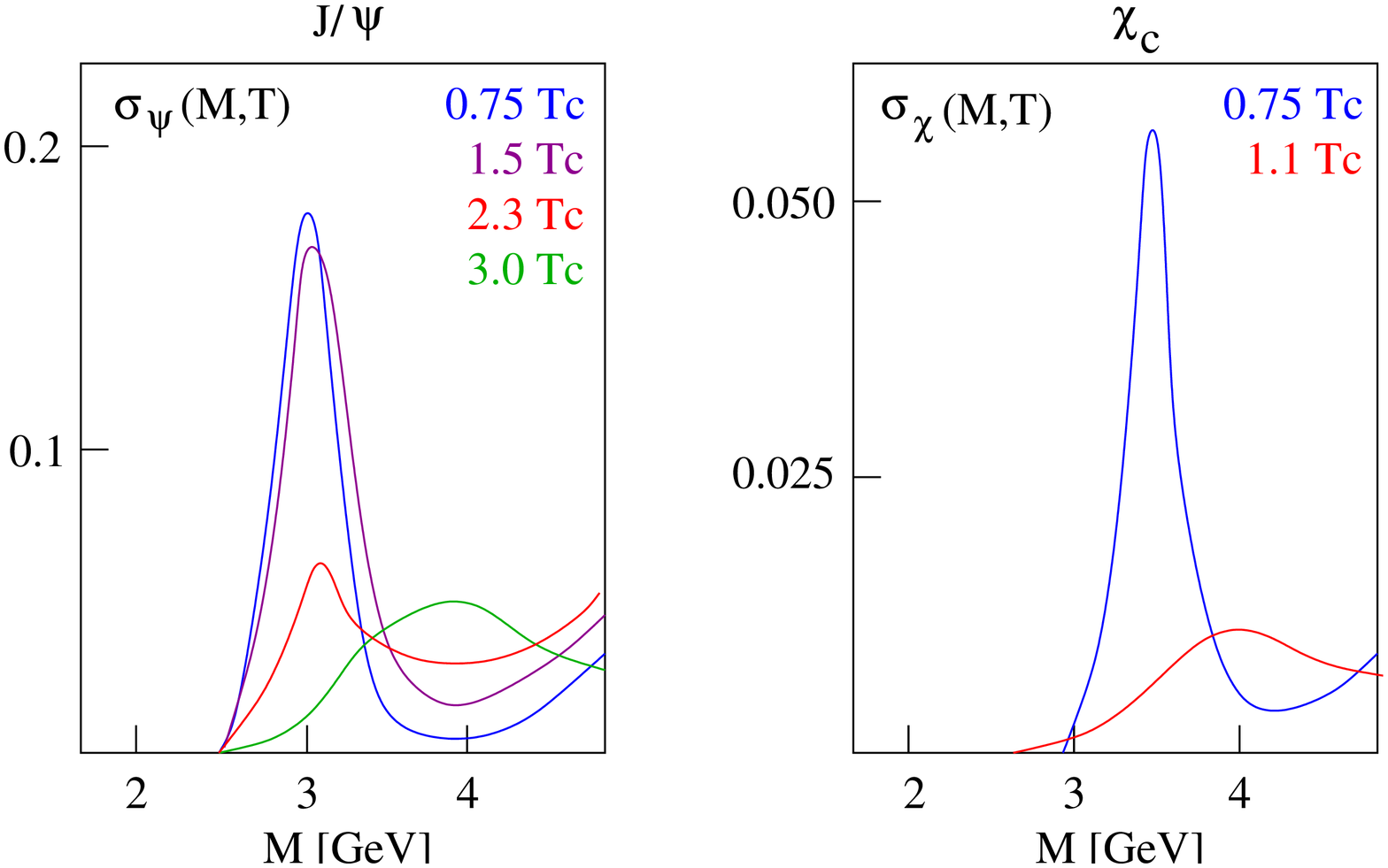,width=9cm}}
\caption{\J~and \X~spectral functions at different temperatures}
\label{spec}
\end{figure}

The direct spectral analysis of charmonia in finite temperature lattice 
has come within reach only in very recent years \cite{lattice-charm}.
It is possible now to evaluate the correlation functions $G_H(\t,T)$ for
hadronic quantum number channels $H$ in terms of the Euclidean time
$\t$ and the temperature $T$. These correlation functions are directly
related to the corresponding spectral function $\sigma_H(M,T)$,
which describe the distribution in mass $M$ at temperature $T$ for
the channel in question. In Fig.\ \ref{spec}, schematic results at different
temperatures are shown for the \J~and the \X~channels. It is seen that
the spectrum for the ground state \J~ remains essentially unchanged
even at $1.5~T_c$. At $3~T_c$, however, it has disappeared; the
remaining spectrum is that of the $\C$ continuum of \J~quantum numbers
at that temperature. In contrast, the \X~is already absent at $1.1~T_c$, 
with only the corresponding continuum present. 

\medskip

These results are clearly very promising: they show that in a foreseeable
future, the dissociation parameters of quarkonia can be determined 
{\sl ab initio} in lattice QCD. For the moment, however, they remain
indicative only, since the underlying calculations were generally
performed in
quenched QCD, i.e., without dynamical quark loops. Since such loops are
crucial in the break-up of quarkonia into light-heavy mesons, final
results require calculations in full QCD. Some first calculations
in two-flavour QCD have just appeared \cite{skullerud} and support
the late dissociation of the \J.  
The widths of the
observed spectral signals are at present determined by the precision of
the lattice calculations; to study the actual physical widths, much
higher precision is needed. Finally, one has so far only first signals
at a few selected points; a temperature scan also requires higher performance
computational facilities. Since the next generation of computers, in the
multi-Teraflops range, is presently going into operation, the next
years should bring the desired results. So far, in view of the mentioned
uncertainties in both approaches, the results from direct lattice studies
and those from the potential model calculations of the previous section
appear quite compatible. 

\vskip0.5cm

\noindent{\large \bf 5.\ Quarkonium Production in Nuclear Collisions}

\vskip0.5cm

The aim of high energy nuclear collisions is to study colour deconfinement
and the resulting quark-gluon plasma in the laboratory. Here we want to 
show how quarkonia can be used as a probe in this study. The medium to be 
probed as well as the quarkonium probes are produced in the collision, so
that we have to address evolution aspects in both cases. After a brief 
section on the evolution of nuclear collisions, we shall first consider 
quarkonium formation in elementary proton-proton collisions and then turn 
to in-medium phenomena of quarkonia in nuclear collisions.

\vskip0.5cm

\noindent{\bf 5.1 Nuclear Collisions}

\bigskip

\begin{figure}[htb]
\centerline{\psfig{file=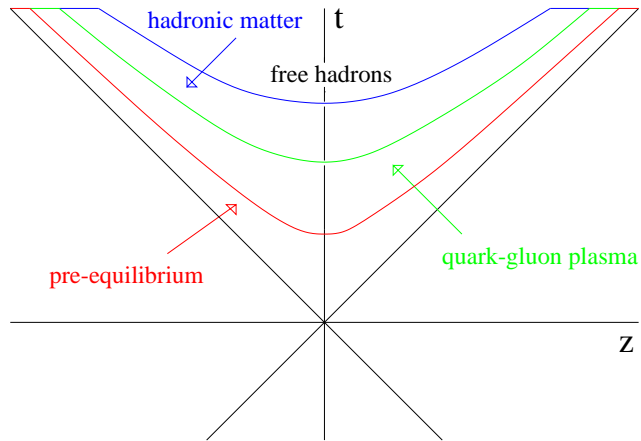,width=8.5cm}}
\caption{Expected evolution of a nuclear collision}
\label{evo}
\end{figure}

Starting from the 
non-equilibrium configuration of two colliding nuclei, the evolution of the 
collision is assumed to have the form shown in Fig.\ \ref{evo}. After the 
collision, there is a short pre-equilibrium stage, in which the primary 
partons of the colliding nuclei interact and thermalize to form the 
quark-gluon plasma. This then expands, cools and hadronizes. 

\begin{figure}[h]
\centering
\begin{tabular}{cccccc}
\resizebox{0.25\textwidth}{!}{%
\includegraphics*{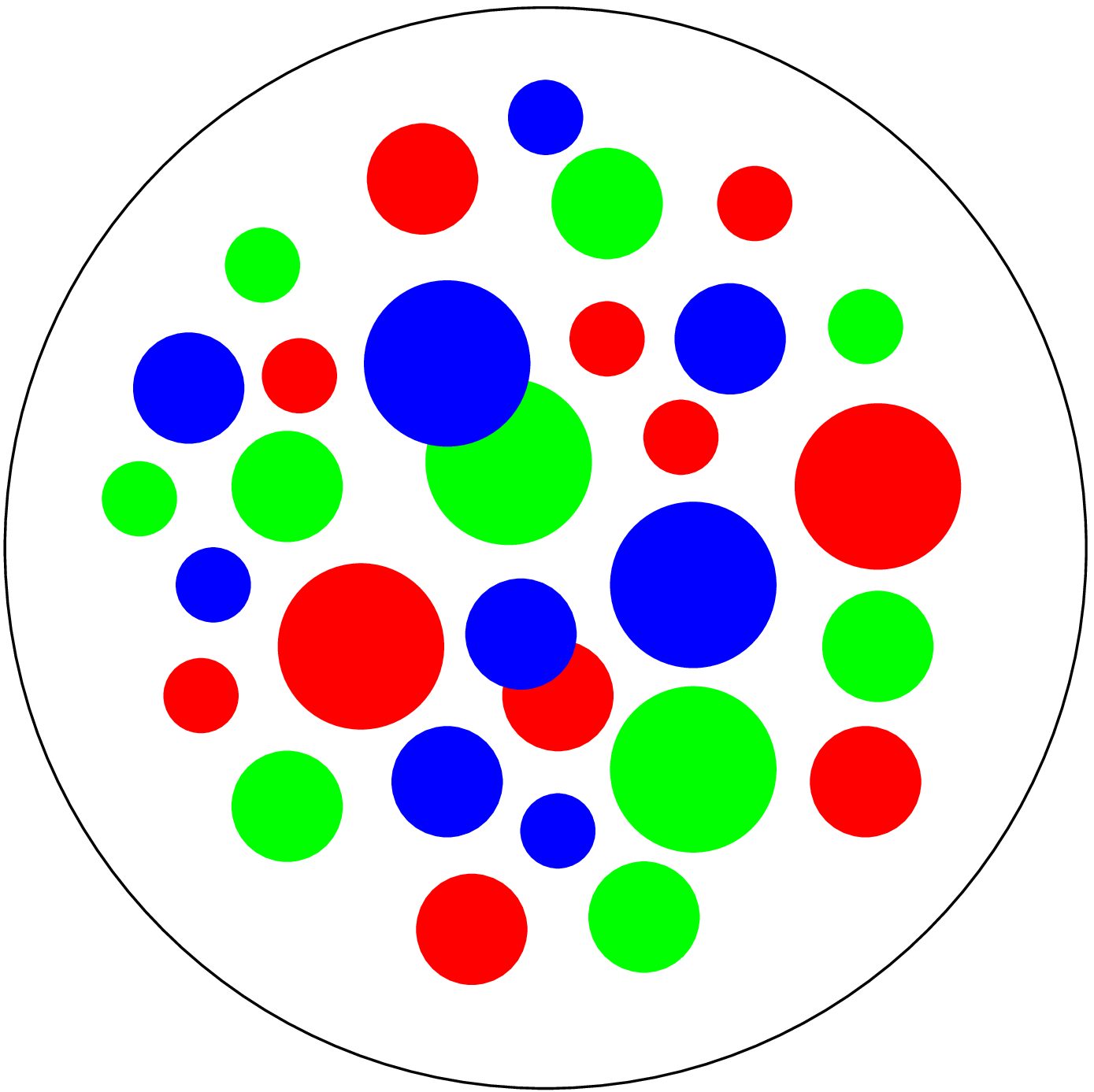}}
& & & & &
\resizebox{0.25\textwidth}{!}{%
\includegraphics*{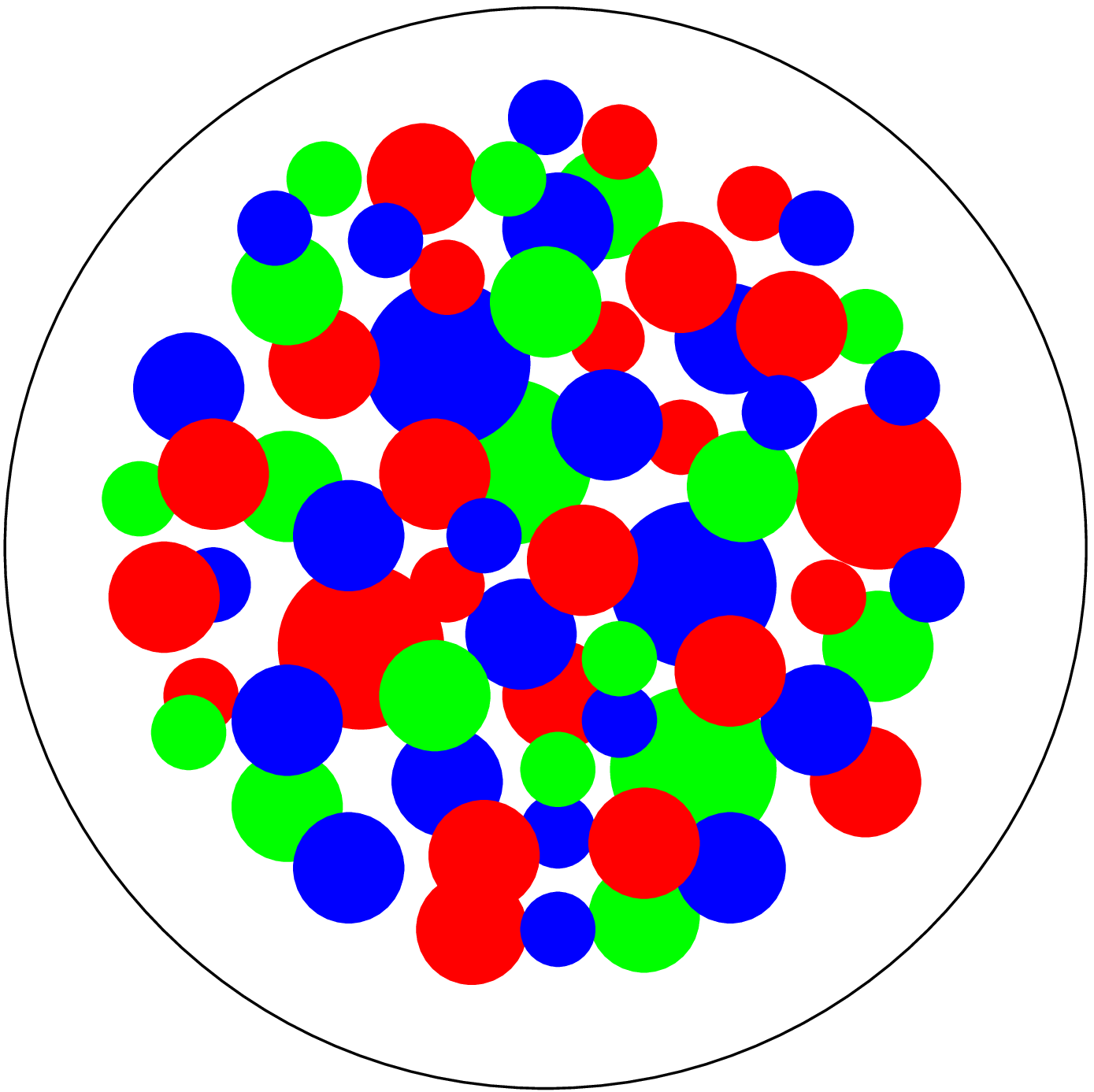}}
\end{tabular}
\caption{Parton distributions in the transverse plane of a
nucleus-nucleus collision}
\label{PP}
\vglue-4mm
\end{figure}

The partonic constituents in the initial state of the
collision are given by the parton distribution functions of the
colliding nuclei. To produce a large-scale thermal system, partons from 
different nucleon-nucleon collisions have to undergo multiple interactions. 
In the center of mass of a high energy collision, the incoming nuclei 
are strongly Lorentz-contracted; the resulting parton distribution in the 
transverse collision plane is schematically illustrated in Fig. \ref{PP}. 
The transverse size of the partons is determined by their intrinsic 
transverse momentum, and the number of partons contained in a nucleon is 
known from deep inelastic scattering experiments. The density of partons 
increases with both $A$ and $\sqrt s$, and at some critical point, parton 
percolation occurs \cite{PP} and global colour connection sets in. In 
the resulting connected medium, partons lose their independent existence
and well-defined origin, so that there is deconfinement, but not yet 
thermalization. In recent years, such partonic connectivity requirements 
(parton saturation, colour glass condensate) have attracted much 
attention \cite{PP,cgc}; they appear to form a prerequisite for
subsequent thermalization.  

\bigskip

\begin{figure}[htb]
\centerline{\psfig{file=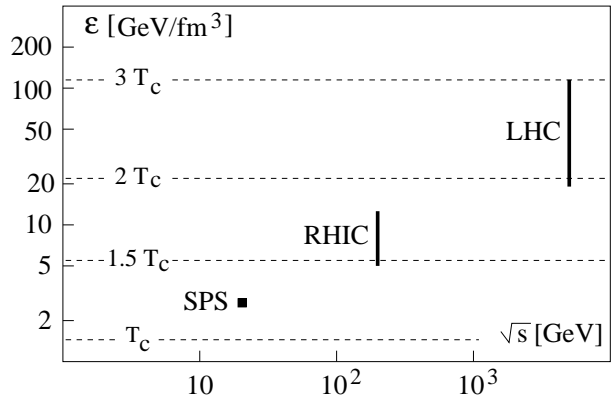,width=8cm}}
\caption{Energy density estimates vs.\ maximum collision energies, for 
different accelerators, compared to corresponding temperature}
\label{accel}
\end{figure}

Assuming that after an equilibration time $\t_0$ a thermalized plasma 
of deconfined (but interacting) quarks and gluons is formed, the 
initial energy density of the medium can be estimated in terms of the
emitted hadrons and the initial interaction volume \cite{Bjorken83} 
\be
\e = \left({dN_h \over dy}\right)_{y=0} {w_h \over \pi R_A^2 \tau_0},
\label{bj}
\ee
where $(dN_h/dy)_{y=0}$ specifies the number of hadrons emitted in the
nuclear collision per unit rapidity at mid-rapidity,
 and $w_h$ their average
energy. The initial transverse size is determined by the nuclear radius 
$R_A$, or the nuclear overlap area in non-central collisions. The initial 
longitudinal extension is governed by the equilibration time, which
was originally taken to be about 1 fm \cite{Bjorken83}. This appears 
quite reasonable in media, in which the intrinsic QCD scale 
$\l^{-1} \simeq $ 1 fm is the relevant measure. For denser partonic 
systems, one can expect a faster thermalization, and so it seems meaningful 
to use the parton density to scale $\t_0$ to higher collision energies.
The number of partons emitted by a colliding nucleon is about two at
SPS energy \cite{MRS}, and thus the parton density in the transverse 
nucleon area is about 1/fm$^2$; hence the above estimate of $\t_0$
appears applicable. The number of partons in a nucleon increases with 
c.m.s.\ collision energy roughly as $s^{1/6}$ \cite{MRS}, so that
the density at RHIC becomes about 2/fm$^2$, at the LHC about 6/fm$^2$.
In Fig.\ \ref{accel}, we show the initial energy densities for central
$A\!-\!A$ collisions with $A=200$; the range shown extends from
the value based on $\tau_0=1$ fm to that obtained if we scale $\t_0$ 
accord to the mentioned parton densities, always with $w_h\simeq 0.5$ GeV.
We note that in all cases the energy densities exceed the deconfinement 
value $\e(T_c) \simeq 0.5 - 1.0$ GeV/fm$^3$. We also indicate which 
temperatures are expected to be accessible at the different facilities, 
using the results of Fig.\ \ref{edens}.

\medskip

In the following sections, we shall try to show how one can probe 
deconfinement in nuclear collisions, based on the analysis of 
charmonium production; the extension to bottomonium is quite
straightforward.
First, the production of the probe must be understood when there is
no bulk medium, so that we begin with charmonium production in elementary 
collisions. Next, one has to understand the modifications arising
when the production occurs in a confined medium, for which $p\!-\!A$
collisions provide the experimental reference. With the probe thus
prepared and gauged in confined matter, it can be applied in an
environment in which there might be deconfinement.

\vskip0.5cm

\noindent{\bf 5.2 Hadroproduction of Charmonia}

\bigskip

\begin{figure}[tbp]
\vskip0.5cm
\centerline{\psfig{file=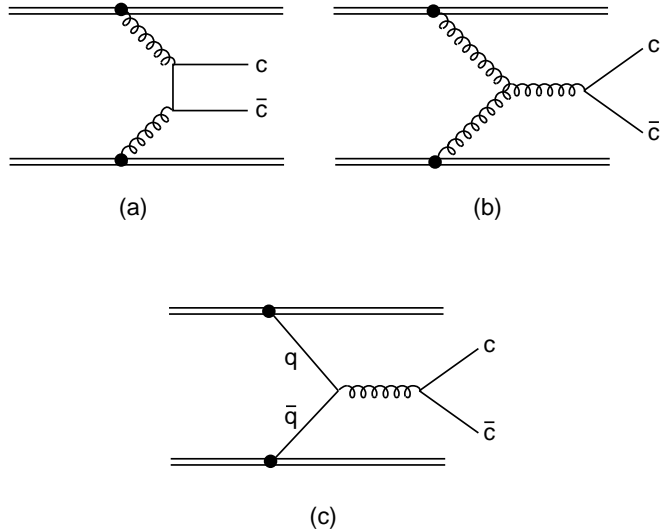,height=70mm}}
\caption{Lowest order diagrams for $\C$ production in hadronic
collisions, through gluon fusion (a,b) and quark-antiquark
annihilation (c).}
\label{5_1}
\end{figure}

The hadroproduction of charmonia occurs in three stages. The first stage
is the production of a $\C$ pair; because of the large quark mass, this
process is well described by perturbative QCD (Fig.\ \ref{5_1}).
A parton from the projectile interacts with one from the target; the
(non-perturbative) parton distributions within the hadrons are
determined empirically in other reactions, e.g.\ by deep inelastic
lepton-hadron scattering. The produced $\C$ pair is in general in
a colour octet state. In the second stage, it neutralises its colour,
leading to the third stage, physical resonances, such as \J, \X~ or \P. 
Colour neutralisation occurs by interaction with the surrounding colour 
field; this and the corresponding resonance binding are presumably of
non-perturbative nature. 

\medskip

On a fundamental theoretical level, colour neutralization is not yet
fully understood. However, the colour evaporation model \cite{CE}
provides a simple and experimentally well-supported phenomenological 
approach. In the evaporation process, the $\C$ can
either combine with light quarks to form open charm mesons ($D$ and
$\bar D$) or bind with each other to form a charmonium state.
The basic quantity in this description is the total sub-threshold
charm cross section $\S$, obtained by integrating the perturbative $\C$
production cross section $\sigma$ over the mass interval from 
$2m_c$ to $2m_D$. 
At high energy, the dominant part of $\S$ comes from gluon fusion (Fig.\
\ref{5_1}a), so that we have
\be
\S(s) \simeq \int_{2m_c}^{2m_D} d\hat s \int dx_1 dx_2~g_p(x_1)~g_t(x_2)~
\sigma(\hat s)~\delta(\hat s-x_1x_2s), 
\label{5.1}
\ee
with $g_p(x)$ and $g_t(x)$ denoting the gluon densities, $x_1$ and $x_2$
the fractional momenta of gluons from projectile and target,
respectively; $\sigma$ is the $gg \to \C$ cross section. In
pion-nucleon collisions, there are also significant quark-antiquark
contributions (Fig.\ \ref{5_1}c), which become dominant at low
energies. The basic statement of the colour evaporation model is
that the production cross section of any charmonium state $i$
is a fixed fraction of the subthreshold charm cross section, 
\be
\sigma_i(s)~=~f_i~\S(s), \label{5.2}
\ee
where $f_i$ is an energy-independent constant to be determined empirically. 
It follows that the energy dependence of the production cross section 
for any charmonium state is predicted to be that of the perturbatively
calculated sub-threshold charm cross section. As a further consequence,
the production ratios of different charmonium states
\be
{\sigma_i(s)\over \sigma_j(s)} = {f_i\over f_j} = {\rm const.}
\label{5.3}
\ee
must be energy-independent. Both these predictions have been compared in 
detail to charmonium and bottomonium hadroproduction data over a wide range
of energies \cite{quarko}; they are found to be well supported, both in the 
energy dependence of the cross sections and in the constancy of the
relative species abundances.  Let us consider in more detail what 
this tells us about the hadronization of charm quarks.

\medskip

We recall that the relative abundances of light hadrons produced in 
hadron-hadron and $e^+e^-$ interactions follow the statistical pattern 
governed by phase space weights \cite{Hagedorn,Becattini}: the relative 
production rates are those predicted by an ideal resonance gas at the 
confinement/deconfinement transition temperature $T_c\simeq 175$ MeV. 
For two hadron species $i$ and $j$ that implies at all (high) collision 
energies
\be
R_{i/j} \simeq {d_i \over d_j} \left({m_i\over m_j}\right)^{3/2}
\exp-\{(m_i-m_j)/T_c\}
\label{rates}  
\ee
for the ratio of the production rates, with $d_i$ for the degeneracy
(spin, isospin) and $m_i$ for the mass of species $i$. For strange
particles, the rates (\ref{rates}) overpredict the experimental data;
this can, however, be accommodated by a common strangeness suppression 
factor $\gamma_s \simeq 0.5$, applied as $\gamma_s^n$ if the produced 
hadron contains $n$ strange quarks.

\medskip

For the hadroproduction of charm, such a statistical description does
not work, as seen in three typical instances: 
\begin{itemize}
\vspace*{-0.1cm}
\item{The total $\C$ cross section increases with energy by about a
factor ten between $\sqrt s = 20$ and 40 GeV, while the light hadron
multiplicity only grows by about 20\%. Hence the ratios for hadrons 
with and without charm are not energy-independent.} 
\vspace*{-0.2cm}
\item{From $p-p$ data one finds for \J~production a weight factor 
$f_{\j} \simeq 2.5 \times 10^{-2}$ \cite{quarko}. Since the subthreshold 
$\C$ cross section is about half of the single $D$ production cross 
section \cite{Quack}, this implies $R_{(\j)/D} \simeq 10^{-2}$;
the ideal resonance gas gives with $R_{(\j)/D} \simeq 10^{-3}$
a prediction an order of magnitude smaller. Of the total charm production,
more goes into the hidden charm sector than statistically allowed.} 
\vspace*{-0.2cm}
\item{ For the production ratio of \P~to \J, which have the same charm 
quark infrastructure, one finds experimentally over a wide range of 
collision energies $R_{\p/(\j)} \simeq 0.23$. This energy-independent 
\P~to \J~ratio can be accounted for in terms of the charmonium masses and 
wave functions; it disagrees strongly with the statistical prediction, which 
gives with $R_{\p/(\j)} \simeq 0.045$ a very much smaller value. The same
holds true for the other measured charmonium states, as seen in Table 5, 
where we
list experimental results obtained in 300 GeV/c $\pi$ and proton 
interactions \cite{Antoniazzi}.
All ratios are given relative to directly produced (1S) \J~states.}
\vspace*{-0.1cm}
\end{itemize} 

While the first two points could be accommodated by an energy-dependent
`charm enhancement' factor, the last rules even this out. In the presently 
available collision energy range, charm production in elementary collisions 
thus does not seem to be of statistical nature. It appears to be determined 
by parton dynamics at an early stage, rather than by the phase space
size at the confinement temperature.

\vskip1cm

\centerline{
\renewcommand{\arraystretch}{1.8}
\begin{tabular}{|c|c|c|c|}
\hline
{\rm ratio}& $\psi'/(J/\psi)$ & $\chi_{c1}/(J/\psi)$ & $\chi_{c2}/(J/\psi)$ \\
\hline
{\rm experimental}&
0.23&
1.06&
1.50\cr
\hline
{\rm statistical}&
0.045&
0.113&
0.148\cr
\hline
\end{tabular}}

\vskip1cm

\centerline{~~Table 5: Charmonium production ratios vs.\ statistical
predictions}

\vskip0.5cm

Although the colour evaporation model provides a viable phenomenological
description of the hadroproduction of quarkonia, leading to correct 
quantitative predictions up to the highest energies under consideration,
it cannot predict the fractions $f_i$ of the hidden charm cross sections, 
and it can even less describe the space-time evolution of colour 
neutralization. For charmonium production in p-A and A-B collisions, 
the latter is crucial, however, and hence a more detailed description 
of colour neutralization is needed. 

\medskip

\begin{figure}[htb]
\centerline{\epsfig{file=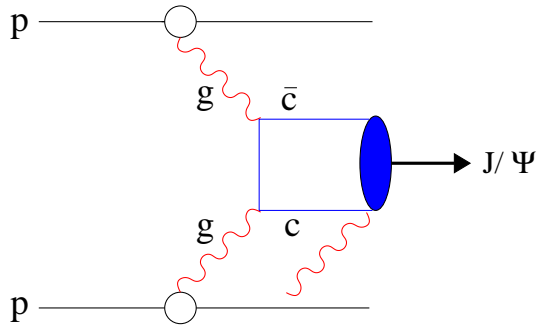,width=7cm}}
\caption{The evolution of \J~production}
\label{Jpsi-evo}
\end{figure}

\begin{figure}[htb]
\centerline{\epsfig{file=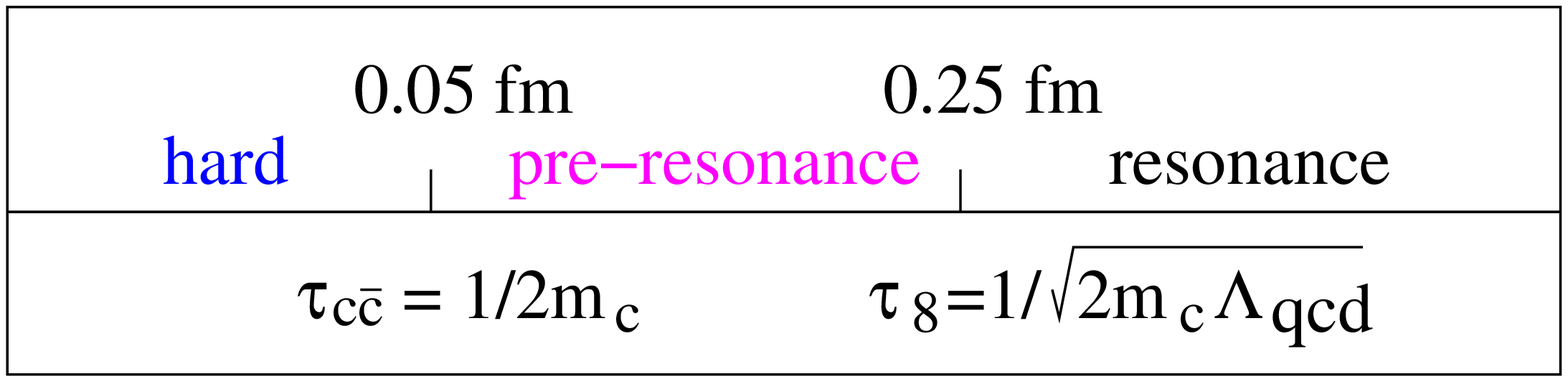,width=8cm}}
\caption{Scales of \J~production}
\label{scales}
\end{figure}

The colour octet model \cite{CO} proposes that the colour octet $\C$ 
combines with a soft collinear gluon to form a colour singlet $(\C\!-\!g)$ 
state. After a rather short relaxation time $\tau_8$, this pre-resonant 
$(\C\!-\!g)$ state turns into the physical $\C$ mode by absorbing 
the accompanying gluon, with similar formation processes for $\chi_c$ and 
\P~production. The colour octet model encounters difficulties if the 
collinear gluons are treated perturbatively, illustrating once more that 
colour neutralization seems to require non-perturbative elements 
\cite{polarize}. However, it does provide a conceptual basis for the 
evolution of the formation process (see Fig.\ \ref{Jpsi-evo}). The colour 
neutralization time $\tau_8$ of the pre-resonant state can be estimated 
\cite{KS6}; it is essentially determined by the lowest momentum possible 
for confined gluons, $\t_8 \simeq (2m_c\l)^{1/2} \simeq 0.25$ fm.
The resulting scales in \J~formation are illustrated in Fig.\ \ref{scales}.
The formation time for the actual physical ground state \J~is presumably
somewhat larger than $\t_8$; although $r_{\j} \simeq \t_8$, the
heavy $c$ quarks do not move with the velocity of light. For the larger 
higher excited states, the formation times will then be correspondingly 
larger.

\medskip

There is one further important feature to be noted for \J~hadroproduction. 
The \J's actually measured in hadron-hadron collisions have three distinct
origins: about 60 \% are directly produced 1S charmonium states, while
about 30 \% come from the decay $\x_c(1P) \to \j +$ anything, and the remaining
10 \% from $\p(2S) \to \j +$ anything \cite{feeddown}. 
Such feed-down also occurs in \U~production \cite{Upsi-feed}. In all cases,
the decay widths of the involved higher excited states are 
extremely small (less than one MeV), so that their life-times are very long. 
The presence of
any medium in nuclear collisions would therefore affect these excited 
states themselves, and not their decay products.

\vskip0.5cm

\noindent{\bf 5.3 Charmonium Production in $p-\!A$ Collisions}

\bigskip

We have seen that in the use of charmonia to study nuclear collisions,
the creation of the medium and the production of the probe lead to
two distinct formation scales, $\t_0$ and $\t_8$. In $p-\!A$ collisions, 
the presence of normal nuclear matter can affect charmonium production. 
Here there is no formation time for the medium, so that such collisions 
provide a tool to probe charmonium production, evolution and absorption 
in confined matter.   

\medskip

Nuclear effects can arise in all the evolution stages of \J~production,
and a number of different phenomena have been studied in considerable
detail.
\begin{itemize}
\item{ The presence of other nucleons in the nucleus can modify the 
initial state parton distribution functions, which enter in the perturbative
$\C$ production process illustrated in Fig.\ \ref{5_1}. This can lead
to a decrease ({\sl shadowing}) or to an increase ({\sl antishadowing})of 
the production rate.}
\item{Once it is produced, the $\C$ pair can suffer {\sl absorption} in 
the pre-resonance as well as in the resonance stage, caused by the
successive interactions with the target nucleons.}
\item{The extent of this absorption can moreover be modified through
{\sl coherence} effects for the interactions on different nucleons,
which can lead to partial cancellation (Landau-Pomeranchuk-Migdal effect).}
\end{itemize} 
In all cases, the crucial quantity is the momentum of the charmonium state
as measured in the nuclear target rest frame. 

\medskip

Since we eventually want to probe the effect which the `secondary' medium 
{\sl produced} by nucleus-nucleus collisions has on charmonium production, 
it is of course essential to account correctly for any effects of the
nuclear medium initially present. Let us therefore first summarize 
the main features observed for charmonium production in $p-\!A$ collision
experiments \cite{Leitch}.

\begin{itemize}
\item{At fixed collision energy, quarkonium production rates per target
nucleon decrease with increasing $A$.}
\item{The production rates decrease for increasing \J~momentum as measured
in the nuclear target rest frame.}
\item{The nuclear reduction at $p-N$ mid-rapidity appears to become weaker
with increasing collision energy.}
\item{For fixed collision energy, mass number $A$ and \J~rapidity, the 
reduction appears to increase with the centrality of the collision.} 
\item{At sufficiently high momentum in the target rest frame, the different 
charmonium states appear to suffer the same amount of reduction, while at 
lower energy, the \P~is affected more than the \J.}
\end{itemize}

At present, there does not seem to exist a theoretical scenario able to
account quantitatively for all these observations. In fact, so far not even 
a common scaling variable for the different effects has been found.
Shadowing would suggest scaling in the fractional target parton 
momentum $x_2$, while absorption of the pre-resonance state would point 
to the fractional beam momentum $x_F$. Neither is in good accord with the 
data. For a further discussion, we therefore refer to recent reviews 
\cite{Leitch,Vogt} and concentrate here on obtaining some operational methods 
to specify nuclear effects in $p-\!A$ collisions in a way that can be 
extended to $A-B$ collisions. This, incidentally, illustrates the crucial 
importance of having $p\!-\!A$ data in order to arrive at an interpretation
of $A\!-\!A$ results; for future LHC experiments, where parton saturation
may play a crucial role, this is even more so the case.

\medskip

Consider the production of a charmonium state in a $p-\!A$ collision
\cite{K-S-ZP}.
If the initial $\C$ production process occurs at a random point inside
the nucleus, the evolving charmonium state will on the average traverse
a path of length $L_A= 3 R_A/4$ through nuclear matter. We have to know
in what stage of its evolution the charmonium will be along this path.
The distance travelled by the charmonium is, in the rest frame of the nucleus, 
given by
\be
d = \t \left({P_A \over M}\right) 
\label{distance}
\ee
where $P_A$ is the charmonium momentum in the nuclear rest frame,
$M$ the mass and $\t$ the proper age of the charmonium, i.e., its 
lifetime in its own rest frame. For simplicity, we consider for the
moment only charmonia produced at rest in the center of mass of a 
proton-nucleon collision ($x_F=0$), where most of the data are taken. 
Then we have
\be
P_A \simeq {M\over 2 m} \sqrt s,
\label{PA}
\ee
with $m$ for the nucleon mass and $\sqrt s$ for the proton-nucleon 
c.m.s.\ collision energy.  
In Table 6, we show as illustration the resulting values for the 
charmonium momenta $P_A$ in the nuclear rest frame at different collision
energies; FNAL refers to the energy of the fixed target program. We
also give the corresponding colour neutralization path lengths 
$d_8 = d(\t=\t_8)$. Note that these are the values for $x_F=0$; 
$P_A$ and $d_8$ depend quite sensitively on $x_F$.

\vskip0.5cm

\centerline{
\renewcommand{\arraystretch}{2.0}
\begin{tabular}{|c|c|c|c|}
\hline
{\rm accelerator}& $\sqrt s~[{\rm GeV]}$ & $P_A ~[{\rm GeV}]$ & 
$d_8 ~[{\rm fm}]$ \\ 
\hline
{\rm SPS}&
17&
29&
2\cr
\hline
{\rm SPS}&
29&
48&
4\cr
\hline
{\rm FNAL, HERA B}&
40&
66&
5\cr
\hline
{\rm RHIC}&
200&
330&
26\cr
\hline
{\rm LHC}&
5500&
9000&
730\cr
\hline
\end{tabular}}

\vskip0.7cm

\centerline{Table 6: Charmonium formation parameters at $x_F=0$
for different collision energies}

\vskip0.5cm

We now follow the charmonium in time along its path through the nucleus.
For times shorter than the colour neutralization time $\t_8$, $\t < \t_8$,
the nucleus is traversed by a pre-resonance state, which can be dissociated 
through interaction with nuclear matter, with a dissociation cross section 
$\sigma_{\rm diss}$. For $\t > \t_8$, the physical resonance passes through 
the remaining nuclear medium. From Table 5 we note that for RHIC and LHC, the 
colour neutralization point lies well outside any nucleus, so that there 
the nuclear medium only sees the pre-resonance state. For FNAL and HERA B,
$d_8$ is approximately equal to the size of the largest nuclei, so that
here as well only pre-resonance effects should play a role. Since up to 
times $\t_8$ the different charmonium states are indistinguishable, 
absorption effects for $\sqrt s > 40$ GeV (at $x_F=0$) should be the
same for all states. At the SPS, both pre-resonance and resonance
absorption can come into play, so that the effect should be stronger
for the \P~than for the \J. Here it should be noted, however, that in 
both cases, the expansion of the $\C$ occurs with less than the speed
of light, so that at $\tau=\tau_8$ the colour singlet is not yet of physical
size. 
   
\medskip

Let us now quantify these considerations \cite{K-L-N-S}.
According to the Glauber formalism of nuclear scattering theory,
a $\C$ pair formed at point $z_0$ in a target nucleus $A$ has a survival 
probability
\be
S_i^A = \int d^2 b~\! dz\! ~\rho_A(b,z)
\exp \left\{ -(A-1)\int_{z_0}^{\infty} dz'~ \rho_A(b,z')~
\sigma^i_{\rm diss} \right\},
\label{5.11}
\ee
where the integration covers the path, at impact parameter $b$,
remaining from $z_0$ out of the nucleus, and where $\sigma^i_{\rm diss}$ 
describes the overall ``absorption'' effect on the observed
charmonium state $i$ along the path. The result is then averaged over
impact parameter and path lengths. The traversed medium of nucleus $A$ 
is parametrized through a Woods-Saxon density distribution $\rho_A(z)$, 
and by comparing $S_i$ to data for different targets $A$, the effective 
dissociation cross section can be obtained for the \J~and \P~absorption 
in nuclear matter. The effect of the charmonium passage through the nucleus 
will arise from a superposition of the different stages; but if
part of the passage is carried out as physical resonance, higher 
excited states should lead to larger absorption cross-sections than the 
much smaller ground state \J. In accord with this, the most recent
SPS analysis, based on $p\!-\!A$ collisions with $Be,~Al,~Cu,~W$ and 
$Pb$ targets, gives \cite{Bordalo} 
\be
\sigma_{\j} = 4.18 \pm 0.35~{\rm mb} \label{5.12}
\ee
for the \J, and 
\be
\sigma_{\p} = 7.3 \pm 1.6~{\rm mb} \label{5.12a}
\ee
for the \P~as nuclear absorption cross sections at $\sqrt s \simeq 29$ GeV. 

\medskip

\begin{figure}[htb]
\centerline{\epsfig{file=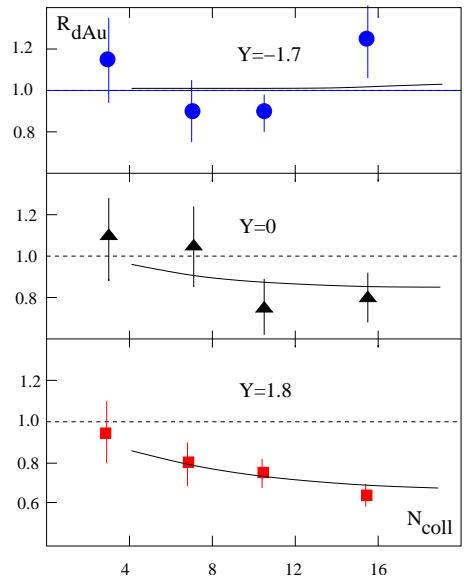,width=6cm}}
\caption{\J~production in $d-Au$ collisions at RHIC \cite{Phenix}}
\label{d-Au}
\end{figure}

Data from the SPS were taken for a number of different nuclear targets,
but so far no information on the centrality-dependence of the results has 
been given. In contrast, RHIC experiments have provided only data for
$d-\!Au$ collisions, but as function of centrality and for three different
rapidity ranges \cite{Phenix}. In the central rapidity range $|y|\leq 0.35$,
the data is obtained from $e^+e^-$ measurements, in the larger rapidity
intervals $1.2 \leq y \leq 2.2$ and $-2.2 \leq y \leq -1.2$ through forward 
and backward $\mu^+\mu^-$ detection. The results are shown in Fig.\ 
\ref{d-Au}, where
\be
R_{d-Au}(y) = {(d\sigma/dy)_{d-Au} \over N_{coll}~ (d\sigma/dy)_{p-p}}
\label{R-dAu}
\ee
measures the nuclear modification in the given $y$ range, relative to 
the corresponding $p-p$ cross section scaled by the number $N_{coll}$ 
of binary nucleon-nucleon collisions. This number is obtained from the
measured number of participant nucleons through a Glauber analysis and
is also used to parametrize the centrality dependence.

So far, within the rather limited statistics available, it is not easy
to quantify these RHIC results enough to obtain a reasonable estimate
of normal nuclear effects. As first attempt, we adopt a similar 
parametrization as used for SPS results and apply the well-known
simplified form of eq.\ (\ref{5.11}),
\be
S \simeq \exp\{-n_0 \sigma_{\rm diss} L\},
\label{path}
\ee
where $L$ denotes the path of the $\C$ in the nuclear medium.
Glauber analysis \cite{K-L-N} provides the relation between impact 
parameter $b$ and the number of collisions $N_{coll}$, and simple 
geometry gives $L^2=R_A^2 - b^2$ in terms of $b$ and the
nuclear radius $R_A$. A fit of Eq.\ (\ref{path}) to the data of 
Fig.\ \ref{d-Au} gives\footnote{In the fit, we neglect the most 
peripheral point at $N_{coll}$, which corresponds to $b > R_{Au}$
and is thus due to nuclear surface rather than medium effects.}  
$$
\sigma_{\rm diss}(y=1.8) = 3.1 \pm 0.2 ~{\rm mb}
$$
\vskip-0.7cm
$$
\sigma_{\rm diss}(y=0) = 1.2 \pm 0.4 ~{\rm mb}
$$
\be
\sigma_{\rm diss}(y=-1.7) = -0.1 \pm 0.2 ~{\rm mb}
\label{cross}
\ee
for the corresponding \J~dissociation cross sections; for $y=-1.7$,
there thus are essentially no nuclear modifications. We note that
here, as for the SPS case, these cross sections are just a global
way to account for whatever nuclear effects can arise. A more detailed
analysis based on shadowing and absorption is given in \cite{ramona-shadow}.

\bigskip

\noindent{\bf 5.4 Charmonium Production in Nuclear Collisions}

\bigskip

\begin{figure}[htb]
\centerline{\epsfig{file=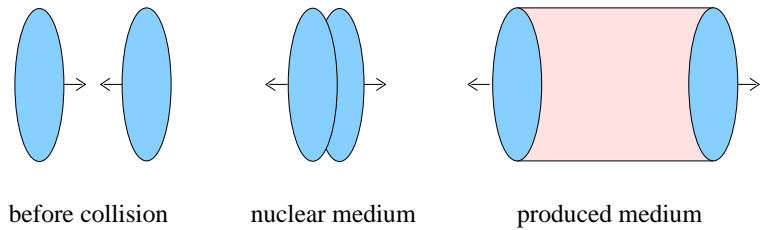,width=10cm}}
\caption{Collision stages}
\label{col-evo}
\end{figure}

The basic assumption in the attempt to create deconfined matter through
nuclear collisions is that the excited vacuum left after the passage of
the colliding nuclei forms a thermal medium. This picture is schematically
illustrated in Fig.\ \ref{col-evo}. A charmonium state produced in such
a collision will in its early stages always be subject to the possible
effects of the nuclear medium, just as it is in $p-\!A$ collisions. 
Knowing the $p-\!A$ behaviour at the corresponding energy is thus a 
necessary baseline for probing the additional effects of the produced 
medium. 

\medskip

\medskip

The Glauber formalism used above to calculate the survival probability of 
an evolving charmonium state in a $p-A$ collision now has to be extended
to $A-B$ interactions \cite{K-L-N-S}. The survival probability at impact 
parameter $b$ now becomes
\vskip-0.2cm
$$
S^{AB}_i(b) = \int d^2s~\!dz~\!dz' \rho_A(s,z) \rho_B(b-s,z') ~\times
$$
\vskip-0.2cm
\be
\exp \left\{ -(A-1)\int_{z_0^A}^{\infty} dz_A~ \rho_A(s,z_A)~
\sigma_i ~-(B-1)\int_{z_0^B}^{\infty} dz_B~ \rho_B(b-s,z_B)~ \sigma_i
\right\},
\label{5.13}
\ee
\vskip0.2cm
as extension of Eq.\ (\ref{5.11}). Here $z_0^A$ specifies the formation
point of the $\C g$ within nucleus $A$, $z_0^B$ its position in $B$.
With the dissociation cross sections $\sigma^i_{diss}$ determined in
$p-A$ collisions, Eqs.\ (\ref{5.12}/\ref{5.12a}), the `normal'
survival probability, i.e., that due to only the nuclear medium,
is fully specified.

\medskip

Since experiments cannot directly measure the impact parameter $b$, 
we have to specify how Eq.\ (\ref{5.13}) can be applied to data.
The Glauber formalism allows us to calculate the number $N^{AB}_w(b)$ of 
participant (`wounded') nucleons for a given collision as function of
centrality. In $A-\!A$ collisions, the number of unaffected `spectator' 
nucleons $N^A_s(b)$ can be determined directly through the use of a 
zero-degree calorimeter, and hence $N^{AA}_w(b)=2A-2N^A_s(b)$ is 
accessible experimentally. For asymmetric $AB$ collisions, the situation 
is somewhat more complex.

\medskip

\begin{figure}[htb]
\centerline{\epsfig{file=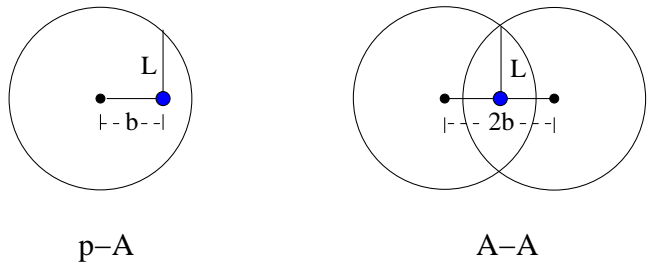,width=8.5cm}}
\caption{Impact parameter relation between $p-\!A$ and $A-\!A$ collisions}
\label{impact}
\end{figure}

At RHIC energy, we make again use of the simplified form (\ref{path}). 
The geometry connecting the impact parameter $b$ and path length $L$
in $p-Au$ and $Au-Au$ collisions is illustrated in Fig.\ \ref{impact};
the relation between $b$ and $N_{part}$ is again given by a Glauber
analysis \cite{K-N}. The resulting survival probability is 
\be
S^{AA}_i(y,N_w)= {R_{AA}(y,N_w)\over \exp\{-n_0 [\sigma_{\rm diss}(y)
+ \sigma_{\rm diss}(-y)] L\}},
\label{S-AA}
\ee
corresponding to the fact that, for $y \not= 0$, the charmonium state
passes one nucleus at rapidity $y$ and the other at rapidity $-y$.

\medskip

\medskip

Eqs.\ (\ref{5.13}) and (\ref{S-AA}) specify the suppression pattern for 
charmonium states as predicted from normal nuclear effects. 
To use quarkonia as probes for the produced medium, we now have to study 
how the behaviour observed in $A-\!A$ collisions differs from this 
predicted pattern. 

\medskip

Several possible and quite different effects have been considered as
consequences of the produced medium on quarkonium production. To be
specific, we again consider the charmonium case for illustration. 
\begin{itemize}
\vskip0.2cm 
\item{Suppression by comover collisions: A charmonium state produced in 
a primary nucleon-nucleon collision can be dissociated through interactions
with the constituents of any medium subsequently formed in the collision.
Such dissociation could occur in a confined \cite{hadro-co}
as well as in a deconfined medium \cite{K-S-OP}.}
\item{Suppression by colour screening: If the produced medium is a hot
QGP, it will dissociate by colour screening the charmonium states produced 
in primary nucleon-nucleon collisions. Due to the rareness of thermal charm 
quarks in the medium, the separated $c$ and $\bar c$ combine at hadronization 
with light quarks to form open charm mesons \cite{M-S}.}
\item{Enhancement by recombination: In the hadronization of the QGP, 
charmonium formation can occur by binding of a $c$ with a $\bar c$ from 
different nucleon-nucleon collisions, as well as from the same. If the 
total number of available $\C$ pairs considerably exceeds their thermal 
abundance, statistical recombination enhances hidden relative to open 
charm production, as compared to hadron-hadron collisions \cite{enhance}.}
\end{itemize}
In addition, the partonic initial state of the colliding nuclei, which 
leads to the formation of the produced medium and to that of 
charmonium states, will change its nature with increasing $A$ and/or 
$\sqrt s$; eventually, parton percolation (colour glass formation) can 
lead to a very different medium, with possible effects on production 
and binding of charmonia \cite{perco}.

\medskip

Is it possible for experiment to distinguish between these different
scenarios? Before turning to the experimental situation, we want
to discuss in some more detail the salient features of each approach. 

\bigskip

{\bf Suppression by Comover Collisions}

\medskip

If the charmonium state moves in a random scattering pattern through
the produced medium, its survival rate is approximately given by
\be
S_i = \exp\{-\sigma_i n \tau_0 \ln [n/n_f]\}
\label{comover}
\ee
with $\sigma_i$ denoting the dissociation cross section, $n$ the
initial density of the medium after a formation time $\tau_0$, and
$n_f$ the `freeze-out' density, at which the interactions stop.

\medskip

Since the cross section for \J~break-up through gluon collisions is 
large \cite{K-S-OP} and the gluon density high, there will be significant 
charmonium suppression in a deconfined medium, even if this is not
thermalized. In an equilibrium QGP, this dissociation is presumably
accounted for by colour screening, provided the effect of the
medium on the width of the surviving states is also calculated.

\medskip

Charmonium dissociation by interaction with hadronic comovers has
received considerable attention in the past \cite{hadro-co}. However,
if one restricts the possible densities to values appropriate to
hadronic matter ($n_h \lsim 0.5$ fm$^{-3}$) and the cross sections
to those obtained in section 3, the effect of hadron dissociation
is negligible. Even a cross section increase to the high energy
limit still leads to less than 10\% effects. More recent analyses
\cite{Maiani} thus conclude that a hadronic medium will not result
in significant suppression.

\medskip

\begin{figure}[h]
\vskip0.5cm
\begin{minipage}[t]{7cm}
\epsfig{file=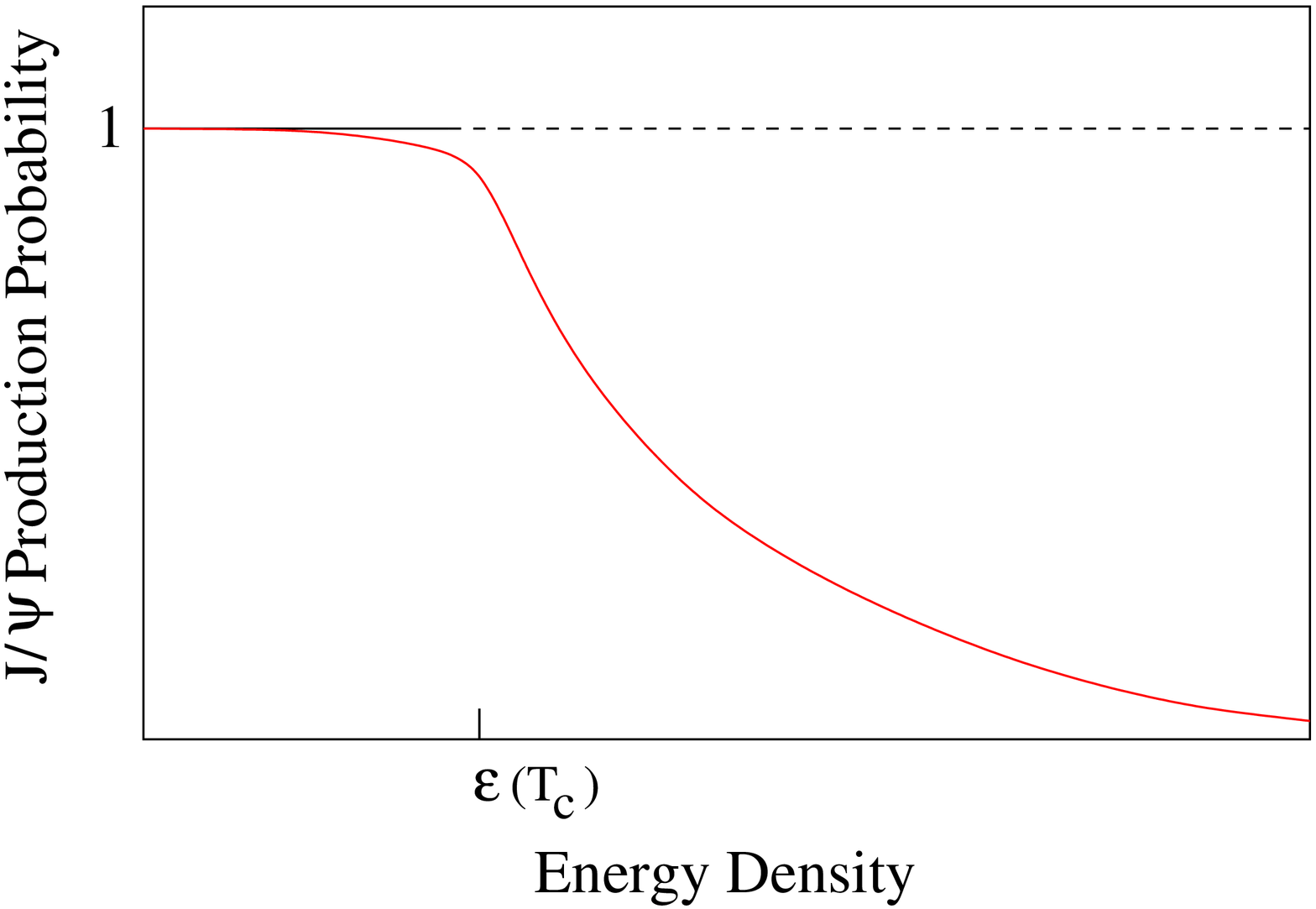,width=6.5cm}
\caption{\J~suppression by comover collisions}
\label{abs-cartoon}
\end{minipage}
\hspace{1.3cm}
\begin{minipage}[t]{7cm}
\vspace*{-4.5cm}
\hskip0.3cm
\vspace*{-0.1cm}
\epsfig{file=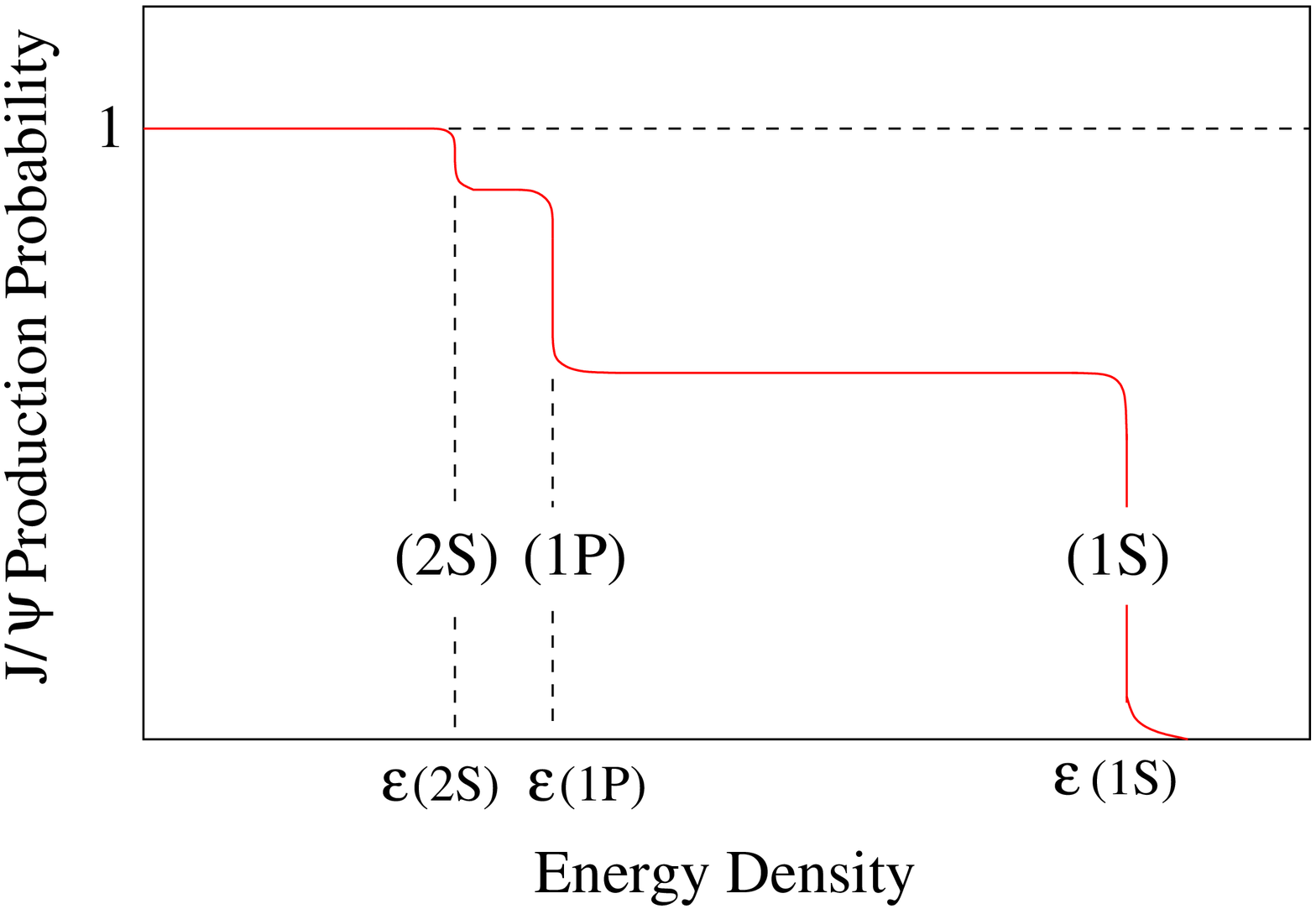,width=6.7cm}
\caption{Sequential \J~suppression by colour screening}
\label{seq-cartoon}
\end{minipage}
\end{figure}

In Fig.\ \ref{abs-cartoon} we illustrate schematically the overall
behaviour expected for \J~dissociation through comover collisions,
assuming that beyond a deconfinement threshold, the comover density 
increases with energy density in a monotonic fashion, with little or
no prior suppression in the hadronic regime. 

\medskip

\bigskip

{\bf Suppression by Colour Screening}

\medskip

The theoretical basis of this effect has been considered in detail in
chapter 5; the colour field between the heavy quarks becomes modified 
due to the presence of a medium of unbound 
colour charges. The results obtained for this effect in statistical QCD 
are as such model-independent, once all calculational constraints are 
removed. What is speculative and model-dependent is its application to 
nuclear collisions: these do not necessarily produce the medium studied 
in thermal QCD, and the different evolution stages in nuclear collisions 
can introduce factors not present in the study of equilibrium thermodynamics, 
such as the oversaturation of $\C$ pairs just mentioned.

\medskip

If the medium produced in high energy nuclear collisions is indeed the
quark-gluon plasma of statistical QCD, and if charmonium production can
be treated as a distinct process within such an environment, then the
effect of colour screening seems clear. The partition of the $\C$ pairs
produced in nucleon-nucleon collisions into hidden and open charm is
non-statistical, favouring the hidden charm sector because of dynamical
binding effects. Colour screening destroys these and hence moves the partition 
more and more towards a statistical distribution, thus suppressing 
charmonium production rates relative to those observed in elementary 
interactions. 

\medskip

A crucial feature of \J~suppression by deconfinement is its sequential
nature \cite{Gupta,D-P-S2}. In the feed-down production of \J, the 
produced medium affects the intermediate excited states, so that 
with increasing temperature or energy density, first the \J's originating
from \P~decay and then those from \X~decay should be dissociated. Only
considerably higher temperatures would be able to remove the directly
produced \J's. Such a stepwise onset of suppression, with specified
threshold temperatures, is perhaps the most characteristic feature 
predicted for charmonium as well as bottomonium production in nuclear
collisions. It is illustrated schematically in Fig.\ \ref{seq-cartoon},
where we have defined the \J~production probability to be unity if
the production rate suffers only the calculated nuclear suppression. 

\bigskip

{\bf Enhancement through Recombination}

\medskip

In charmonium hadroproduction, \J's are formed because some of the 
$\C$ pairs produced in a given collision form a corresponding bound state.
In a collective medium formed through superposition of many nucleon-nucleon
(NN) collisions, such as a quark-gluon plasma, a $c$ from one NN collision
can in principle also bind with a $\bar c$ from another NN collision. 
This `exogamous' charmonium formation can lead to enhanced \J~production,
provided the overall charm density of the medium at hadronization is 
sufficiently high and provided the binding probability between charm 
quarks from different sources is large enough.

\medskip

The production of $\C$ pairs in primary collisions is a hard process and
thus grows in $A-\!A$ interactions with the number of nucleon-nucleon
collisions; in contrast, the multiplicity of light hadrons grows roughly 
as the number of participant nucleons. Hence the relative abundance
of charm to non-charm quarks will be higher in $A-\!A$ than in $p-\!p$
collisions. Moreover, the $\C$ production cross section increases faster
with energy than that for light hadron production. The two effects
together imply that in the medium produced in energetic $A-\!A$ collisions, 
the ratio of charm to non-charm quarks is initially much higher than 
in a equilibrated QGP. 
If this oversaturation of charm is preserved in the thermalization
process, the combination of random $c$ and $\bar c$ quarks from different 
primary nucleon-nucleon interactions becomes more and more likely with 
increasing energy.

\medskip

Whether or not such an enhancement becomes significant depends on two
factors. On one hand, the initial charm oversaturation must be preserved,
so that the total charm abundance in non-thermal. On the other hand, it is
necessary that the `recombination' of the pairs into charmonia is
sufficiently strong. Here it is generally assumed that the final 
hadronization occurs according to the available phase space. Thus the 
number of statistically recombined \J's has the form 
$N_{\j} \sim N_{\C}^2 /N_h$, growing quadratically in the number of 
produced $\C$ pairs; here $N_h$ denotes the number of light hadrons 
(or quarks). This implies that the hidden to open charm ratio, e.g.,
$N_{\j} / N_{D} \sim N_{\C} /N_h$, increases with energy, in contrast 
to the energy independent form obtained for a fully equilibrated QGP, 
or to the decrease predicted by colour screening.

\medskip

\begin{figure}[htb]
\centerline{\epsfig{file=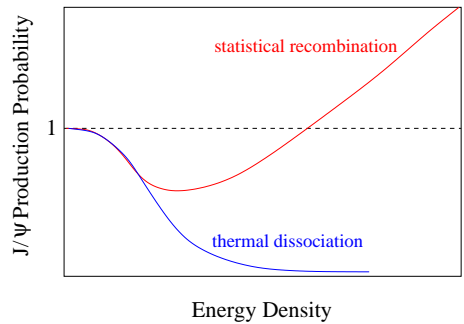,width=6cm}}
\caption{\J~enhancement by recombination} 
\label{recom}
\end{figure}

Several studies of this effect have led to different noticeable enhancement 
factors \cite{Bob}, in the strongest form even predicting an overall 
enhancement of \J~production in $A-\!A$ collisions relative to $p-\!p$ 
results scaled by binary collisions. A crucial prediction of the
approach is the {\sl increase} of the enhancement with centrality, as
shown in Fig.\ \ref{recom}, because of the corresponding increase in 
the number of collisions and hence of the number of $\C$ pairs.

\bigskip

{\bf Initial State Suppression}

\medskip

For two colliding nuclei at large enough $A$ or $\sqrt s$, the density 
of partons in the transverse plane becomes so large (see Fig. \ref{PP}) 
that the partons percolate, producing an interconnected network. If the 
resolution scale of the partons in this network is sufficient to 
resolve charmonia, it could also dissociate them. This scenario is in
many ways similar to suppression by colour screening, but it is on
the whole more model-dependent. Given the average transverse size of 
the partons in terms of the resolution scale of the parton distribution 
function, it is possible to calculate the percolation point; but its effect 
on the different charmonium states is not {\sl a priori} evident and
can only be addressed in models \cite{perco,D-F-S}.

\medskip

In other words, the relation between the onset of charmonium dissociation 
and the percolation point remains so far of phenomenological nature. In 
contrast to thermal QCD, it does not seem possible here to evaluate 
dissociation points from basic theory. One promising approach
might be the study of quarkonium dissociation in the strong classical
colour fields created in the percolating medium through the onset of
colour glass formation.  

\vskip0.8cm

\noindent{\large \bf 6.\ Experimental Results from Nucleus-Nucleus Collisions} 

\vskip0.5cm

Over the past twenty years, great experimental efforts have produced a
wealth of data on charmonium production in nucleus-nucleus collisions.
The pioneering work of the NA38 and NA50 collaborations at the SPS
led to the second generation experiment NA60, which is just now 
announcing its first results. At the same time, RHIC is providing us
with a first look at how things change when the collision energy is 
increased by a factor ten. It is certainly an exciting time; but 
although the results may give some indications, more information is 
necessary in order to reach final conclusions. So let us try to summarize 
what we know today and then try to point out what that could mean and 
what more is needed before we can be sure.

\vskip0.5cm

\noindent{\bf 6.1 SPS Results}

\bigskip

\begin{figure}[htb]
\begin{minipage}[t]{7cm}
\epsfig{file=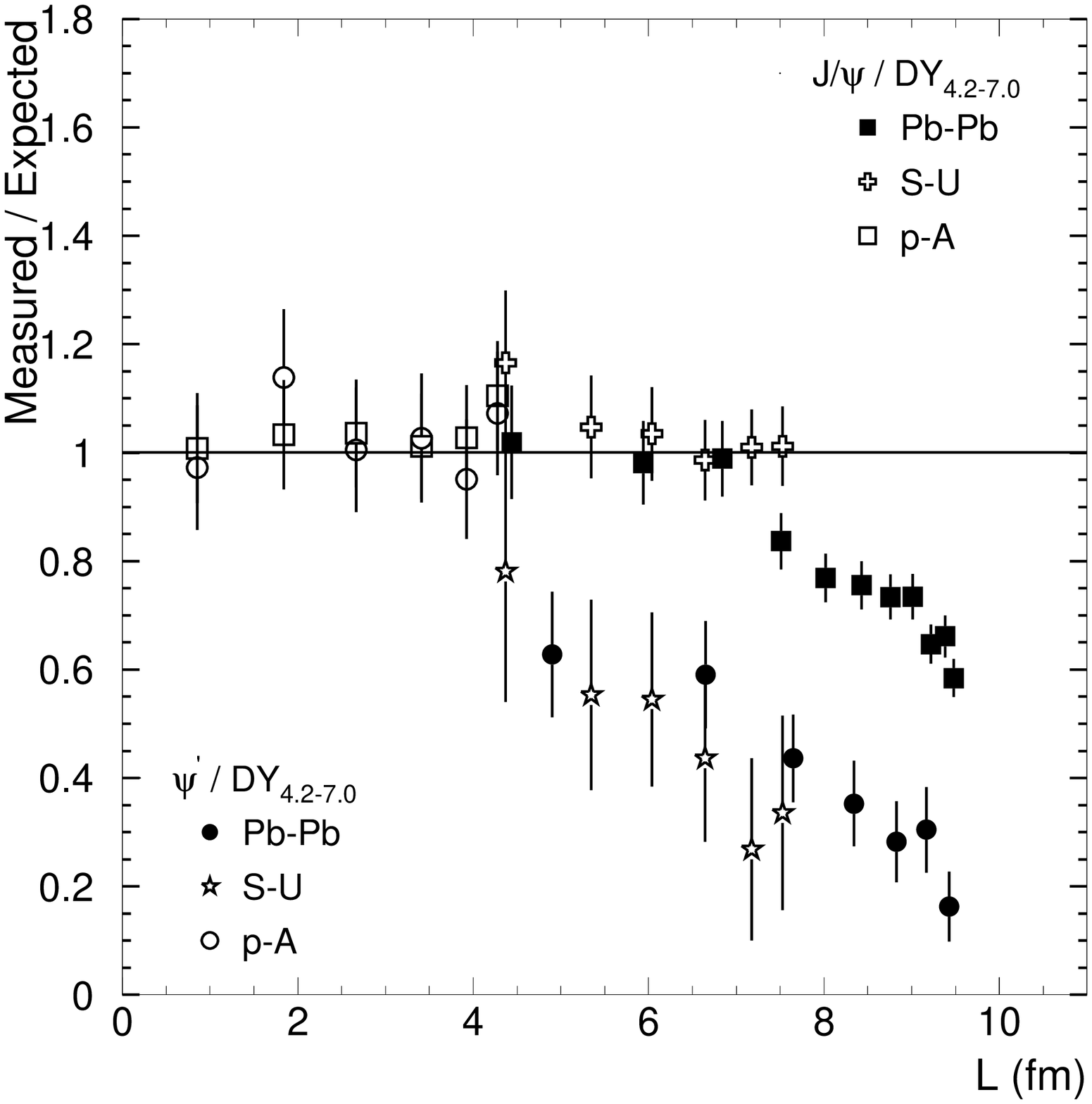,width=7.2cm}
\vspace*{-0.5cm}
\caption{\J~and \P~production from NA38/NA50 \cite{Kluberg}}
\label{Borges}
\end{minipage}
\hspace{1.3cm}
\begin{minipage}[t]{7cm}
\vspace*{-7.3cm}
\hskip-0.3cm
\epsfig{file=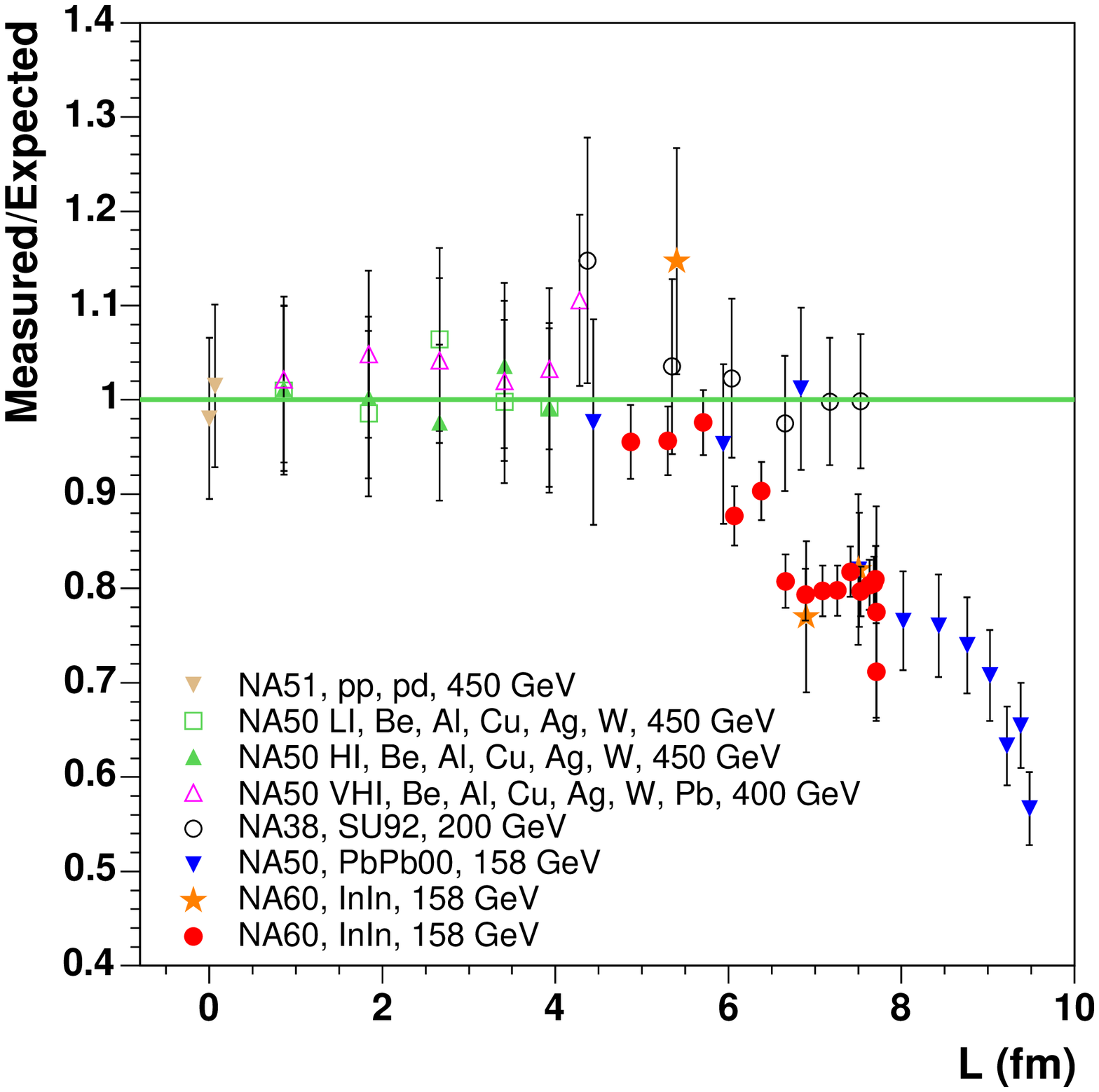,width=7.8cm}
\vspace*{-0.7cm}
\caption{\J~production from NA38/NA50/NA51/NA60 \cite{arnaldi-qm}}
\label{Ar2}
\end{minipage}
\end{figure}

Extensive data was taken for $S\!-\!U$ collisions (NA38) at $\sqrt s = 19.4$ 
GeV and for $Pb\!-\!Pb$ (NA50) collisions at $\sqrt s = 17.3$ GeV, for 
\J~\cite{Kluberg,arnaldi-qm} and for \P~production 
\cite{Sitta}, as well as the mentioned reference data for 
$p\!-\!A$ interactions. The main aspects of the NA38/NA50 results 
are shown in Fig.\ \ref{Borges}. Here the charmonium production rates are 
measured relative to Drell-Yan pair production in the same collision 
configuration, and the resulting ratio is then normalized to the value 
expected when normal nuclear absorption is taken into account, with the 
cross sections of eqs.\ (\ref{5.12}/\ref{5.12a}). The centrality is 
experimentally determined either by a zero degree calorimeter, which 
measures the overall spectator energy ($E_{ZDC}$) and thus leads directly 
to the number of participant nucleons, or through the transverse
energy ($E_T$) or multiplicity of the produced hadrons. The results 
are shown as 
function of centrality as given by the average length $L$ of nuclear 
matter traversed, which can be  determined through a Glauber analysis 
of either $E_{ZDC}$ or $E_T$ data.

\medskip

In summary, we note the following behaviour:

\begin{itemize}
\vspace*{-0.2cm}
\item{The rates for \J~and \P~production in $p\!-\!A$ collisions are 
correctly accounted for by normal nuclear absorption.}
\vspace*{-0.2cm}
\item{\J~production in $S\!-\!U$ collisions at $\sqrt s=20$ GeV 
is compatible with normal nuclear absorption. 
\P~production in the same $S\!-\!U$ collisions, however, shows a further 
reduction. This is the first instance of an `anomalous' suppression, 
decreasing the normal rate by up to a factor five in the most central 
collisions.}
\vspace*{-0.2cm}
\item{\J~production in $Pb\!-\!Pb$ collisions at $\sqrt s \simeq 17$ GeV 
as function of centrality shows an onset of anomalous suppression. While 
peripheral results again agree with extrapolated $p-A$ results, beyond 
some centrality there is an additional reduction by some 20 - 40 \%.
The anomalous \P~suppression in $Pb\!-\!Pb$ interactions starts for more 
peripheral collisions than that for \J, and as function of the 
available medium it is compatible to the $S\!-\!U$ results.}
\vspace*{-0.2cm}
\end{itemize}


The onset of anomalous \J~suppression in $Pb\!-\!Pb$ collisions and its absence
in $S-U$ interactions suggested studying a lighter $A-\!A$ combination,
to obtain further information about this phenomenon. The NA60 collaboration 
has now presented first results on \J~production in $In-\!In$ collisions
\cite{arnaldi-qm}; they are shown in Fig.\ \ref{Ar2}, together with the
other \J~data taken at the SPS.

\begin{itemize}
\vspace*{-0.2cm}
\item{In $In\!-\!In$ collisions there is again an onset of anomalous 
\J~suppression, which here, however, appears to occur at a centrality at 
which the corresponding $S\!-\!U$ data do not show any such effect.
The $Pb\!-\!Pb$ data is within errors compatible with an earlier as 
well as a later onset position.}
\vspace*{-0.2cm}
\end{itemize}


Finally we note that also the transverse momentum behaviour of \J~production
in $p\!-\!A$ and $A\!-\!A$ collisions was studied at the 
SPS \cite{transv-Pb,transv-In}.

\begin{figure}[h]
\begin{minipage}[t]{7cm}
\epsfig{file=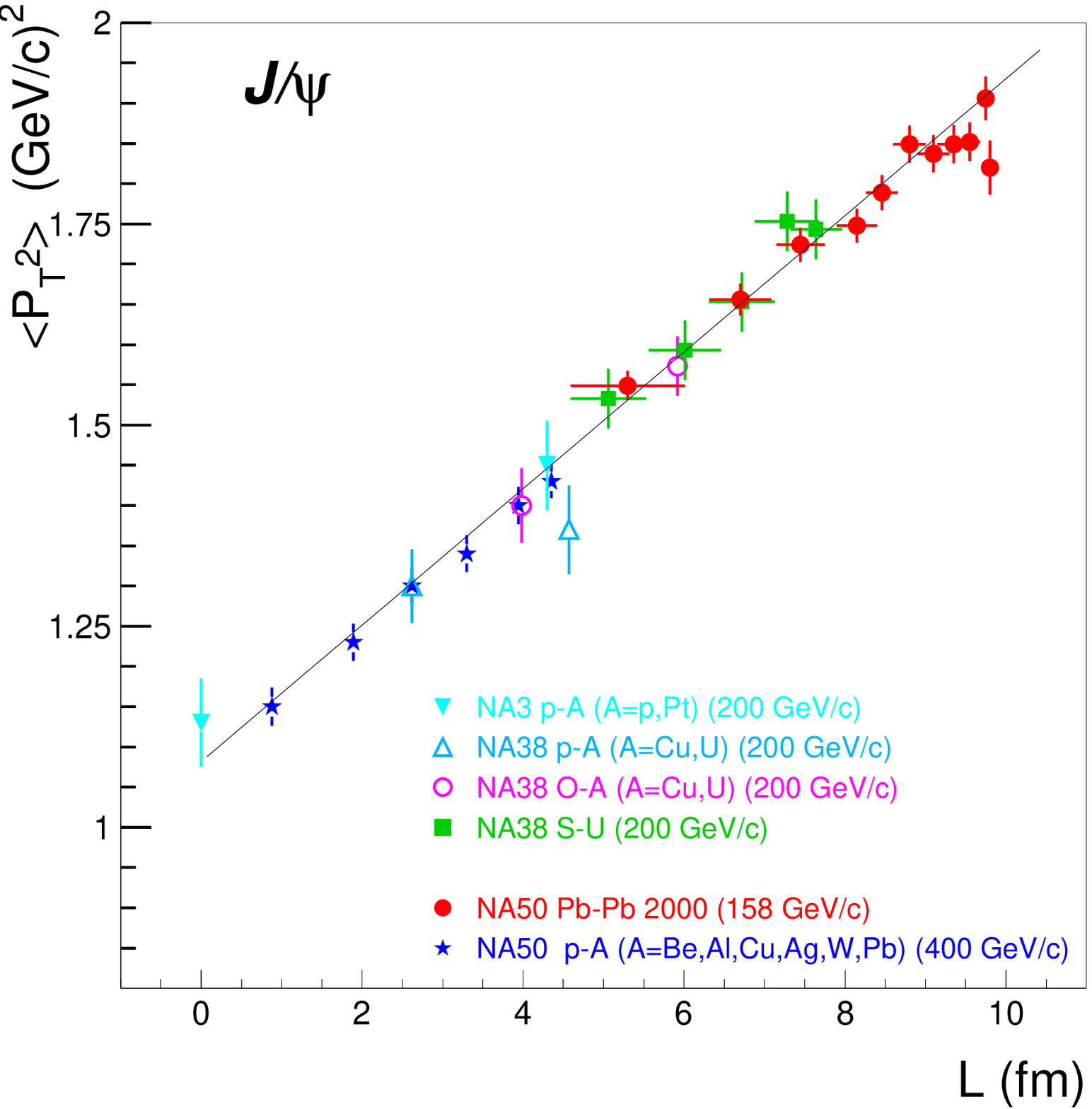,width=7.5cm,height=5.5cm}
\vspace*{-0.5cm}
\caption{Transverse momentum behaviour of \J~production \cite{transv-Pb}}
\label{PT}
\end{minipage}
\hspace{1.3cm}
\begin{minipage}[t]{7cm}
\vspace*{-4.85cm}
\hskip-0.3cm
\epsfig{file=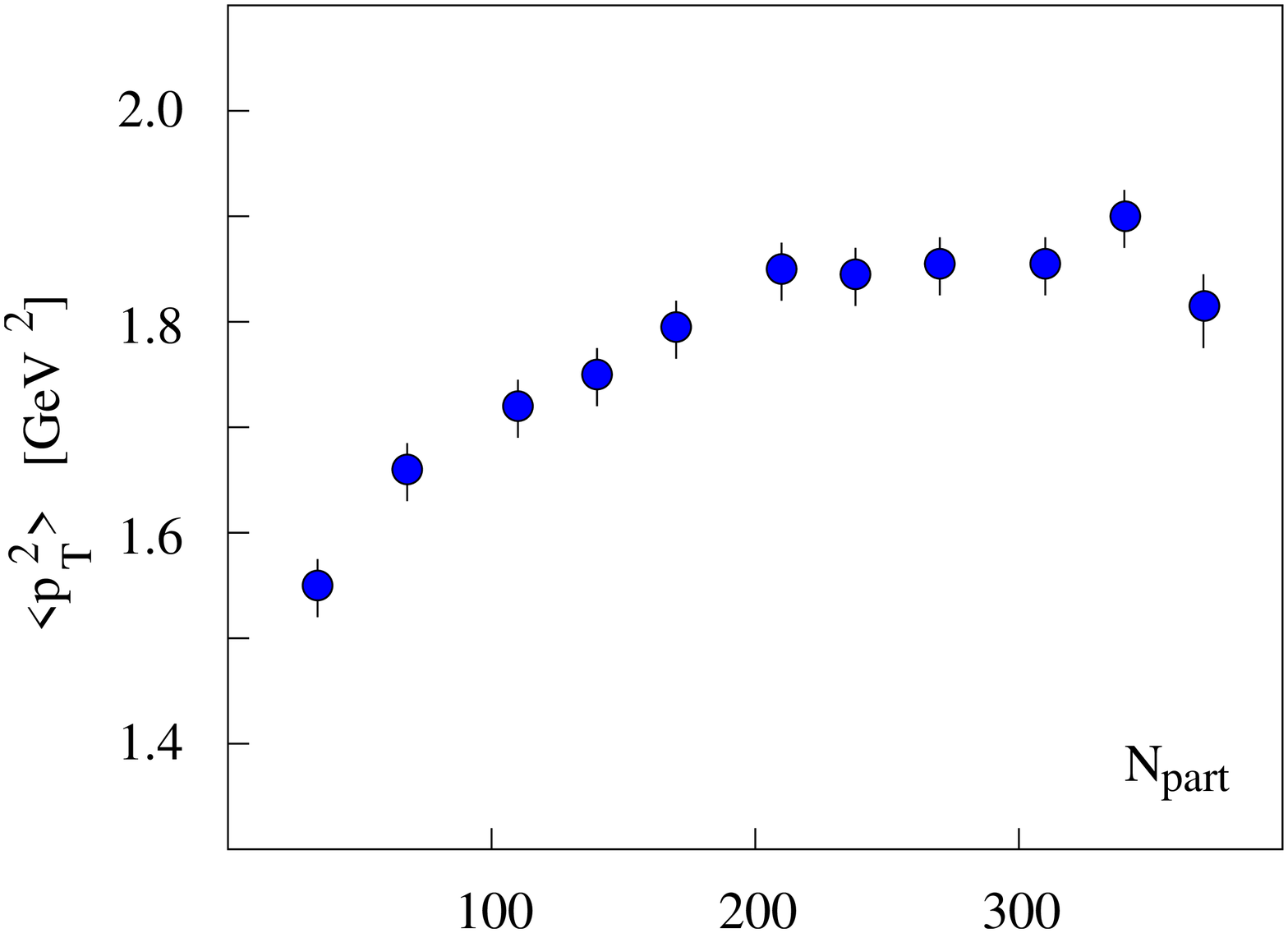,width=7.2cm,height=4.7cm}
\vspace*{0.2cm}
\caption{\J~transverse momentum behaviour in $Pb-Pb$ collisions 
\cite{transv-Pb}}
\label{PT-Pb}
\end{minipage}
\end{figure}

\begin{itemize}
\vspace*{-0.2cm}
\item{As shown in Fig.\ \ref{PT}, the average (squared) \J~transverse 
momentum increases with the amount of nuclear matter. A detailed 
view of the $Pb\!-\!Pb$ results in Fig.\ \ref{PT-Pb} suggests a reduced
broadening rate for more central collisions.}
\vspace*{-0.2cm}
\end{itemize}


\vspace*{-0.3cm}

\vskip0.5cm

\noindent{\bf 6.2 RHIC Results}

\bigskip

In the past months, the PHENIX collaboration at RHIC has presented 
\J~production results at $\sqrt s =200$ GeV for $d\!-\!Au,~Au\!-\!Au$ and 
$Cu\!-\!Cu$ collisions \cite{costa}. We should note that the increase 
of a factor ten in collision energy corresponds to an increase by a factor 
two to four in energy density, depending on the value of $\tau_0$, so that
the new data smoothly extend the regime studied so far. The data are
generally presented in the form of a nuclear modification factor
$R_{AA}$, giving the \J~production rate in $Au\!-\!Au$ collisions
relative to the $p\!-\!p$ rate scaled up by the number of binary collisions.

\medskip

\begin{figure}[htb]
\vspace*{0.2cm}
\centerline{\epsfig{file=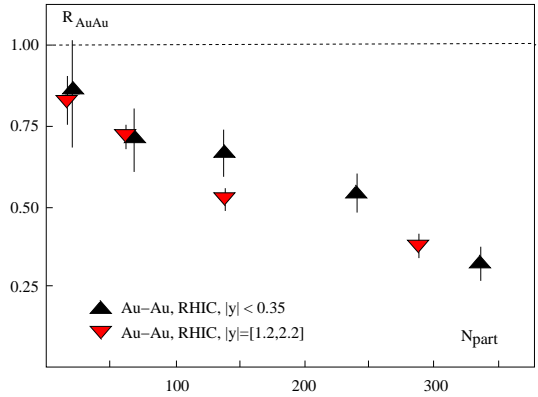,width=7cm}}
\caption{\J~production rates at RHIC \cite{costa}} 
\label{AArates}
\end{figure}

The results for $Au\!-\!Au$ collisions\footnote{We shall not consider the
results of the $Cu\!-\!Cu$ interactions here, since at present we do not 
have access to the corresponding Glauber studies needed for comparison.} 
at $\sqrt s = 200$ GeV are given for two rapidity ranges, $|y|\leq 0.35$ 
and $|y|\in [1.2,2.2]$, as was the case for the reference $d\!-\!Au$ 
collisions (Fig.\ \ref{d-Au}. The resulting rates $R_{AA}$ are shown 
in Fig.\ \ref{AArates}.

\medskip
 
\begin{figure}[h]
\begin{minipage}[t]{7cm}
\hskip0.2cm
\epsfig{file=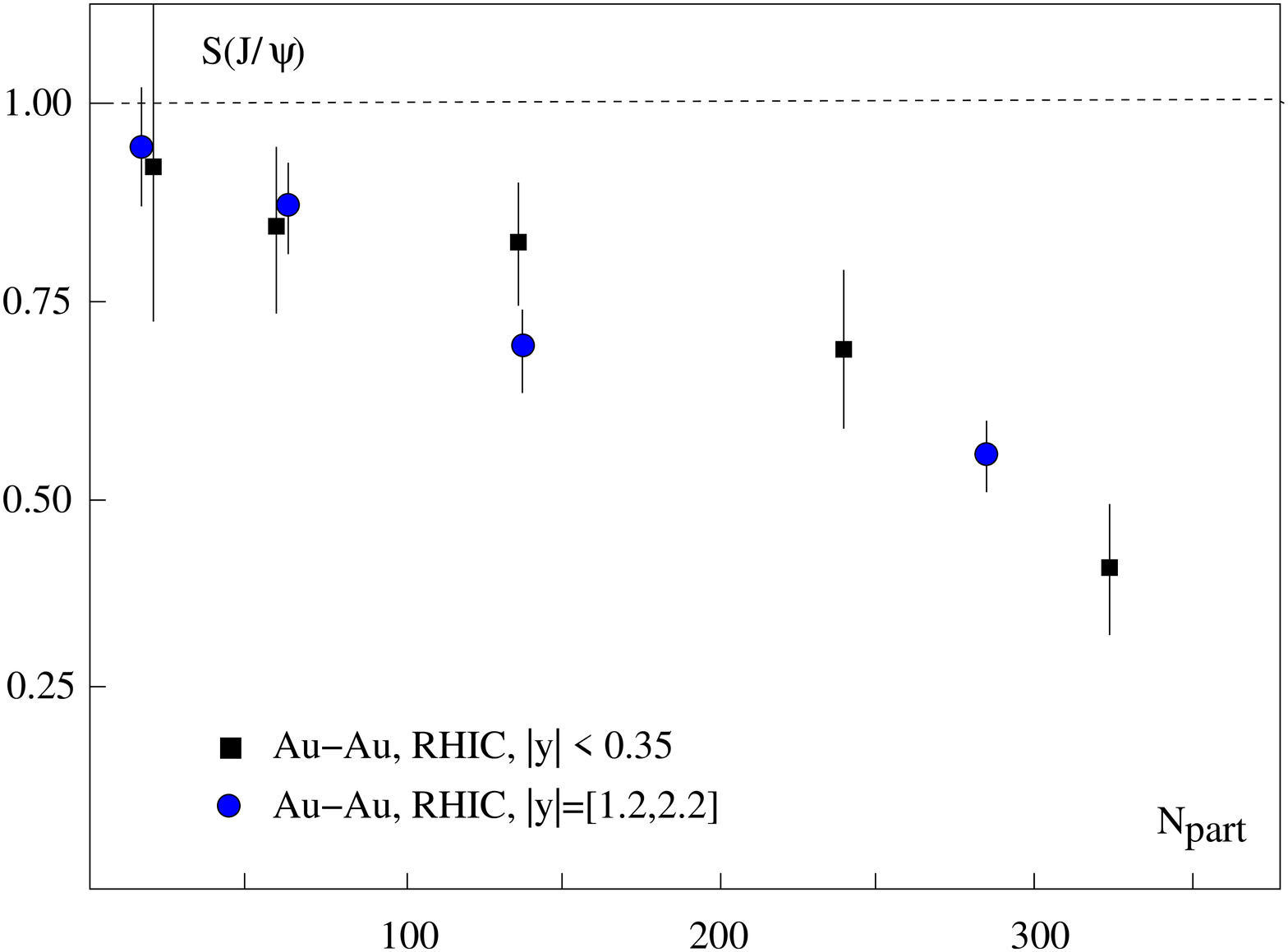,width=6.5cm}
\caption{\J~survival probability at RHIC}
\label{npart}
\end{minipage}
\hspace{1.3cm}
\begin{minipage}[t]{7cm}
\vspace*{-4.8cm}
\hskip0.3cm
\epsfig{file=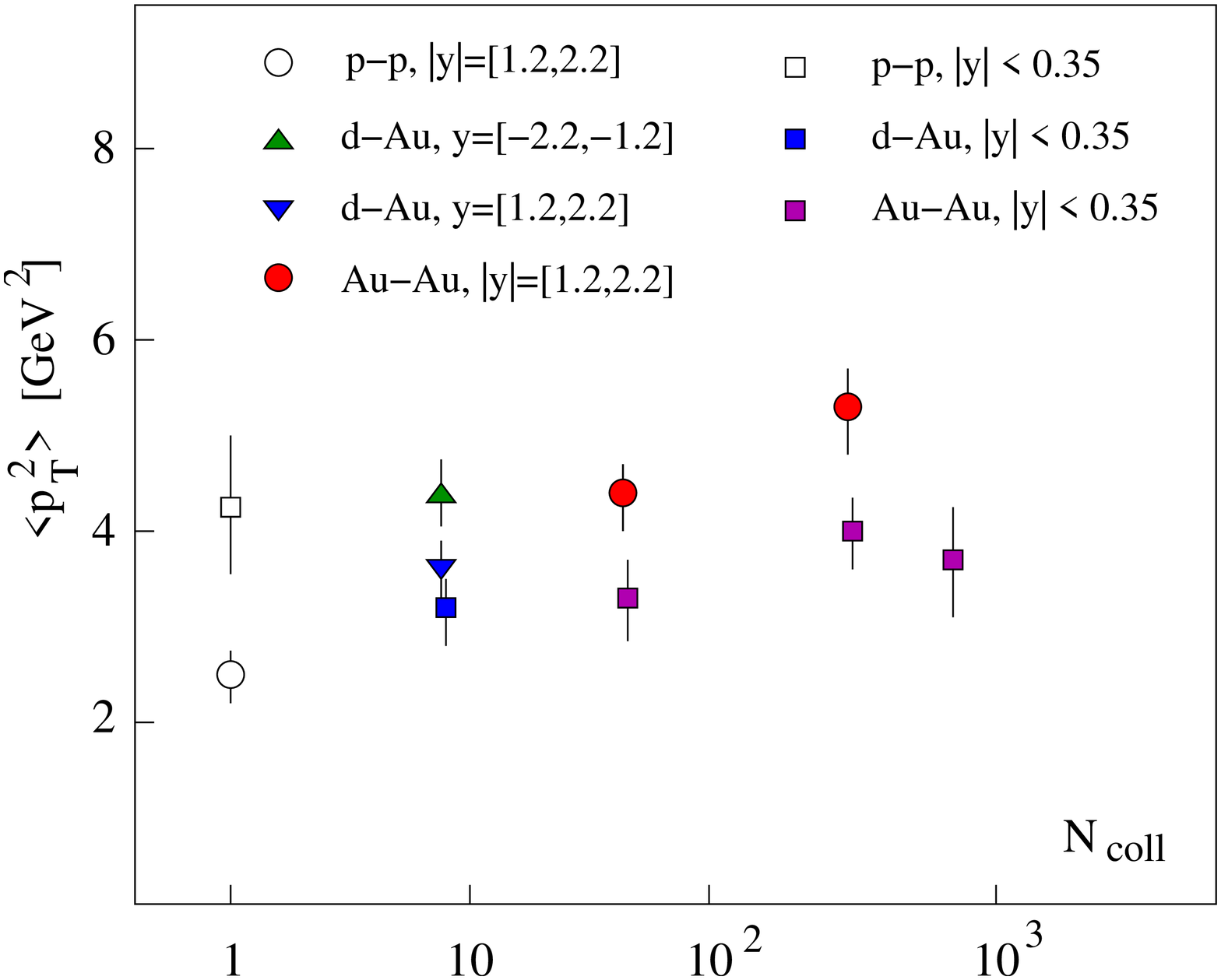,width=6.4cm,height=4.7cm}
\vspace*{0.1cm}
\caption{\J~transverse momentum behaviour at RHIC \cite{costa,Gunji}}
\label{pT-rhic}
\end{minipage}
\end{figure}

The corresponding survival probabilities can now be determined by eq.\ 
(\ref{S-AA}), making use of the nuclear medium modification through
the simplified Glauber form (\ref{path}). Since the resulting cross
sections are so far known with very limited accuracy, we do not include 
the additional error arising from $R_{dAu}$. In Fig.\ \ref{npart}, we 
show the \J~survival probability in $Au\!-\!Au$ collisions for the two 
rapidity intervals, as function of the number of participants $N_{\rm part}$. 
The suppression is seen to increase with centrality, up to a survival 
rate of 40 - 60\% remaining for the most central $Au\!-\!Au$ interactions. 

\medskip

Preliminary RHIC data on the transverse momentum behaviour of \J~production
have also been presented \cite{Gunji}. As seen in Fig.\ \ref{pT-rhic}, the
muon arm results are similar to those obtained at the SPS, with increasing
broadening from $p\!-\!p$ to $d\!-\!Au$ and on to $Au\!-\!Au$ as function 
of centrality. The $e^+e^-$ data have so far very limited statistics and, 
in particular, do not (yet?) show a broadening when going from $p\!-\!p$ 
to $d\!-\!Au$.

\bigskip

\noindent{\bf 6.3 What does it all mean?}

\bigskip

High energy nuclear collision data thus provide a considerable amount of
information on the in-medium behaviour of charmonium production. We recall
that the {\it raison d'\^etre} for the experimental programme was to study 
in the laboratory the deconfinement transition and the new deconfined state 
of matter predicted by statistical QCD. The crucial question in the present
context is thus whether the pattern of charmonium suppression observed in
heavy ion studies is in accord with that calculated in finite
temperature lattice QCD. Here we should keep in mind that it is not 
{\sl a priori} obvious that the medium produced in nuclear collisions
is in fact the equilibrium QGP studied in statistical QCD. A quantitative
relation between the charmonium suppression features in nuclear collisions
and those calculated in lattice QCD would thus provide excellent support
for QGP production. So where are we now? Does a combination of all the 
different aspects lead to some first conclusions? In this section, we want 
to address several issues that come up if we try to find a coherent 
interpretation. 

\medskip

We begin by recalling that if we want to compare data from different nuclear 
collisions and from different collision energies, we have to define some
more or less `universal' scale variable. We had shown SPS results as 
function of the effective path length $L$ through nuclear matter, since much 
of the data was presented that way. From an experimental point of view, 
$E_{ZDC}$ data gives the number of participants as the most directly 
measurable quantity, while from a theoretical point of view, density 
variables, such as participant density, energy density or parton density, 
are conceptually preferable. To compare different $A\!-\!B$ configurations 
at the same energy, the participant density $n_{part}$ is still meaningful, 
but for different $A\!-\!B$ at different collision energies, one has to 
resort to energy or parton densities in order to accommodate the resulting 
modifications.

\begin{figure}[h]
\begin{minipage}[t]{7cm}
\epsfig{file=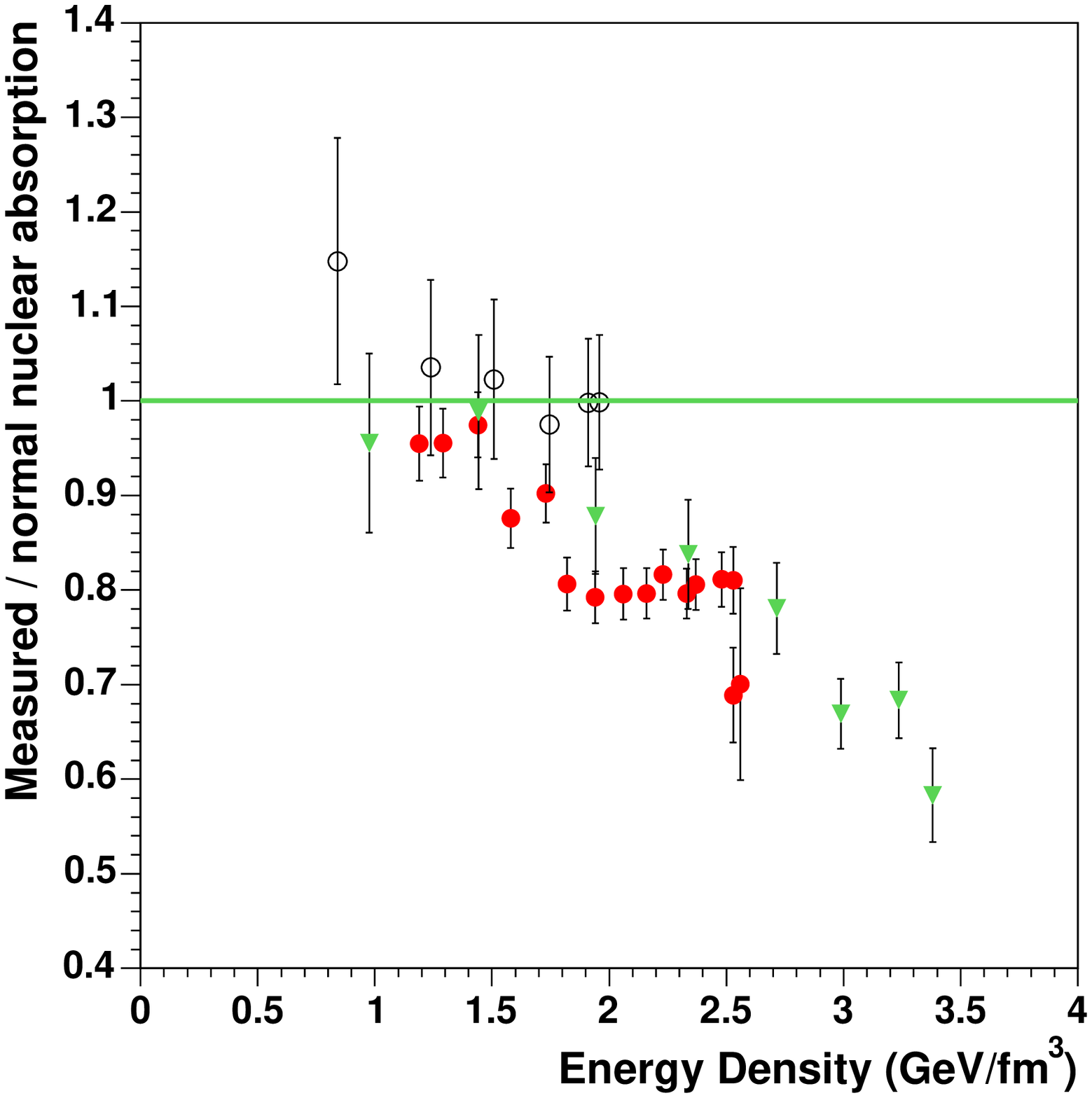,width=6.5cm}
\vspace*{-0.5cm}
\caption{\J~production in $In\!-\!In$, $Pb\!-\!Pb$ collisions ($E_{ZDC}$ data) 
and in $\!S-\!U$ collisions ($E_T$ data) \cite{arnaldi-qm}}
\label{Ar5}
\end{minipage}
\hspace{1.3cm}
\begin{minipage}[t]{7cm}
\vspace*{-6.25cm}
\hskip0.3cm
\epsfig{file=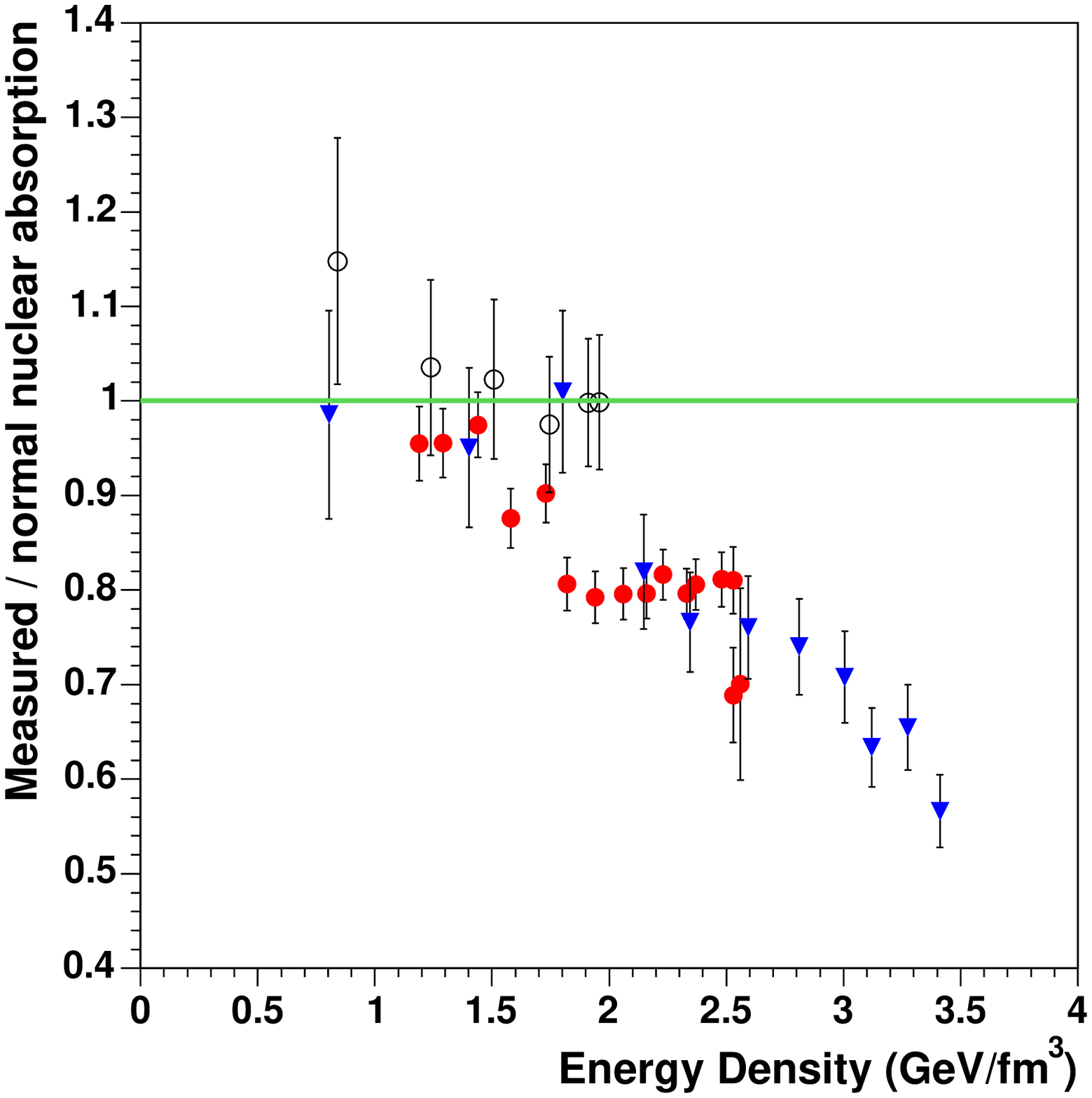,width=6.5cm}
\vspace*{-0.5cm}
\caption{\J~production in $Pb\!-\!Pb$, $S\!-\!U$ collisions ($E_T$ data) and in
$In\!-\!In$ collisions ($E_{ZDC}$ data) \cite{arnaldi-qm}}
\label{Ar3}
\end{minipage}
\end{figure}

\medskip

The first question that arises by looking at SPS results concerns the 
onset of anomalous \J~suppression in $S\!-\!U$, $Pb\!-\!Pb$ and $In\!-\!In$ 
collisions. Do the data as shown in Fig.\ \ref{Ar2} imply different 
onsets in the different reactions? In Figs.\ \ref{Ar5} and \ref{Ar3} we see 
that the discrepancy between $S\!-\!U$ and $In\!-\!In$ remains, though less 
pronounced, when we determine the centrality through the energy density. We 
also note, however, that different experimental methods of determining the 
centrality can lead to slight differences. 
The onset of anomalous suppression in the $Pb\!-\!Pb$ data 
agrees very well with that found in the $In\!-\!In$ data when both are 
shown in terms of the $E_{ZDC}$ determined centrality (Fig.\ \ref{Ar5}), 
while an apparently later $Pb\!-\!Pb$ onset, in accord with the $S\!-\!U$ 
behaviour, is found for an $E_T$-based analysis (Fig.\ \ref{Ar3}). Finally
we also note that a 10\% shift of the $S\!-\!U$ normalization would remove 
all differences. Further study here would thus seem very appropriate, with
the aim of comparing only data in which the centrality determination
is carried out in the same way (ideally, $E_{ZDC}$ determined centrality) 
for different collision configurations and for both \J~and \P.

\begin{figure}[htb]
\vspace*{0.2cm}
\centerline{\epsfig{file=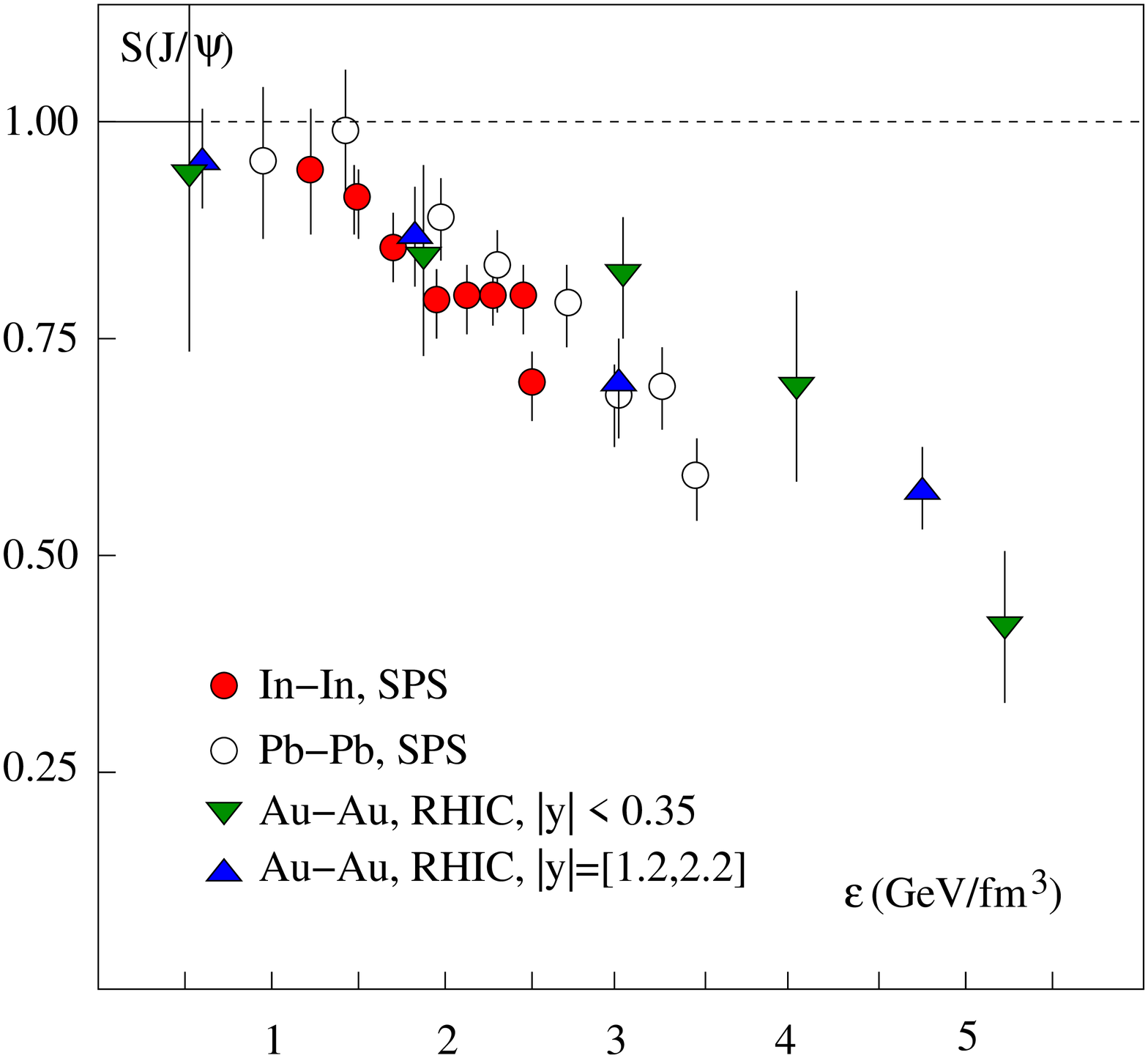,width=8cm,height=6.5cm}}
\caption{\J~suppression as function of energy density}
\label{in-rhic}
\end{figure}

\medskip

In Fig.\ \ref{in-rhic}, we show the \J~survival probability $S_{\j}$
for a combination of SPS and RHIC data as function of the energy density,
with all data determined through $E_{ZDC}$ measurements. For the RHIC data, 
the energy density values for the different centralities are obtained 
by a Glauber analysis based directly on the PHENIX data, with $\tau_0=1$ fm 
\cite{phenix-bj}.

\medskip

The suppression pattern shown in Fig.\ \ref{in-rhic} is fairly
consistent with that obtained in chapter 4 from finite temperature 
lattice QCD. We had found there a suppression of \X~and \P~just
above $T_c$, which implies around $\e \simeq 1$ GeV/fm$^3$, while the
\J(1S) should survive up to $\e\simeq 10$ GeV/fm$^3$ or more. For
\J~production in nuclear collisions, this means a suppression of
the 40\% coming from \X~and \P~decay starting around 1 GeV/fm$^3$,
while the 60\% directly produced \J~remain unaffected until much
higher $\e$. The onset of the anomalous suppression in Fig.\ \ref{in-rhic} 
occurs indeed at the expected energy density, and the \J~survival 
probability converges towards some 50 - 60\%. Before considering the 
possibility of a next suppression onset at large $\e$, we should 
recall that the energy density values for the RHIC data were obtained 
with a formation time $\tau_0=1$ fm. As mentioned, $\tau_0$ could 
well be smaller (parton densities suggest $\tau_0=0.5$ fm), which 
would move the RHIC points to correspondingly higher $\e$. However, 
given the present statistics, an interpretation of the large $\e$ 
behaviour as onset would in any case seem premature. 

\begin{figure}[htb]
\vskip0.5cm
\centerline{\epsfig{file=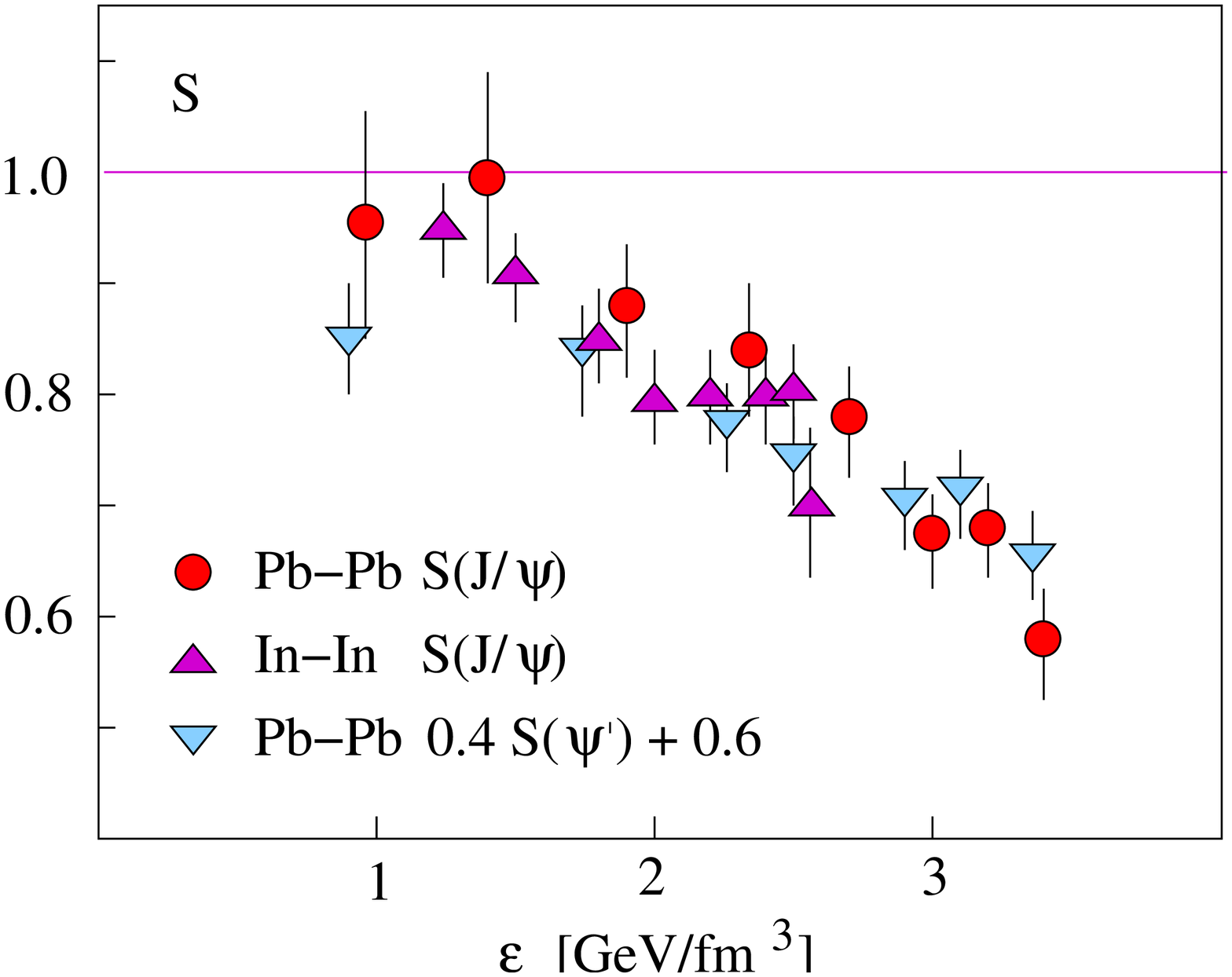,width=7cm}}
\caption{Universal \J~and \P~suppression \cite{KKS}}
\label{scaling}
\end{figure}

\medskip

If the interpretation in terms of sequential suppression is correct, 
the data for \P~suppression can be used to form $0.4~S_{\p} + 0.6$, and 
this should then coincide with the measured \J~data \cite{KKS}. We 
show in Fig.\ \ref{scaling} the SPS \J~data from $Pb\!-\!Pb$ and $In\!-\!In$
collisions together with the mentioned \P~form. Within the 
present errors, this universal scaling relation seems to hold quite well.
Note that here the centrality of the \J~data is determined by $E_{ZDC}$,
that for the \P~by $E_T$. It is obviously very important to obtain also the
corresponding \P~data with $E_{ZDC}$ as the centrality variable.

\medskip

The sequential suppression pattern expected from statistical QCD means 
that in most of the RHIC data as well as in the central SPS data, what 
is seen should be the surviving directly produced \J(1S). This also has 
consequences on the transverse momentum behaviour, which thus provide 
a way to check the conclusion \cite{KKS}. The main effect leading to a 
broadening of transverse momentum spectra in nuclear interactions is the 
collision broadening of the incident gluons which fuse to make $\C$ pairs. 
A standard random walk analysis gives for the squared transverse momentum 
of the produced \J~\cite{K-N-S,early}
\be
\langle p_T^2 \rangle_{pA} = \langle p_T^2 \rangle_{pp} + N_c^A \delta_0
\label{pTpA}
\ee
in $p\!-\!A$ and to
\be
\langle p_T^2 \rangle_{AA} = \langle p_T^2 \rangle_{pp} + N_c^{AA} \delta_0
\label{pTAA}
\ee
in $A\!-\!A$ collisions. 
Here $N_c^A$ denotes the average number of collisions of a projectile parton 
in the target nucleus $A$, and $N_c^{AA}$ the sum of the average number of 
collisions of a projectile parton in the target and vice versa, at the given 
centrality. The parameter $\delta_0$ specifies the average ``kick'' which 
the incident parton receives in each subsequent collision. 
The crucial parameters are thus the elementary $\langle p_T^2 \rangle_{pp}$ 
from $p\!-\!p$ interactions and the value of $\delta_0$, determined by 
corresponding $p\!-\!A$ data; both depend on the collision energy. 
The $A$-dependence of $N_c^A$ as well as the behaviour of $N_c^{AA}$ as 
function of centrality can be obtained through a Glauber analysis, thus
specifying the ``normal'' centrality dependence of 
$\langle p_T^2 \rangle_{AA}$. 

\medskip

In $p\!-\!A$ interactions, the average number of collisions of the 
projectile nucleon in the nuclear target can be estimated as 
\cite{K-N-S,early} 
\be
N_c \simeq (3/4)~\!2\!~\!R_A\! ~\!n_0\!~\! \sigma,   
\label{partoncol}
\ee
where $\sigma$ denotes the nucleon-nucleon cross section and the factor 
(3/4) accounts for the (spherical) nuclear geometry. If the parton fusion 
leading to $\C$ production occurs on the average in the center of the target 
nucleus, $N_c^A \simeq (N_c-1)/2$. However, the presence of normal nuclear 
absorption of the charmonium shifts the fusion point further 
``down-stream'', increasing the parton path and hence the number of 
collisions in the target up to a maximum of $N_c^A \simeq N_c-1$.
A systematic analysis would require Glauber calculations making use 
of the absorption cross sections measured at the corresponding energies. 
The variation of $\sigma$ and of the charmonium absorption cross sections 
\cite{K-N-S,early} with energy lead to an estimate $N_c^A \simeq 3 \pm 1$. 

\medskip

\begin{figure}[htb]
\vskip0.5cm
\centerline{\epsfig{file=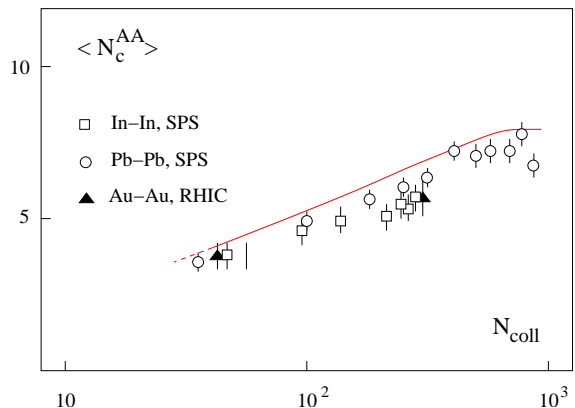,width=7.5cm}}
\caption{Transverse momentum broadening at SPS and RHIC \cite{KKS}}
\label{pTcurve}
\end{figure}

If RHIC as well as central SPS data basically show the surviving direct 
\J, then one should observe the $p_T$ behaviour as given by eqs.\ (\ref{pTpA}) 
and (\ref{pTAA}). This again leads to a relation between the different 
data sets from SPS \cite{transv-Pb,transv-In} and RHIC \cite{Gunji}. From 
the RHIC $\mu^+\mu^-$ data, we obtain $<p_T^2>_{pp} = 2.51 \pm 0.21$ GeV$^2$
and $<p_T^2>_{dAu} = 3.96 \pm 0.28$ GeV$^2$; for the latter value, we 
have taken the average of the positive and negative rapidity ranges. 
With $N_c^A \simeq 3$, we then get

\be
\delta_0^{RHIC} \simeq 0.5~{\rm GeV}^2
\label{d-rhic}
\ee
for the parton broadening at RHIC as shown in the large rapidity data.
We note here that each kinematic data set requires a separate analysis,
which for the muon arm data appears possible. The central $e^+e^-$ data
so far give (with large errors) a $\langle p_T^2 \rangle$ which for $p\!-\!p$
is larger than that for $p\!-\!A$, so that here more statistics are
needed. -- At SPS energy, we have \cite{Badier}
$\langle p_T^2 \rangle_{pp} = 1.25 
\pm 0.05$  GeV$^2$ and $\langle p_T^2 \rangle_{pU} = 1.49 \pm 0.05$ GeV$^2$.
Using again $N_c^A \simeq 3$, we get
\be
\delta_0^{SPS} \simeq 0.08~{\rm GeV}^2
\label{d-sps}
\ee
for the corresponding parton broadening. With all parameters thus determined,
we obtain as a characteristic of the transverse momentum broadening the 
average number of collisions of a projectile and a target nucleon in an $AA$ 
collisions at the given centrality,
\be
\langle N_c^{AA} \rangle = \{\langle p_T^2 \rangle_{AA} - 
\langle  p_T^2 \rangle_{pp}\} / \delta_0.
\label{pTmeasure}
\ee
If our interpretation of sequential suppression is correct and the
observed $\langle p_T^2 \rangle$ due to the unaffected direct \J's, 
then the SPS and the RHIC muon data should follow a common curve. 
We see in Fig. \ref{pTcurve} that this is in fact the case. Once the
corresponding broadening pattern for the RHIC electron data is determined,
it should also follow this pattern. -- The normal centrality dependence of 
$\langle N_c^{AA} \rangle$ has also been calculated directly in a Glauber 
analysis based on the measured normal nuclear suppression
\cite{K-N-S}; the resulting curve is included in Fig.\ \ref{pTcurve}. --
Contributrions from \J~formation through exogamous $\C$ recombination
should weaken the increase with centrality for RHIC data compared 
to that from the SPS \cite{bob-recom}.

\medskip

We thus find that the overall view of the charmonium data today is more than
before in accord with the results from statistical QCD. The crucial
elements for this are 
\begin{itemize}
\vspace*{-0.2cm}
\item{the common onset of anomalous suppression for \J~and \P~at the
dissociation value $\e \simeq 1$ GeV/fm$^3$ expected for higher excited
charmonium states; and}
\vspace*{-0.2cm}
\item{the apparent suppression saturation at about 50 -- 70\%, following
the initial onset, suggesting a survival of directly produced \J's up
to much higher energy densities.}
\vspace*{-0.6cm} 
\end{itemize}
The first observation revises prior experimental indications for distinct 
onsets for \P~and \J~suppression, at $\e \simeq$ 1 GeV/fm$^3$ and 2.5 
GeV/fm$^3$, respectively. The second observation becomes meaningful because 
of the revision of the theoretical claim of a direct \J~suppression just 
slightly above $T_c$.

\medskip

It seems clear what is further needed to make these indications conclusive.
\begin{itemize}
\vspace*{-0.2cm}
\item{More precise \P~data at SPS energy (and with $E_{ZDC}$ determined
centrality) could substantiate the scaling behaviour of Fig.\ \ref{scaling}.}
\vspace*{-0.2cm}
\item{More precise \J~data in the ``saturation'' regime $\e \geq 3$ GeV/fm$^3$
could substantiate the direct (1S) survival rate ($\sim$ 60\%).}
\vspace*{-0.2cm}
\item{An eventual further suppression onset at much higher $\e$ (LHC?)
could
confirm the sequential suppression pattern predicted by statistical QCD.}  
\vspace*{-0.2cm}
\item{Finally, corresponding studies of the bottomonium spectrum with
its different suppression onsets could provide decisive substantiation.}
\end{itemize}

Given the present state of the data, any conclusions must necessarily remain
tentative or even speculative. How could the picture just presented be
experimentally falsified, and what would be a possible alternative account?
The most frequently considered alternative is based on the possibility of
\J~formation by ``recombination'', i.e., by the binding of $c$ and $\bar c$
quarks originating from different nucleon-nucleon collisions. This 
approach \cite{enhance} assumes two distinct mechanisms affecting 
\J~production in nuclear collisions \cite{rapp}:  
\begin{itemize}
\vspace*{-0.2cm}
\item{A charmonium state produced in a given nucleon-nucleon interaction
can be dissociated either by comover collisions or by screening
effects in the hot medium. This leads to the decreasing survival
probability which was labelled ``thermal dissociation'' in Fig.\ \ref{recom}.}
\vspace*{-0.2cm}
\item{Any two $c$ and $\bar c$ in the resulting medium, whatever their
origin, can then at hadronization combine to form a charmonium state.
An estimate of this recombination possibility leads in Fig.\ \ref{recom}
to the curve labelled ``statistical recombination''.}
\vspace*{-0.7cm}
\item{Since the charm production rate as a hard process grows faster than 
the rate for light quarks, the recombination probability will increase with 
centrality as well as with collision energy. This will eventually lead
in $A\!-\!A$ collisions to \J~production rates, relative to overall charm
production, which are larger than those in $p\!-\!p$ collisions.}
\vspace*{-0.2cm}
\end{itemize}
At present RHIC energy and centralities, the decreasing rate for ``direct''
production (i.e., in a given nucleon-nucleon collision), due to thermal
dissociation, is taken to be just compensated by the increasing rate of 
charmonia formed by combination of charm quarks from different 
nucleon-nucleon collisions, thus leading to an approximately constant 
``sa\-tu\-ration'' regime, as seen in Figs.\ \ref{npart} or \ref{in-rhic}.

\medskip

We should here note two specific aspects inherent in this approach. The
assumed form for the decreasing thermal dissociation pattern is, as
far as we know today (see section 4.) not that obtained for a QGP 
in statistical QCD, with its survival of \J(1S) up to much higher 
temperatures; instead, it is a phenomenological extrapolation of SPS
data. As mentioned, it is not at all obvious that nuclear
collisions do lead to the equilibrated QGP studied in lattice QCD,
and it is also not obvious that even in such a medium, collision effects
on a short-time scale could not modify the purely thermal behaviour found
there (see the remarks at the end of sections 4.3 and 4.4). Nevertheless,
the form used at present in the recombination approach \cite{rapp} has no 
direct quantitative QCD basis. -- Secondly,
we had seen that the survival probability appears to converge to roughly
the fraction (60\%) of the remaining directly produced \J's. This 
certainly needs to be checked in more detail, but, if it were true, 
it would need to find some explanation in a recombination approach. 

\medskip

The ultimate tests to falsify one or the other approach seem in any case
quite clear. The most characteristic feature of the recombination approach
is an eventual {\sl increase} of the survival probability with centrality.
If observed, this would rule out a direct relation between the experimental 
pattern and that obtained in statistical QCD. On the other hand,
if higher centralities lead to {\sl further suppression}, and if moreover
a similar pattern is found for bottomonium production, this would
seem to rule out recombination of exogamous pairs as significant charmonium
formation process. It thus appears that the LHC, with its higher
energy density access, will still play a decisive role in obtaining
the final answer to the use of quarkonium spectra as probes of the 
strongly interacting matter produced in nuclear collisions.

\vskip0.8cm

\noindent{\large \bf 7.\ Outlook}

\vskip0.5cm

The theoretical analysis of the in-medium behaviour of quarkonia has 
greatly advanced in the past decade. Potential model studies based on
lattice results for the colour-singlet free energy as well as direct
lattice calculations appear to be converging, and within a few more
years, the dissociation temperatures for the different quarkonium
states will be calculated precisely. Through corresponding calculations
of the QCD equation of state, these temperatures provide the energy
density values at which the dissociation occurs. In statistical QCD,
quarkonia thus allow a spectral analysis of the quark-gluon plasma.

\medskip

The application of this analysis in high energy nuclear collisions is
so far less conclusive. There exists a wealth of data from different
interactions at the SPS, and first RHIC results are now also available.
These results indicate the production of a hot, dense, strongly 
interacting medium. It is not yet clear, however, to what extent this
medium is indeed the quark-gluon plasma studied in statistical QCD. Is
it possible to identify dissociation onsets for different charmonium
states? Does the initial (non-thermal) overabundance of charm survive a
subsequent thermalization? Is the apparent \J~suppression saturation
from SPS up to fairly central RHIC collisions the survival of the
directly produced \J's as seen in lattice QCD studies, or is it the
production of additional \J's through exogamous $\C$ pairings?

\medskip

It is undoubtedly very challenging to construct models incorporating
as many of the observed features as possible. Nevertheless it seems
worthwhile to note that measurements of the relative dissociation  
points of the different quarkonium states might be our only chance to 
compare quantitative {\sl ab initio} QCD predictions directly to data. 

\vskip1cm

\centerline{\large \bf Acknowledgements}

\bigskip

In the preparation of this report, I have received stimulation and support
from more colleagues than I can possibly list; but I am certainly very
grateful to all of them. Particular thanks go to R.\ Arnaldi, S.\ Digal, 
F.\ Karsch, D.\ Kharzeev, C.\ Louren{\c c}o, M.\ Nardi, R.\ L.\ Thews and
R.\ Vogt, whose help was really essential.

\end{document}